\documentclass[pre, twocolumn, showpacs, reprint, nofootinbib, superscriptaddress, floatfix]{revtex4-2}
\usepackage{amsfonts}
\usepackage{amssymb}
\usepackage{amsmath}
\usepackage{graphics}
\usepackage{graphicx}
\usepackage{soul}
\usepackage{natbib}
\usepackage{bm}
\usepackage{mathtools}
\usepackage{stmaryrd}
\usepackage{epstopdf}
\usepackage{color}
\usepackage{lipsum}
\usepackage{balance}
\usepackage{enumitem}
\usepackage{amsmath}
\usepackage{verbatim}
\usepackage{empheq}
\usepackage{mathrsfs}

\usepackage[colorlinks=true,citecolor=black,linkcolor=black,urlcolor=black,filecolor=black]{hyperref}
\usepackage{orcidlink}
 
\usepackage{acronym}
\newacro{PDF}{probability distribution function}
\newacroplural{PDF}[PDFs]{probability distribution functions}
\newcommand{\PDF}{\ac{PDF}}

\newacro{DF}{distribution function}
\newacroplural{DF}[DFs]{distribution functions}
\newcommand{\DF}{\ac{DF}}

\newacro{LDP}{Large Deviation Principle}
\newcommand{\LDP}{\ac{LDP}}
\newacro{LD}{Large Deviation}

\newacro{LDT}{Large Deviation Theory}
\newcommand{\LDT}{\ac{LDT}}
\newacro{LDK}{Landau--Dean--Kawasaki}
\newcommand{\LDK}{\ac{LDK}}
\newacro{SPDE}{Stochastic Partial Differential Equation}
\newcommand{\SPDE}{\ac{SPDE}}
\newacro{PDE}{Partial Differential Equation}
\newcommand{\PDE}{\ac{PDE}}

\newcommand{\p}{\partial}

\newcommand{\re}{\mathrm{e}}
\newcommand{\rd}{\mathrm{d}}
\newcommand{\ri}{\mathrm{i}}
\newcommand{\rL}{\mathrm{L}}
\newcommand{\rNL}{\mathrm{NL}}
\newcommand{\rS}{\mathrm{S}}
\newcommand{\rTr}{\mathrm{Tr}}

\newcommand{\bv}{\mathbf{v}}
\newcommand{\bvp}{\mathbf{v}^{\prime}}
\newcommand{\bT}{\boldsymbol{\theta}}
\newcommand{\bTp}{\boldsymbol{\theta}^{\prime}}
\newcommand{\bJ}{\mathbf{J}}
\newcommand{\bk}{\mathbf{k}}
\newcommand{\deltaD}{\delta_{\mathrm{D}}}
\newcommand{\bw}{\mathbf{w}}
\newcommand{\Mtot}{M_{\mathrm{tot}}}
\newcommand{\half}{\tfrac{1}{2}}
\newcommand{\rT}{\mathrm{T}}
\newcommand{\mS}{\mathcal{S}}
\newcommand{\mF}{\mathcal{F}}
\newcommand{\bW}{\mathbf{W}}
\newcommand{\mO}{\mathcal{O}}
\newcommand{\bvpp}{\bv^{\prime \prime}}
\newcommand{\ba}{\mathbf{a}}
\newcommand{\bu}{\mathbf{u}}
\newcommand{\bI}{\mathbf{I}}
\newcommand{\bxi}{\boldsymbol{\xi}}
\newcommand{\bbV}{\mathbb{V}}
\newcommand{\bbP}{\mathbb{P}}
\newcommand{\sB}{\mathsf{B}}

\newcommand{\omL}{\overline{\mathcal{L}}}
\newcommand{\mLL}{\mathcal{L}_{\mathrm{L}}}
\newcommand{\sD}{\mathsf{D}}
\newcommand{\kmin}{k_{\mathrm{min}}}
\newcommand{\kmax}{k_{\mathrm{max}}}

\newcommand{\mL}{\mathcal{L}}
\newcommand{\mN}{\mathcal{N}}
\newcommand{\mQ}{\mathcal{Q}}
\newcommand{\mK}{\mathcal{K}}
\newcommand{\Tdyn}{t_{\mathrm{dyn}}}
\newcommand{\Trelax}{t_{\mathrm{relax}}}
\newcommand{\Tfin}{t_{\mathrm{fin}}}
\newcommand{\mW}{\mathcal{W}}
\newcommand{\bZero}{\mathbf{0}}
\newcommand{\sP}{\mathsf{P}}
\newcommand{\sZero}{\mathsf{0}}
\newcommand{\sI}{\mathsf{I}}
\newcommand{\Ekin}{E_{\mathrm{kin}}}
\newcommand{\bPtot}{\mathbf{P}_{\mathrm{tot}}}
\newcommand{\DT}{\texttt{DT}}
\newcommand{\NV}{\texttt{NV}}
\newcommand{\DV}{\texttt{DV}}
\newcommand{\VMAX}{\texttt{VMAX}}
\newcommand{\sG}{\mathsf{G}}
\newcommand{\bFL}{\mathbf{F}_{\mathrm{L}}}
\newcommand{\ip}{\mathbf{i}^{\prime}}
\newcommand{\jp}{\mathbf{j}^{\prime}}
\newcommand{\kp}{\mathbf{k}^{\prime}}
\newcommand{\perm}{\mathfrak {S}}
\newcommand{\bbM}{\mathbb{M}}
\newcommand{\barF}{\overline{F}}
\newcommand{\barC}{\overline{C}}
\newcommand{\sDL}{\mathsf{D}_{\mathrm{L}}}
\newcommand{\mA}{\mathcal{A}}
\newcommand{\mG}{\mathcal{G}}
\newcommand{\tPhi}{\widetilde{\Phi}}
\newcommand{\tbbV}{\widetilde{\mathbb{V}}}
\newcommand{\tbbM}{\widetilde{\mathbb{M}}}
\newcommand{\tchi}{\widetilde{\chi}}

\newcommand{\deltaf}{\delta f}
\newcommand{\deltaF}{\delta F}
\newcommand{\sA}{\mathsf{A}}
\newcommand{\sC}{\mathsf{C}}
\newcommand{\bF}{\mathbf{F}}
\newcommand{\bphi}{\boldsymbol{\phi}}
\newcommand{\bpsi}{\boldsymbol{\psi}}

\newcommand{\SDK}{S^{\mathrm{LDK}}}
\newcommand{\SNB}{S^{\mathrm{NB}}}
\newcommand{\MDK}{M^{\mathrm{LDK}}}

\newcommand{\Smin}{S^{\mathrm{M}}}
\newcommand{\ST}{S^{\mathrm{T}}}

\newcommand{\DTref}{\texttt{DT}_{\mathrm{ref}}}
\newcommand{\NVref}{\texttt{NV}_{\mathrm{ref}}}
\newcommand{\KNB}{K^{\mathrm{NB}}}

\newcommand{\1}{\ensuremath{\mathbf{1}}}
\newcommand{\2}{\ensuremath{\mathbf{2}}}
\newcommand{\3}{\ensuremath{\mathbf{3}}}
\newcommand{\4}{\ensuremath{\mathbf{4}}}
\newcommand{\5}{\ensuremath{\mathbf{5}}}
\newcommand{\6}{\ensuremath{\mathbf{6}}}

\providecommand{\a}{}
\renewcommand{\a}{\ensuremath{\mathbf{a}}}
\providecommand{\b}{}
\renewcommand{\b}{\ensuremath{\mathbf{b}}}
\providecommand{\c}{}
\renewcommand{\c}{\ensuremath{\mathbf{c}}}
\providecommand{\d}{}
\renewcommand{\d}{\ensuremath{\mathbf{d}}}
\providecommand{\e}{}
\renewcommand{\e}{\ensuremath{\mathbf{e}}}
\let\ioriginal\i
\providecommand{\dli}{}
\renewcommand{\dli}{\textbf{\ioriginal}}
\providecommand{\i}{}
\renewcommand{\i}{\ensuremath{\mathbf{i}}}
\providecommand{\j}{}
\renewcommand{\j}{\ensuremath{\mathbf{j}}}
\providecommand{\k}{}
\renewcommand{\k}{\ensuremath{\mathbf{k}}}
\providecommand{\l}{}
\renewcommand{\l}{\ensuremath{\mathbf{l}}}
\providecommand{\m}{}
\renewcommand{\m}{\ensuremath{\mathbf{m}}}
\providecommand{\n}{}
\renewcommand{\n}{\ensuremath{\mathbf{n}}}

\usepackage{soul}
\usepackage{xcolor}

\begin{document}
 
\title{A deterministic theory of kinetic fluctuations in long-range interacting homogeneous systems without collective effects}
 
\author{Anwar El Rhirhayi\,\orcidlink{0009-0001-0405-4730}}
\affiliation{Institut Denis Poisson, Universit\'{e} d'Orl\'{e}ans, CNRS Universit\'{e} de Tours, Orl\'{e}ans, France}
\affiliation{Institut d'Astrophysique de Paris, UMR 7095, 98 bis Boulevard Arago, F-75014 Paris, France}
 
\author{Jean-Baptiste Fouvry\,\orcidlink{0000-0002-0030-371X}}
\affiliation{Institut d'Astrophysique de Paris, UMR 7095, 98 bis Boulevard Arago, F-75014 Paris, France}
 
\author{Julien Barr\'{e}\,\orcidlink{0000-0002-7460-8524}}
\affiliation{Institut Denis Poisson, Universit\'{e} d'Orl\'{e}ans, CNRS Universit\'{e} de Tours, Orl\'{e}ans, France}
 
\begin{abstract}
Finite-$N$ fluctuations govern the slow relaxation of long-range interacting systems, yet accessing their full statistics demands huge ensembles of costly simulations.
We show that, in homogeneous systems without collective effects, this statistical problem collapses onto just three deterministic kinetic equations.
Starting from the Landau--Dean--Kawasaki equation, i.e.\ the stochastic counterpart of the large deviation principle for the Landau equation, we derive the infinite hierarchy of $n$-point moments of the empirical distribution function at leading order in $1/N$.
We prove that it is solved exactly by a quasi-Gaussian closure.
The resulting minimal Landau fluctuation hierarchy is formed by the Landau equation, a Lyapunov equation for the two-point moment, and a kinetic equation for the three-point moment.
From these three equations, all higher-order moments follow algebraically.
The hierarchy exactly preserves the collisional invariants and their fluctuations, and predicts the full probability distribution of observables in the central fluctuation regime.
Benchmarked against $N$-body realizations, it reproduces the mean, variance and higher-order moments with an accuracy that improves systematically with $N$.
A single deterministic time-integration of three coupled partial differential equations thus replaces large ensembles of simulations, for any large $N$.
\end{abstract}
\maketitle

\section{Introduction}
\label{sec:Introduction}

The long-term evolution of long-range interacting systems
is a fundamental topic of interest across diverse branches of physics, including astrophysics~\cite{Heggie2003,BinneyTremaine2008}, plasma physics~\citep{Nicholson1992}, geophysical flows~\cite{Bouchet+2012}
and many others~\citep[see, e.g.\@,][]{Dauxois+2002}.
Driven by violent relaxation and phase mixing~\cite{Lynden+1967},
long-range interacting $N$-particle systems settle rapidly,
in a few dynamical times $\Tdyn$,
into a quasistationary state.
This state is dynamically frozen for the smooth mean-field dynamics. Later, on the (slower) relaxation time $\Trelax$, finite-$N$ fluctuations induce an irreversible collisional relaxation, ultimately driving the system closer to thermodynamic equilibrium.

By taking the large-$N$ limit and retaining ${ 1/N }$ corrections to account for (weak) cumulative two-body deflections~\cite{Chandrasekhar1942}, the late-stage evolution of the system's average \DF\@, $F$, can be described by the Balescu--Lenard kinetic equation~\citep{Balescu1960,Lenard1960}.
When collective effects are neglected
(i.e.\ the system does not amplify its own self-generated perturbations),
this kinetic equation reduces to the Landau equation~\citep{Landau1936, Landau+1981}.
Yet for any finite $N$, the empirical \DF\@, $F_N$, remains a stochastic object, fluctuating around $F$ on the relaxation timescale $\Trelax$.
 
Though challenging, quantifying these finite-$N$ fluctuations is important both theoretically and numerically.
Indeed, fluctuations encode the granularity of the phase-space distribution, drive the long-term evolution and ultimately determine the statistical uncertainty of any observable measured from a single realization of the system. 
For instance, they determine the intrinsic variability in the time of core collapse of globular clusters~\citep{Lynden+1968},
or the onset and evolution of stellar bars~\cite{Sellwood2009,Weinberg2007}.

\LDT\@~\citep[see, e.g.\@,][]{Touchette2009} is a natural theoretical tool to quantify the full statistics of macroscopic fluctuations. The use of \LDT\@ around deterministic long-range kinetic theories has seen major advances recently: a \LDP\@ was first derived for the homogeneous Landau equation~\citep{Feliachi+2021}, and subsequently generalized to include collective effects~\citep{Feliachi+2022} or spatially inhomogeneous configurations~\citep{Feliachi+2024}.   

Such an \LDP\@ expresses the probabilities of fluctuations through the infimum
of a rate functional over an appropriate functional space.
Despite these theoretical successes, exploiting in practice the \LDP\@ remains particularly arduous.
Similarly, running direct $N$-body simulations
to characterize fluctuations is computationally intractable, because of the triple curse
of (i) long-range interactions,
(ii) large timescale separation between the fast dynamical time and the slow relaxation time,
and (iii) the need for large samples.
Recently, \cite{Elrhirhayi+2026} derived a set of efficient Langevin stochastic differential equations able to mimic the slow relaxation of a self-gravitating system, including the statistics of finite-$N$ fluctuations.
However, this approach does not alleviate the need for a large ensemble of realizations to reconstruct the full statistics of the dynamics.

In this paper, we shift to an alternative paradigm.
We derive a deterministic hierarchy for the centered $n$-point moments of the fluctuating empirical \DF\@ and compute the leading order of this hierarchy in the large-$N$ limit.
We show that this leading-order hierarchy is equivalent to a closed set of only three kinetic equations, supplemented by an algebraic closure.
We refer to it as the minimal Landau fluctuation hierarchy, henceforth the minimal hierarchy.
Its three kinetic equations are: (i) the usual Landau equation for the mean \DF\@, (ii) the Lyapunov covariance equation for the second moment $S_2$, and (iii) an evolution equation for the third moment $S_3$.
Once $S_2$ and $S_3$ have been determined, the leading large-$N$ contribution to every fixed-order higher moment, $S_{n}$, is reconstructed algebraically through a quasi-Gaussian moment closure. 
The minimal hierarchy therefore provides a deterministic description of the secular evolution of the variance and all higher-order moments in homogeneous self-gravitating systems, without requiring averages over a large number of stochastic or $N$-body realizations. 
Note, however, that it does not provide the leading order of the cumulants beyond order three.

The paper is organized as follows.
Section~\ref{sec:LDK} introduces the system and recasts the Langevin dynamics derived in~\cite{Elrhirhayi+2026} as a stochastic evolution equation for the empirical distribution.
This provides the starting point for deriving the hierarchy of finite-$N$ correlations.
Section~\ref{sec:Lyapunov_hierarchy} presents the resulting minimal hierarchy, including the kinetic equations for $S_2$ and $S_3$ and the closure yielding $S_n$ at leading order for any ${n \!\geq\! 4}$. 
Section~\ref{sec:prop_hierarchy} discusses its main properties: observables, central \PDF, conservation laws.
Section~\ref{sec:Application} presents numerical applications and comparisons with $N$-body simulations. Section~\ref{sec:Conclusion_perspectives} concludes and gives some perspectives. Throughout the main text, technical details are kept to a minimum and deferred to Appendices or to relevant references. 

\section{Landau--Dean--Kawasaki equation}
\label{sec:LDK}

We are interested in the long-term evolution of a long-range interacting Hamiltonian system in dimension $d$, defined on a spatial
domain periodic in each direction.
This is the same system as the one described in~\cite{Elrhirhayi+2026}, whose key elements are briefly reproduced here.
We write the phase space coordinates as ${ \bw \!:=\! (\bT,\bv) }$.
The system is composed of $N$ particles of individual mass ${ m \!:=\! \Mtot / N }$
with $\Mtot$ the system's total mass.
In addition to their kinetic energy,
particles interact through
a (truncated) gravitational potential,
${ U(\bT , \bTp) \!:=\! \sum_{\bk} \psi_{\bk} \, \re^{\ri \bk \cdot (\bT - \bTp)} }$,
with ${ \psi_{\bk} \!:=\! - G / |\bk|^{2} }$
for ${ \kmin \!\leq\! |\bk| \!\leq\! \kmax }$.
We also suppose that the system is
(i) dynamically hot, in order to ignore collective effects~\cite[see, e.g.\@,][]{Chavanis2013};
(ii) of infinite extent, i.e.\ ${ \kmin \!\gg\! 1 }$,
so as to simplify the collision kernel $\sB$ (see Appendix B of~\cite{Elrhirhayi+2026}). 

At any given time, the full microscopic state of the system is exactly characterized
by all the particles' phase-space coordinates ${ \bw_{i} (t) \!:=\! (\bT_i(t), \bv_i(t)) }$.
To bridge this exact microscopic dynamics with macroscopic kinetic theories, we rely on a fundamental timescale separation between the (fast) dynamical time $\Tdyn$, which corresponds to the typical crossing time, and the (slow) relaxation time, ${\Trelax \!:=\! N \Tdyn }$.
Assuming spatial homogeneity and keeping only the slow variables, i.e.\ the velocities, the system is described by the empirical distribution function
\begin{equation}
    F_N(\bv, \tau) := m \sum_{i = 1}^{N} \deltaD [\bv \!-\! \bv_{i} (\tau)],
    \label{eq:def_FN}
\end{equation}
where, henceforth, we use ${ \tau \!:=\! t/N }$ to refer to the slow time.
Here, $F_N$ is a stochastic object whose dynamics is quite well understood: in the absence of collective effects, its mean follows the Landau equation at leading order in $1/N$, and its fluctuations are described by the \LDP\@ derived in~\cite{Feliachi+2021}.
In principle, this \LDP\@ provides an appropriate description of the fluctuations.
Yet, in practice, it is (very) challenging to effectively use,
since the computation of any relevant probability requires an optimization in a functional space.
Recently, Ref.~\cite{Elrhirhayi+2026} showed that the fluctuations of $F_N$ can be captured by a set of $N$ effective coupled Langevin equations for the particles' velocities reading\footnote{Compared to~\cite{Elrhirhayi+2026}, Eq.~\eqref{eq:Langevin} here is formulated in terms of the macroscopic time $\tau$.}
\begin{equation}
    \rd \bv_{i} = \sum_{j \neq i} \sqrt{m} \, \sB^{1/2} (\bv_i,\bv_j) \circ \rd \bW_{ij, \tau} .
    \label{eq:Langevin}
\end{equation}
In that expression, ${ \rd \bW_{ij, \tau} }$ are antisymmetric Wiener processes (${ \rd \bW_{ij, \tau} \!=\! - \rd \bW_{ji, \tau} }$) and ${ \sB(\bv,\bvp) \!:=\! \sB(\bv \!-\! \bvp) }$ is the collision kernel, whose detailed expression is reproduced in Appendix~\ref{app:Collision_kernel}.\footnote{To avoid singularity,
we adopt the no-self-interaction prescription ${ \sB(\bZero) \!:=\! \sZero }$ and ${ (\p_{\bu} \!\cdot\! \sB)(\bZero) \!:=\! \bZero }$.
This ensures the absence of self-interactions, as in the underlying $N$-body dynamics.} 
As shown in~\cite{Elrhirhayi+2026}, Eq.~\eqref{eq:Langevin} preserves exactly the total energy and momentum of the system, reproduces the correct averaged behavior (Landau equation) and captures the fluctuation statistics.

The Langevin rewriting from Eq.~\eqref{eq:Langevin}
offers a direct route to characterize the finite-$N$ fluctuations and correlations of $F_N$.
To do so, we apply It\^o's calculus to the empirical distribution $F_N$ (Eq.~\ref{eq:def_FN}).
In the weak distributional sense (i.e.\ testing the empirical distribution against arbitrary smooth functions), the empirical distribution $F_N$ evolves according to the \LDK\@ \SPDE\@~\cite{Dean1996,Kawasaki1994,Feliachi+2024}.
In It\^o form, it reads
\begin{equation}
    \rd F_N(\bv) = \mLL[F_N, F_N](\bv) \, \rd \tau + \tfrac{1}{\sqrt{N}} \rd \mW_\tau[F_N](\bv) .
    \label{eq:LDK}
\end{equation}
The derivation of Eq.~\eqref{eq:LDK} is detailed in Appendix~\ref{app:Langevin_to_LDK}.
We stress that Eq.~\eqref{eq:LDK} is not the standard Dean--Kawasaki equation for a system of Brownian particles, either independent or interacting in a mean-field way~\cite{Kawasaki1994,Dean1996,Cornalba+2023}.
Here, the dependence of the stochastic terms on the empirical \DF\@ is more involved, as shown below.

In Eq.~\eqref{eq:LDK},
the (deterministic) drift is the nonlinear Landau collision operator~\citep[see, e.g.\@,][for a review]{Chavanis2013}.
It reads
\begin{equation}
    \mLL[F, F](\bv) \!:=\! \half \frac{\p}{\p \bv} \cdot \!\!\int\!\! \rd \bvp \sB(\bv,\bvp) \!\bigg[\! \frac{\p F}{\p \bv} F(\bvp) - F(\bv) \frac{\p F}{\p\bvp} \!\bigg] .
    \label{eq:Landau_operator}
\end{equation}
The stochastic increment ${\rd\mW_\tau[F_N]}$ in Eq.~\eqref{eq:LDK} is, conditional on $F_N(\tau)$, a Gaussian multiplicative noise.
It encodes finite-$N$ fluctuations and satisfies ${\big\langle\rd\mW_\tau[F_N](\bv)\big\rangle=0}$,
with covariance
\begin{equation}
    \big\langle \rd \mW_\tau[F_N](\bv) \, \rd \mW_\tau[F_N](\bvp) \big\rangle =  C[F_N](\bv, \bvp) \, \rd \tau ,
    \label{eq:noise}
\end{equation}
where the average ${ \langle \, \cdot \, \rangle }$ is conditional on $F_N$ at time $\tau$.
The covariance kernel $C$ reads~\cite{Feliachi+2024}
\begin{align}
    C[F_N](\bv, {} & \bvp) \!:=\! \Mtot \frac{\p }{\p \bv} \!\cdot\! \bigg\{ \frac{\p}{\p \bvp} \!\cdot\! \big[ \!-\! F_N(\bv)F_N(\bvp) \sB(\bv, \bvp) \nonumber \\
    + \, {} & \deltaD(\bv \!-\! \bvp) F_N(\bv) \!\!\int\!\! \rd\bvpp F_N(\bvpp) \sB(\bv, \bvpp) \big] \bigg\} .
    \label{eq:covariance_kernel}
\end{align}
In that expression, the first term describes the correlated kicks
received by two distinct particles during a binary encounter,
while the second (diagonal) term
represents the self-contribution of a particle
diffusing against the background $F_N$.

In the limit ${ N \!\to\! + \infty }$,
fluctuations are suppressed in Eq.~\eqref{eq:LDK}.
Then, the macroscopic state is entirely described by the large-$N$ mean \DF\@ ${ F(\bv,\tau) }$, which is the solution of the deterministic homogeneous Landau equation (see Eq.~\ref{eq:Landau} below).
For a fixed value of $N$, performing an ensemble average over the initial conditions,
\LDK\@ reduces, at leading order in ${ 1/N }$,
to the same Landau equation.

As shown by~\cite{Feliachi+2024}, \LDK\@ is the stochastic-field counterpart of the \LDP\@ derived in~\cite{Feliachi+2021}: it reproduces the exact same dynamical large deviation Hamiltonian.
In principle, \LDK\@ offers an alternative way to access the fluctuations.
Indeed, \LDK\@ transfers the computational bottleneck from the particle number $N$ to the resolution of a \SPDE\@.
Yet, the direct simulation of Eq.~\eqref{eq:LDK} is delicate, because it involves the multiplicative noise ${ \rd \mW_{\tau}[F_N] }$~\cite{Cornalba+2023}.
This will be the topic of future investigation.

In the present work, we take a different route: rather than simulating \LDK\@ (Eq.~\ref{eq:LDK}), we use it as the starting point to derive the minimal hierarchy, a closed set of deterministic equations.
As we will show, this hierarchy gives direct access to all the moments of the empirical \DF\@,
through a single time-integration
of a few coupled partial differential equations.

\section{The Landau fluctuation hierarchy}
\label{sec:Lyapunov_hierarchy}

\subsection{Statistical moments}
\label{subsec:Moments}

We now introduce the hierarchy of centered moments of the empirical \DF\@ governed by \LDK\@ (Eq.~\ref{eq:LDK}).
We define the centered fluctuation by ${ \delta F := F_N \!-\! \langle F_N\rangle }$, where here $\langle\cdot\rangle$ denotes the ensemble average over both the initial conditions and the trajectory noise.
In the following, a bold label denotes a velocity variable, for example ${ \1:=\bv_1 }$ and we write ${ \delta F_{\1} (\tau) \!=\! \delta F (\1 , \tau) }$.
The $n$-th (centered) moment 
${ \MDK_{\1...\n} (\tau) \!:=\! \MDK_{n} (\1,...,\n; \tau) }$ is defined by
\begin{equation}
    \MDK_{\1... \n} (\tau) :=  \big\langle \delta F_{\1}(\tau) ...  \delta F_{\n}(\tau) \big\rangle ,
    \label{eq:def_Mn}
\end{equation}
where the superscript refers to the moments of $F_N$ evolving according to \LDK\@ (Eq.~\ref{eq:LDK}).

By construction, ${ \MDK_1 \!=\! 0}$.
The other moments do not share the same magnitude in $N$. Indeed, as shown in Appendix~\ref{app:IC}, their initial values satisfy ${ \MDK_n \!\thicksim\! N^{-\left\lceil n/2 \right\rceil} }$.
Following these scalings, we introduce the rescaled moments
\begin{equation}
    \SDK_n := N^{\left\lceil n/2 \right\rceil} \MDK_n ,
    \label{eq:def_Sn}
\end{equation}
making them all of order unity w.r.t.\ $N$,
i.e.\@ ${ \SDK_n \!=\! \mO(1) }$.
In particular, the first two nonzero moments read
\begin{subequations}
\begin{align}
    \SDK_{\1\2} &:= N \big\langle \delta F_{\1} \delta F_{\2} \big\rangle \thicksim 1,
    \label{eq:def_S2}
    \\
    \SDK_{\1\2\3} &:= N^2 \big\langle \delta F_{\1} \delta F_{\2} \delta F_{\3} \big\rangle \thicksim 1 ,
    \label{eq:def_S3}
    \end{align}
    \label{eq:def_S2_S3}\end{subequations}
with ${ \SDK_{\1\2} \!:=\! \SDK_{2} (\1,\2) }$ and similarly for $\SDK_{3}$.
In the following, to avoid any confusion with the moments of \LDK\@, we denote by $\SNB_n$ the rescaled moments of $F_N$ (Eq.~\ref{eq:def_FN}) when the particles follow the full $N$-body Hamiltonian dynamics in phase space.
In each case, the moments are centered on the corresponding finite-$N$ ensemble mean.

\LDK\@ is expected to capture the leading fluctuation statistics of the $N$-body system. 
More precisely, we conjecture that the \LDK\@ moments satisfy
\begin{equation}
    \SDK_n-\SNB_n = \mO \big( \tfrac{1}{N} \big), \quad n \geq 2.
    \label{eq:main_conjecture}
\end{equation}
Such a result is not immediately apparent.
Indeed, Landau's \LDP\@~\citep{Feliachi+2021}, from which \LDK\@ derives, only accounts for weak pairwise deflections.
It neglects strong encounters or higher-order irreducible many-particle dynamical effects.
Although \LDK\@ generates connected correlations of all orders, Eq.~\eqref{eq:main_conjecture} implies that the dynamical effects neglected by \LDK\@ do not alter the leading-order behavior of any moment $\SNB_n$.

In practice, estimating $\SDK_n$ directly requires a large ensemble of
realizations of \LDK\@.
To remove the need for numerous realizations,
in Appendix~\ref{app:Sn_equation},
using It\^o calculus, we derive evolution equations for $\SDK_{n}$.
These form the full Landau fluctuation hierarchy.
Yet, although they are deterministic, these equations are not closed:
the dynamics of any $\SDK_{n}$ is coupled to higher-order moments.

To circumvent this problem,
we then retain only the leading-order terms in ${1/N}$
of the \LDK\@ full fluctuation hierarchy.
This leads to the truncated Landau fluctuation hierarchy, henceforth the truncated hierarchy (Appendix~\ref{app:truncated_hierarchy_dynamical}).
As discussed later, this truncated hierarchy is equivalent to the minimal hierarchy presented in the introduction.
We denote by ${ \Smin_n }$ the moments obtained from this minimal hierarchy (with the superscript ``$\mathrm{M}$'' standing for minimal).
They correspond to the leading-order contribution 
in ${ 1/N }$ to $\SDK_n$.

As shown in Appendices~\ref{app:truncated_hierarchy_dynamical}, \ref{app:S4_closure}, \ref{app:S5_closure}, \ref{app:S6_Sn_closure},
this truncation satisfies, by construction,
\begin{equation}
    \SDK_n - \Smin_n = \mO \big( \tfrac{1}{N} \big), \quad n \geq 2.
    \label{eq:SLDK_vs_Smin}
\end{equation}
There are a couple of crucial gains with this new minimal hierarchy.

First, combining Eq.~\eqref{eq:main_conjecture} with Eq.~\eqref{eq:SLDK_vs_Smin},
we have
\begin{equation}
    \Smin_n - \SNB_n = \mO \big( \tfrac{1}{N} \big), \quad n \geq 2.
\end{equation}
Phrased differently, at leading order in ${ 1/N }$, the minimal hierarchy is as accurate as \LDK\@ for every fixed-order moment.
We stress, however, that \LDK\@ contains more information than the minimal hierarchy.
Indeed, since \LDK\ reproduces the large deviation Hamiltonian, it is expected to capture every cumulant $\KNB_n$ at its own leading order,
i.e.\ ${ \KNB_n \!\thicksim\! N^{1-n} }$ (where $\KNB_n$ denotes the $n$-th cumulant of $F_N$ under the $N$-body dynamics).
By contrast, the minimal hierarchy only captures $\KNB_2$ and $\KNB_3$, while the higher-order cumulants, which are subleading compared to the moments, are discarded by the minimal hierarchy.

Second, the minimal hierarchy is explicitly closed.
Indeed, as detailed in the following section,
this hierarchy is composed of only three self-consistent kinetic equations: the Landau equation for $F$, and the kinetic equations for the two- and three-point moments, $\Smin_2$ and $\Smin_3$.
All higher-order moments $\Smin_n$, with ${ n \!\geq\! 4 }$, are then reconstructed algebraically from ${ (\Smin_2,\Smin_3) }$
using an explicit quasi-Gaussian closure.
Contrary to the full and truncated hierarchies,
this removes the need to integrate \PDE\@
of ever higher dimensionality.

In the following, for simplicity,
we write $S_n$ for $\Smin_n$.
The coupled equations satisfied by ${ (F,S_{2},S_{3}) }$
constitute the central result of this paper.
In the remainder of this section,
we write out explicitly these evolution equations
(Sections~\ref{subsec:Landau_equation},~\ref{subsec:S2_equation} and~\ref{subsec:S3_equation}),
while we present the quasi-Gaussian closure in Section~\ref{subsec:Sn_equation}.

\subsection{Landau equation}
\label{subsec:Landau_equation}

In the minimal hierarchy,
the mean \DF\@, $F$, evolves according to the Landau equation
which describes the relaxation toward equilibrium.
It reads
\begin{equation}
\p_{\tau} F = \mLL [F, F] ,
\label{eq:Landau}
\end{equation}
where ${ \mLL [F, F] }$ is the nonlinear Landau collision operator (Eq.~\ref{eq:Landau_operator}). It scales as ${ \mLL \!\thicksim\! \mO(1) }$.
We note that Eq.~\eqref{eq:Landau} is closed since it only involves $F$.

\subsection{Kinetic equation for \texorpdfstring{$S_2$}{S2}}
\label{subsec:S2_equation}

The hierarchy equation for the two-point moment, $S_2$, is a Lyapunov covariance equation~\cite[see, e.g.\@,][]{Bouchet+2013, Douet+2026}. As derived in Appendix~\ref{app:S2_derivation},
it is given by
\begin{equation}
\p_{\tau} S_{\1\2} = \big[ \p_{\tau} S_{\1\2} \big]_{\rL} \!\!+ C[F]_{\1\2} .
\label{eq:Lyapunov_S2}
\end{equation}
The linear contribution reads
\begin{align}
    \big[ \p_{\tau} S_{\1\2} \big]_{\rL} & {} \!\!= \big[ \mL_{\1} \!+\! \mL_{\2} \big] S_{\1\2} .
    \label{eq:linear_S_2}
\end{align}
Here, $\mL_{\1}$ is the linearized Landau operator acting on the first velocity variable ${ \1 \!:=\! \bv_{1} }$.
It satisfies
\begin{align}
    \mL_{\1}S_{\1\2} := \half \p_{\1} \!\cdot\!\!\int\!\! \rd \3 \, \sB_{\1\3} \big[ {} & \p_{\1} F \, S_{\3\2} \!+\! F_{\3} \, \p_{\1} S_{\1\2}
    \nonumber
    \\
    - \, {} & \p_{\3} F \, S_{\1\2} \!-\! F_{\1} \, \p_{\3} S_{\3\2} \big] ,
    \label{eq:def_mL_1}
\end{align}
with ${ \sB_{\1\2} \!:=\! \sB (\bv_{1}, \bv_{2}) }$.
The operator $\mL_{\2}$ applies the same operation to the second variable $\2$.
Equation~\eqref{eq:Lyapunov_S2} also involves
the source term, ${ C[F] }$,
as already introduced in Eq.~\eqref{eq:covariance_kernel}.
Physically, this term captures the finite-$N$ granularity:
it continuously generates two-point moments
on the secular timescale.

Equation~\eqref{eq:Lyapunov_S2} involves only $F$ and $S_{2}$, and no higher-order moments.
In addition, Eq.~\eqref{eq:Lyapunov_S2} also guarantees the symmetry ${S_{\1\2} \!=\! S_{\2\1} }$.
We refer to Appendix~\ref{app:IC_S2} for the initial condition.

\subsection{Kinetic equation for \texorpdfstring{$S_3$}{S3}}
\label{subsec:S3_equation}

As derived in Appendix~\ref{app:S3_derivation},
the minimal hierarchy equation for $S_3$ reads 
\begin{equation}
    \p_{\tau} S_{\1\2\3} = \big[ \p_{\tau} S_{\1\2\3} \big]_{\rL} \!+\! \big[ \p_{\tau} S_{\1\2\3} \big]_{\rS} \!+\! \big[ \p_{\tau} S_{\1\2\3} \big]_{\rNL} .
    \label{eq:Lyapunov_S3}
\end{equation}
In that expression, the linear term is
\begin{equation}
    \left[ \p_{\tau} S_{\1\2\3} \right]_{\rL} = \big[ \mL_{\1} \!+\! \mL_{\2} \!+\! \mL_{\3} \big] S_{\1\2\3} ,
    \label{eq:linear_S_3}
\end{equation}
with $\mL_{\1}$ already introduced in Eq.~\eqref{eq:def_mL_1}.
This linear term describes the advection of $S_{3}$
by the linear Landau dynamics around $F$.

In Eq.~\eqref{eq:Lyapunov_S3}, ${ [ \p_{\tau} S_3 ]_{\rS} }$ is a source term.
It depends on both $F$ and $S_2$ and reads
\begin{align}
    \left[ \p_{\tau} S_{\1\2\3} \right]_{\rS} = {} & \!\!\int\!\! \rd \4 \, \frac{\delta C_{\2\3}}{\delta (F_{\4})} \, S_{\1\4} + \circlearrowright_{\1\2\3} ,
    \label{eq:def_p_tS_3_source}
\end{align}
where the symbol $\circlearrowright_{\1\2\3} $ denotes the cyclic summation.\footnote{It reads
${ A_{\1\2\3} + \circlearrowright_{\1\2\3} \,:= \! A_{\1\2\3} \!+\! A_{\2\3\1} \!+\! A_{\3\1\2} }$.}
Equation~\eqref{eq:def_p_tS_3_source}
involves ${ \delta C / \delta F }$,
the differential of the covariance from Eq.~\eqref{eq:covariance_kernel}.
It reads
\begin{align}
    \frac{\delta C_{\2\3}}{\delta (F_{\4}) } = \Mtot \p_{\2}\!\cdot\! \p_{\3} \!\cdot\!\bigg\{ {} & \delta_{\2\3} \bigg[  F_{\2} \sB_{\2\4} + \delta_{\2\4}  \!\!\int\!\! \rd \5 \, F_{\5} \sB_{\2\5} \bigg]
    \nonumber
    \\
    - \, {} & \big[ \delta_{\2\4} F_{\3} \!+\! \delta_{\3\4} F_{\2} \big] \sB_{\2\3} \bigg\} ,
    \label{eq:source_S3}
\end{align}
with the shortened notation ${ \delta_{\bv \bvp} \!:=\! \deltaD(\bv - \bvp) }$.
This source term is linear in $F$ and $S_{2}$.
Physically, it describes how local two-point fluctuations modulate the finite-$N$ noise,
hence altering the rate at which three-point moments are generated.

Finally, Eq.~\eqref{eq:Lyapunov_S3} also involves a nonlinear contribution, ${[ \p_{\tau} S_3 ]_{\rNL} }$.
At first, as detailed in Appendix~\ref{app:S3_derivation},
this term appears as a function of $S_{4}$.
In practice, this dependence is resolved through
the Wick closure
\begin{align}
    S_{\1\2\3\4} = {} & S_{\1\2} S_{\3\4} + S_{\1\3} S_{\2\4} + S_{\2\3} \, S_{\1\4} ,
    \label{eq:closure_S4}
\end{align}
so that it becomes a function of $S_{2}$.
Overall, the nonlinear term reads
\begin{align}
    [ \p_{\tau} S_{\1\2\3} ]_{\rNL} = \mK_{\1;\2\3} \,+\, \circlearrowright_{\1\2\3} , 
    \label{eq:NL_S3}
\end{align}
where we introduced
\begin{align}
    \mK_{\1;\2\3} := \mQ_{\1;\2\3} + \mQ_{\1;\3\2} ,
    \label{eq:NL_S3_123}
\end{align}
as well as the function
\begin{equation}
    \mQ_{\1;\2\3} \!:=\! \half \p_{\1} \!\cdot\! \bigg\{ \bigg[ \!\int\!\! \rd \4 \sB_{\1\4} S_{\4\2} \bigg] \p_{\1} S_{\1\3} \!-\! S_{\1\3} \!\!\int \!\! \rd \4 \sB_{\1\4} \p_{\4} S_{\4\2} \bigg\} ,
    \label{eq:def_M123}
\end{equation}
which is quadratic in $S_{2}$.
Physically, it captures the evolution of $S_{3}$
driven by the interaction of two distinct correlated pairs.
We refer to Appendix~\ref{app:IC_S3} for the initial condition
of $S_{3}$.

\subsection{Minimal hierarchy for higher-order moments \texorpdfstring{$S_n$}{Sn}}
\label{subsec:Sn_equation}

For higher-order moments,
the minimal hierarchy relies on the quasi-Gaussian closure
\begin{equation}
S_n(\tau) = \mG_n[S_2, S_3](\tau) \quad \text{for } n \geq 4 ,
\label{eq:Closure_generic}
\end{equation}
where $\mG_n$ is a functional depending only on $S_2$ and $S_3$.
More precisely, for even $n$,
the closure is Wick's formula
\begin{align}
S_{\1\cdots\mathbf{2k}} & {} = \mS \big[ S_{\1\2} \!\times\! \cdots \!\times\! S_{(\mathbf{2k-1})(\mathbf{2k})} \big] ,
\label{eq:Closure_2k}
\end{align}
while for odd moments, it takes the form
\begin{align}
S_{\1\cdots\mathbf{2k+1}} = \mS \big[ S_{\1\2\3} \!\times\! S_{\4\5} 
    \!\times\! {} & \cdots \!\times\! S_{(\mathbf{2k})(\mathbf{2k+1})} \big] .
    \label{eq:Closure_2k+1}
\end{align}
In both expressions, $\mS$ is the symmetrization operator, i.e.\ the sum over all distinct partitions (see Eq.~\ref{eq:closure_explicit}).
We refer to Appendix~\ref{app:S6_Sn_closure} for more details.
With such a closure, even moments, $S_{2k}$,
are entirely built from $S_{2}$,
while odd ones, $S_{2k+1}$,
depend linearly on $S_{3}$.

Together, Eqs.~\eqref{eq:Landau}, \eqref{eq:Lyapunov_S2},
and~\eqref{eq:Lyapunov_S3}, supplemented by the algebraic closures from Eq.~\eqref{eq:Closure_generic},
constitute the minimal hierarchy. 
This is the central result of this work.

The present fluctuation hierarchy offers
a substantial computational advantage
over ensemble-based approaches.
First, the integration timestep is a fraction of the relaxation time, compared to $N$-body simulations that must resolve the dynamical time. 
Second, unlike the Langevin dynamics from~\cite{Elrhirhayi+2026}, the hierarchy is deterministic:
it does not require any large ensemble of stochastic realizations.
Third, when expressed in the slow time ${\tau \!=\! t/N}$,
the evolution equations for ${ (F, S_{2}, S_{3}) }$,
as well as their initial conditions,
do not depend on $N$.
Consequently, the hierarchy needs to be integrated only once.
We now discuss further some of the properties of this hierarchy.

\section{Properties}
\label{sec:prop_hierarchy}

\subsection{Observables}
\label{sec:observables}

Given some scalar function, ${ \phi_{\1} \!:=\! \phi (\1) }$,
we define the corresponding macroscopic one-particle observable ${ \Phi \!:=\! \Phi [F_N] }$ via
\begin{equation}
    \Phi := \tfrac{1}{\Mtot} \!\!\int\!\! \rd \1 \, \phi_{\1} \, F_N(\1) .
    \label{eq:def_observable}
\end{equation}
Its $n$-th centered moment, ${\bbM_n \!:=\! \langle (\Phi \!-\! \langle \Phi \rangle)^{n} \rangle }$, is
\begin{equation}
    \bbM_n (\tau) = \tfrac{1}{N^{\lceil n/2\rceil}\Mtot^n} \!\!\int\!\! \rd \1 ... \rd \n\, \phi_{\1} \, ... \, \phi_{\n} S_{\1...\n} (\tau) ,
        \label{eq:def_Mn_observable}
\end{equation}
where the prefactor follows from Eq.~\eqref{eq:def_Sn}.\footnote{More generally, for a two-particle observable, its $n$-th centered moment would depend on moments up to $2n$, rather than only up to order $n$.}
For ${ n \!=\! 2 }$, $\bbM_n$ reduces to the variance,
${ \bbV \!:=\! \bbV [\Phi] }$.
For ${ n \!\geq\! 4 }$, inserting the quasi-Gaussian closures
from Eq.~\eqref{eq:Closure_generic},
we find
\begin{subequations}
\begin{align}
    \bbM_{2k} {} & = \tfrac{(2k)!}{2^k k!} \bbV^{k} ,
    \label{eq:observable_even_moments}
    \\
    \bbM_{2k+1} {} & = \tfrac{(2k \!+\! 1)!}{3!\,2^{k-1}(k - 1)!} \, \bbM_3 \, \bbV^{k-1} ,
    \label{eq:observable_odd_moments}
\end{align}
\label{eq:observable_moments}\end{subequations}
where the combinatorial factors account
for the different partitions of the velocity arguments (see Appendix~\ref{app:S6_Sn_closure}).
At leading order in ${1/N}$,
all the moments of a scalar observable
are reconstructed directly from the variance and third moment
predicted by the minimal hierarchy.

\subsection{Central fluctuation probability distribution}
\label{sec:central_fluctuation}

The minimal hierarchy determines
all centered moments of the fluctuations of an observable. 
We now go one step further and show that
it can therefore predict the \PDF\@ of an observable
as in Eq.~\eqref{eq:def_observable}
up to the first non-Gaussian correction.

As detailed in Appendix~\ref{app:PDF},
using the quasi-Gaussian closure from Eq.~\eqref{eq:Closure_generic},
and introducing the rescaled fluctuation
${ \delta \tPhi \! :=\! (\Phi \!-\! \langle \Phi \rangle) / \sqrt{\bbV} }$,
we find that its \PDF\@ reads
\begin{align}
     \bbP (\delta \tPhi) \!=\! \tfrac{1}{\sqrt{2\pi}} \exp \!\big(\! \!-\! \tfrac{\delta \tPhi^{2}}{2} \big) \big[ 1 \!+\! \tfrac{\bbM_3}{6 \bbV^{3/2}} \big( \delta \tPhi^3 \!-\! 3 \delta \tPhi \big) \big] \!+\! \mO \big( \tfrac{1}{N} \big) .
    \label{eq:PDF}
\end{align}

Equation~\eqref{eq:PDF} is an asymptotic expansion valid in the central fluctuation regime.
It accounts for the first non-Gaussian correction,
present for large but finite $N$.\footnote{
Because of the asymmetry induced by the $\bbM_3$ term, this \PDF\@ is not positive in the far tails.}
Equation~\eqref{eq:PDF} does not describe rare fluctuations:
these require the full large deviation rate function.
In Section~\ref{sec:Application}, we validate Eq.~\eqref{eq:PDF} numerically.

\subsection{Conservation laws}
\label{sec:cons_law}

For a given \DF\@ $F$, we define the total mass, momentum and kinetic energy as
\begin{subequations}
    \begin{align}
    M[F] & := \!\!\int\!\! \rd\bv \, F(\bv) && \!\!\!\! \text{(Mass)} ,
    \\
    \bPtot[F] & := \!\!\int\!\! \rd\bv \, \bv \, F(\bv) && \!\!\! \text{(Momentum)} ,
    \\
    \Ekin[F] & := \!\!\int\!\! \rd \bv \, \half |\bv|^2 F(\bv) && \!\!\!\!  \text{(Kinetic energy)} .
    \end{align}
    \label{eq:def_invariants}\end{subequations}
Each invariant ${ \Phi \!\in\! \{ M, \bPtot, \Ekin \} }$ is linear in $F$, and is associated with the one-particle function ${ \phi_{\bv} \!:=\! \delta \Phi / \delta F_{\bv} \!\in\! \{ 1, \bv, \half |\bv|^2 \} }$.

These quantities are conserved by the underlying dynamics: ${ \Phi[F_N] }$ is constant along every individual realization of \LDK\@ (see Appendix~\ref{app:conservation_invariants_proof}).
As a consequence, its mean and all its fluctuation moments are exactly conserved.
Because the minimal hierarchy is obtained by truncation, it is not obvious that it inherits these conservation laws.
We now show that it does.

First, the mean value is conserved by the Landau equation (Eq.~\ref{eq:Landau}), namely
\begin{equation}
    \frac{\rd}{\rd \tau} \!\!\int\!\! \rd \1 \, \phi_{\1} F_{\1} (\tau) = 0 .
    \label{eq:conservation_F}
\end{equation}
Second, the minimal kinetic equations for $\Smin_2$ and $\Smin_3$ (Eqs.~\ref{eq:Lyapunov_S2} and~\ref{eq:Lyapunov_S3}) exactly preserve the variance and the third moment of each invariant, namely
\begin{subequations}
\begin{align}
    {} & \frac{\rd}{\rd \tau} \!\!\int\!\! \rd \1 \rd \2 \, \phi_{\1} \, \phi_{\2} \, S_{\1\2} (\tau) = 0 ,
    \label{eq:conservation_S_2}
\\
    {} & \frac{\rd}{\rd \tau} \!\!\int\!\! \rd \1 \rd \2 \rd \3 \, \phi_{\1} \, \phi_{\2} \, \phi_{\3} \, S_{\1\2\3} (\tau) = 0 .
    \label{eq:conservation_S_3}
    \end{align}
    \label{eq:conservation_S_2_S_3}\end{subequations}
Finally, through the quasi-Gaussian closure (Eq.~\ref{eq:Closure_generic}), these properties extend to all the higher-order moments $S_n$.
The minimal hierarchy therefore preserves the collisional invariants, together with all their fluctuation moments.
The proofs are given in Appendix~\ref{app:conservation_invariants_proof}.
In Appendix~\ref{app:Lyapunov_numerical_integration}, we also show how a careful discretization scheme ensures that these conservation laws are satisfied up to round-off errors in the numerical integration of the minimal hierarchy.

\section{Application}
\label{sec:Application}

We now turn to the numerical validation of the minimal hierarchy.
To do so, we compare its predictions with $N$-body simulations. 

The initial conditions of our benchmark case are:
(i) homogeneous in 2D space;
(ii) anisotropic Gaussian in velocities, imposing ${ T_x \!=\! 3 \, T_y }$,
with ${ T_{x} \!=\! N^{-1} \sum_{i} (v^{x}_{i})^{2} }$ the dynamical temperature.
Further details concerning the $N$-body simulations are given
in Appendix~{A} of~\cite{Elrhirhayi+2026} and Appendix~\ref{app:Parameters_numerical_convergence}
of the present work.

To integrate the minimal hierarchy,
we use FFTs to perform convolutions
as in Eq.~\eqref{eq:Landau_operator}.
In addition, gradients and divergences are computed via a second-order finite difference scheme, tailored to preserve invariants (up to round-off and summation errors).
We refer to Appendix~\ref{app:Lyapunov_numerical_integration}
for details on our numerical implementation.
All these elements are implemented in an efficient \texttt{julia} code,
which is publicly available in~\cite{github}.

\begin{figure*}[htbp!]
\includegraphics[width=0.95 \textwidth]{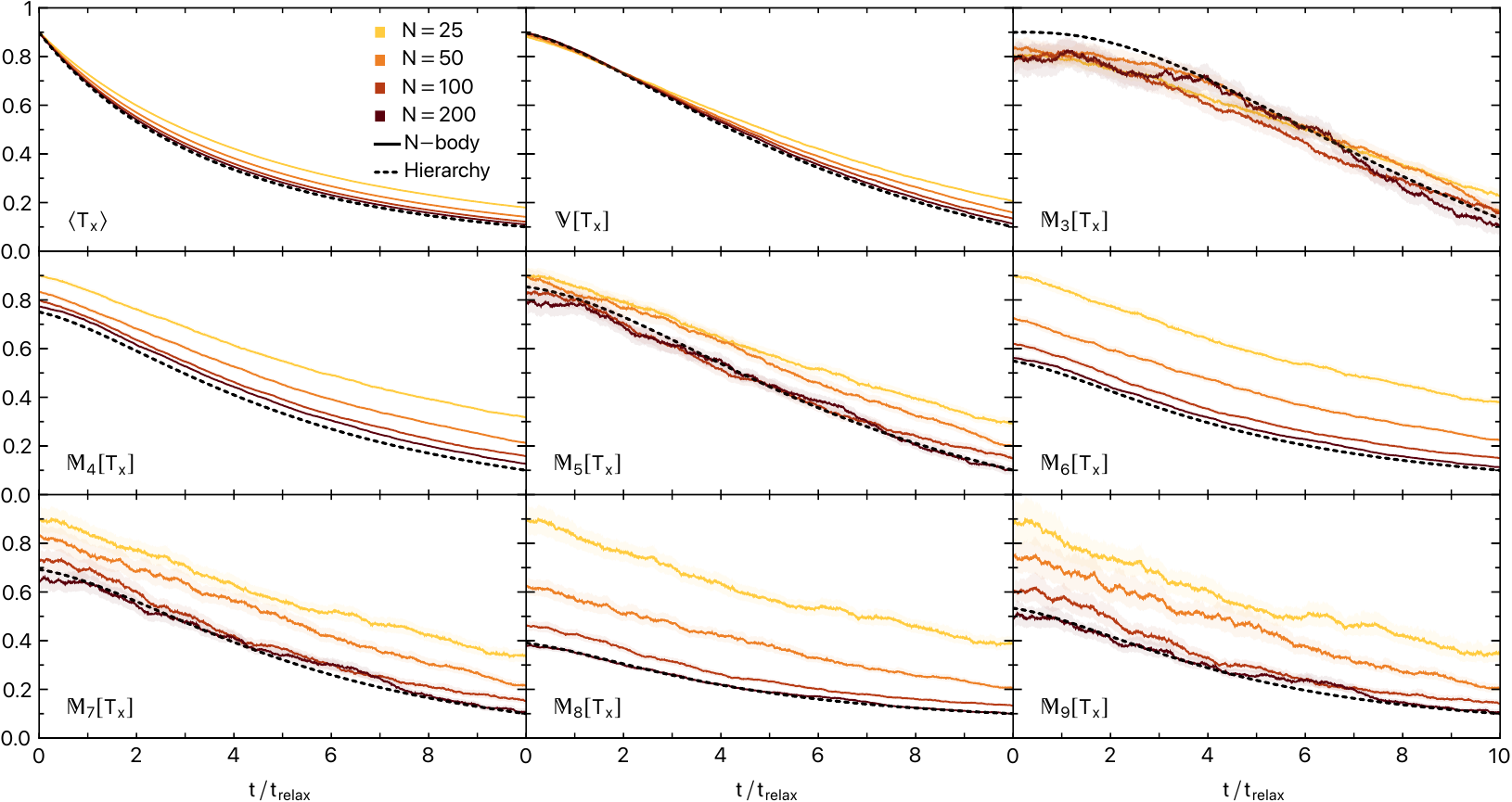}
\caption{
Evolution in time of the mean, variance, and higher-order moments
of the temperature $T_x$.
The dashed black lines are the predictions from the minimal hierarchy. The solid colored lines display the $N$-body measurements,
performed over 256\,000 realizations.  
Shaded bands around the $N$-body lines are statistical uncertainties, estimated via bootstrap resampling across the realizations.
For visual clarity, all plotted quantities have been appropriately rescaled.
We refer to Appendix~\ref{app:Parameters_numerical_convergence} for details.
As $N$ increases, the minimal hierarchy reproduces the average and all higher-order moments with increasing accuracy.
}
\label{fig:money_plot}
\end{figure*}

The main result of this comparison is presented in Fig.~\ref{fig:money_plot}.
In that figure, we compare the average, variance and all higher-order moments up to $\bbM_9$ of the temperature $T_x$ as predicted by the minimal hierarchy, with measurements from 256\,000 $N$-body realizations for ${N \!=\! 25, 50, 100, 200}$. As $N$ increases, the agreement between the hierarchy and the simulations improves steadily. This supports our claim that the minimal hierarchy captures, at leading order in ${1/N}$: (i) the average dynamics; (ii) its variance; (iii) the first non-Gaussian correction as encompassed by the third moment; and (iv) all the higher-order moments generated by the quasi-Gaussian closure.
Figure~\ref{fig:money_plot} is a key result of this work.
In Fig.~\ref{fig:money_plot},
one can note an initial offset visible in the $\bbM_3$ panel.
We believe that this is only a consequence of our finite number
of realizations.
In Appendix~\ref{app:additional_validations},
using the same set of realizations,
we show that no comparable offset is present for another observable.

\begin{figure}[htbp!]
\centering
\includegraphics[width=0.45 \textwidth]{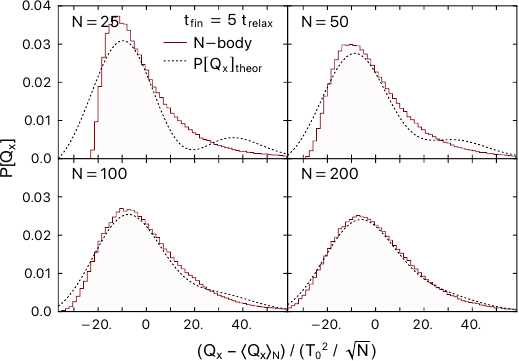}
\caption{Histogram of the observable ${ Q_{x}}$, at ${ t \!=\! 5 \, \Trelax }$,
using the same realizations as in Fig.~\ref{fig:money_plot}.
The dotted line is the \PDF\@ from Eq.~\eqref{eq:PDF}.
As $N$ increases, the $N$-body distribution converges to the predicted \PDF\@.
}
\label{fig:histograms}
\end{figure}

To further support this result, in Fig.~\ref{fig:histograms}, we compare the $N$-body histograms of the quartic observable ${Q_{x} \!=\! N^{-1}\sum_i(v_i^x)^4}$ with the \PDF s predicted by Eq.~\eqref{eq:PDF}.
We chose this observable because its $N$-body histogram shows more pronounced departures from a Gaussian shape. As $N$ increases, the predicted \PDF\@ captures these features in the central fluctuation regime.
As expected, the far tails of the histogram lie outside the validity of Eq.~\eqref{eq:PDF}.

Overall, these comparisons show that the minimal hierarchy captures the leading-order fluctuation statistics of the $N$-body dynamics, not only for the variance, but also for the first non-Gaussian correction and the higher-order moments.
This provides numerical support for the conjecture from Eq.~\eqref{eq:main_conjecture}: although \LDK\@ neglects irreducible three-body correlations, these do not appear to affect the leading-order moments.
We also stress the computational gain offered by the hierarchy.
The $N$-body measurements presented here required several weeks of computation on a cluster for each value of $N$.
Conversely, the hierarchy is integrated only once, on the slow timescale, and its solution applies to any sufficiently large $N$, after restoring the scaling of Eq.~\eqref{eq:def_Sn}.

\section{Conclusion and perspectives}
\label{sec:Conclusion_perspectives}

\subsection{Relation to previous work}
\label{sec:previous_works}

The Lyapunov covariance equation for $S_2$ (Eq.~\ref{eq:Lyapunov_S2}) has long been known~\cite{Sarkka2019}.
Indeed, an equation with a similar structure has appeared in several other physical contexts, always as the second-moment counterpart
of an underlying large deviation or fluctuating-hydrodynamic description. 
For example, in two-dimensional turbulence,
such an equation governs the covariance of the fast eddies~\cite{Bouchet+2013}.
However, in that context, it is assumed that the turbulent eddies
evolve rapidly enough so that their covariance reaches a steady state (i.e.\ ${ \p_{\tau} S_2 \!=\! 0 }$).
Phrased differently, the fast fluctuations are entirely determined
by the slow mean flow. 
More recently, and closer to our setting, Ref.~\cite{Douet+2026} derived a Lyapunov covariance equation of the very same form for the two-point cumulant of the spectral density in local weak wave turbulence.
It was obtained by linearizing a \LDK\@-type stochastic kinetic equation,
similarly to our approach here.
In Ref.~\cite{Douet+2026}, the focus was on the stationary,
long-range correlations sustained by out-of-equilibrium boundary conditions.
This differs from the questions of interest in the present work:
(i) we are interested in the non-stationary long-term evolution of the fluctuation statistics;
(ii) we track how the statistics of these fluctuations
is self-consistently coupled to the mean field $F$;
(iii) we include the contribution from the finite-$N$ source ${ C[F] }$ (Eq.~\ref{eq:Lyapunov_S2}) rather than from an external forcing. 

\subsection{Conclusion}
\label{sec:conclusion}

In this work, we investigated the long-term evolution
of spatially homogeneous, long-range interacting $N$-body systems
in the dynamically hot regime (i.e.\ when the collective amplification is weak).
Starting from \LDK\@
(Eq.~\ref{eq:LDK}),
we derived the minimal hierarchy. 
It consists of three deterministic kinetic equations
for the mean distribution $F$ (Eq.~\ref{eq:Landau}),
as well as the (rescaled) second-order moment $S_{2}$ (Eq.~\ref{eq:Lyapunov_S2})
and third-order moment $S_{3}$ (Eq.~\ref{eq:Lyapunov_S3}), supplemented by a quasi-Gaussian closure
(Eq.~\ref{eq:Closure_generic})
from which all higher-order moments follow.
Together, this hierarchy predicts the leading large-$N$ contribution
to the centered moments of the empirical \DF\@, $F_N$.
It does so deterministically, removing the need for sampling a large ensemble of realizations.
Finally, we numerically validated this hierarchy
by comparing it extensively against $N$-body simulations,
showing excellent agreement, improving as $N$ increases.

\subsection{Perspectives}
\label{sec:perspectives}

Of course, the present work is only one additional step toward an increasingly detailed understanding
of the statistical properties of the long-term
evolution of long-range interacting systems.
We now conclude with a few possible avenues for future work.

\textit{Landau--Dean--Kawasaki equation}. As shown in~\cite{Feliachi+2024}, the \LDP\@ for $F_{N}$ can be recast
into \LDK\@ (Eq.~\ref{eq:LDK}).
Following an approach similar to~\cite{Silvano+2025},
it would be worthwhile to directly integrate \LDK\@.
A key advantage of \LDK\@
is that it should allow one to consider extremely large values of $N$.

\textit{Multi-periodicity}. Here, we focused
on homogeneous systems of infinite extent.
In that case, the two-body scattering kernel
${ \sB (\bv , \bvp) }$
is a smooth function of $\bv$ and $\bvp$.
For a multi-periodic system,
this kernel is replaced with a discrete sum over resonance vectors $\bk$ involving sharp Dirac deltas.
At face value, this makes the Langevin equation (Eq.~\ref{eq:Langevin}) ill-defined. However, such sharp resonances are not expected to be a concern
for the integration of the minimal hierarchy.
This deserves further investigation.

\textit{Inhomogeneity}. In inhomogeneous systems,
the Landau equation (Eq.~\ref{eq:Landau})
is replaced by its inhomogeneous generalization
that involves angle-action coordinates~\citep[see, e.g.\@,][and references therein]{Chavanis2013}. Even though the large deviation Hamiltonian
for such inhomogeneous setups has been recently derived~\citep{Feliachi+2024}, deriving the fluctuation hierarchy for that setup is still challenging,
because of the intricate orbital structure.

\textit{Hamilton--Jacobi equation}. In this work, we derived the minimal hierarchy from \LDK\@, assuming that the scaling of initial moments
is conserved by the dynamics. 
Inspired by~\cite{Bouchet+2016,Bouchet+2022},
we expect that one can derive this hierarchy directly from the Hamilton--Jacobi equation associated with the \LDP\@.
Such a more generic approach could prove powerful to include, for example, irreducible three-body correlations that were neglected here.

\textit{Collective effects}. Here, we placed ourselves in the dynamically hot limit,
hence neglecting collective amplification.
For systems with lower dynamical temperature, the Landau equation
becomes the (homogeneous) Balescu--Lenard equation~\cite{Balescu1960,Lenard1960}.
The associated large deviation Hamiltonian is non-quadratic
in the conjugate field~\citep{Feliachi+2022}.
As a consequence, fluctuations cannot be represented by a stochastic equation with a Gaussian multiplicative noise.
Extending the present fluctuation hierarchy to the regime of the Balescu--Lenard equation is an open problem.

\textit{Kinetic blocking}. In ${ d \!=\! 1 }$,
the Landau collision operator vanishes exactly (Eq.~\ref{eq:Landau_operator}). This is the archetype of a ``kinetic blocking''~\citep[see, e.g.\@,][]{Eldridge+1963,Dawson1964,Rouet+1991,Miller1996,Chavanis2001,Rocha+2014,Fouvry+2020}.
As a result, relaxation in 1D is driven by the (much) slower three-body interactions.
One should derive the associated large deviation Hamiltonian,
whose large deviation rate would probably be in ${1/N^{2}}$.
From it should ultimately follow a generalized Landau fluctuation hierarchy
that accounts for three-body scatterings.

\begin{acknowledgments}
This work is partially supported by the grants GALBAR ANR-25-CE 31-4684 and BEYOND-BL ANR-25-CE57-2626 of the French Agence Nationale de la Recherche.
This project has received financial support from the CNRS through the MITI interdisciplinary programs.
This work has made use of the Infinity Cluster hosted by Institut d'Astrophysique de Paris, partially funded by IDF-DIM-ORIGINES-2023-4-11.
We thank St\'ephane Rouberol for the smooth running of the
Infinity cluster.
Generative AI tools were used to polish the manuscript,
improve the formatting of the figures,
and assist in refining the code's implementation.
\end{acknowledgments}

\medskip
\noindent{\em Data availability} --- The data that support the findings of
this article are openly available~\cite{github}.
 
\appendix

\section{Collision kernel}
\label{app:Collision_kernel}

In this Appendix, we detail the collision kernel $\sB$
at the heart of the Landau equation (Eq.~\ref{eq:Landau_operator}).
Following Appendix~{B} of~\cite{Elrhirhayi+2026},
the collision kernel takes the compact form
\begin{equation}
\sB (\bu) = f (u) \, \sP(\bu) ,
\label{eq:def_sB_generic}
\end{equation}
where ${ \sP (\bu) \!:=\! u^2 \, \sI \!-\! \bu \!\otimes\! \bu }$
is proportional to the projector orthogonal to ${\bu := \bv \!-\! \bvp}$.
In Eq.~\eqref{eq:def_sB_generic}, the prefactor is ${ f (u) \!:=\! \kappa / u^3 }$ with
\begin{equation}
    \kappa := 2 \pi G^2 \Mtot \ln \Lambda \, \frac{1}{d \!-\! 1}\frac{2\pi^{(d-1)/2}}{\Gamma [ \tfrac{d-1}{2} ]} .
\label{eq:def_kappa}
\end{equation}
The geometric factor $\ln \Lambda$ depends on the spatial dimension $d$.
It reads
\begin{equation}
    \ln \Lambda =
    \begin{cases}
        \displaystyle \ln [\kmax / \kmin] {} & \text{if} \quad d = 3 ,
        \\[1.0ex]
        \displaystyle \big[ \kmax^{d-3} - \kmin^{d-3}\big] / (d \!-\! 3) {} & \text{otherwise} ,
    \end{cases}
    \label{eq:lnLambda_generic}
\end{equation}
with ${ (\kmin,\kmax) }$ the minimum and maximum harmonics involved in the interaction.

Compared to~\cite{Elrhirhayi+2026},
Eq.~\eqref{eq:def_kappa} involves
an additional factor $\Mtot$.
This is because Eq.~\eqref{eq:Langevin}
uses the slow time, ${ \tau \!=\! t / N }$,
while Ref.~\cite{Elrhirhayi+2026} wrote the same Langevin equation
using the (fast) time $t$.

\section{From Langevin to Landau--Dean--Kawasaki}
\label{app:Langevin_to_LDK}

In this Appendix, we show that the empirical \DF\@ $F_N$ (Eq.~\ref{eq:def_FN}),
when driven by the $N$ coupled Langevin equations from Eq.~\eqref{eq:Langevin}, satisfies \LDK\@ (Eq.~\ref{eq:LDK}).
We follow the strategy from Dean~\cite{Dean1996}, adapted here to the pairwise antisymmetric structure of the noise.

\subsection{It\^o form of the Langevin dynamics}
\label{app:Ito_Langevin}

The Langevin equations (Eq.~\ref{eq:Langevin}) are written in the Stratonovich convention. To use standard stochastic calculus,
we first convert them to the It\^o convention. Following~\cite{Elrhirhayi+2026}, this reads
\begin{equation}
    \rd \bv_{i} = \sum_{j \neq i} \bigg[ m \, \ba_{\bv_i\bv_j} \, \rd \tau + \sqrt{m} \, \sB^{1/2}_{\bv_i\bv_j} \, \rd \bW_{ij, \tau} \bigg] ,
    \label{eq:Langevin_Ito}
\end{equation}
with the shortened notations
${\ba_{\bv\bvp} \!:=\! \ba (\bv \!-\! \bvp)}$
and
${ \sB_{\bv\bvp} \!:=\! \sB (\bv \!-\! \bvp) }$.
In Eq.~\eqref{eq:Langevin_Ito},
the drift reads
\begin{equation}
    \ba_{\bu} := \p_{\bu} \!\cdot\! \sB_{\bu} ,
    \label{eq:def_drift_a}
\end{equation}
together with the no-self-interaction prescription ${ \ba(\bZero) \!=\! \bZero }$.
We also recall that the kernel $\sB$ is symmetric and even, i.e.\ ${ \sB_{-\bu} \!=\! \sB_{\bu} }$.

\subsection{Dynamics of the empirical distribution function}
\label{app:dynamics_FN}

We consider $\phi$ an arbitrary smooth test function.
Applying It\^o's lemma to ${ \phi (\bv_i) }$,
with $\bv_{i}$ driven by Eq.~\eqref{eq:Langevin_Ito},
we find that ${ \rd \phi (\bv_{i}) }$ satisfies the \SPDE\@
\begin{align}
    \rd \phi(\bv_i) = {} & \!\!\int\!\! \rd \bv \, f_{\bv}^{i} (\tau) \bigg\{ \p_{\bv} \phi \!\cdot\! \rd \bxi_{\bv,\tau}^{i} \!+ \p_{\bv} \phi \!\cdot\! \bigg[ \sum_{j \neq i} m \, \ba_{\bv\bv_j} \bigg] \rd \tau 
    \nonumber
    \\
    & {} + \half (\p_{\bv} \p_{\bv} \phi) \!:\! \bigg[ \sum_{j \neq i} m \, \sB_{\bv\bv_j} \bigg] \rd \tau \bigg\} ,
    \label{eq:Ito_O}
\end{align}
where ${ f_{\bv}^{i} (\tau) \!:=\! \deltaD[\bv \!-\! \bv_i(\tau)] }$ is the microscopic single-particle density.
In Eq.~\eqref{eq:Ito_O},
we introduced the one-particle noise increment,
${ \rd \bxi_{\bv,\tau}^{i} }$, reading
\begin{equation}
    \rd \bxi_{\bv,\tau}^{i} := \sum_{j \neq i} \sqrt{m} \, \sB^{1/2}_{\bv\bv_j} \, \rd \bW_{ij, \tau} .
    \label{eq:def_dxi_i}
\end{equation}
In Eq.~\eqref{eq:Ito_O}, ${ (\p_{\bv}\p_{\bv}\phi) }$ denotes the Hessian matrix of $\phi$, and the symbol ``$:$'' is the full contraction of this Hessian with the diffusion matrix enclosed in square brackets.\footnote{Equivalently, for two matrices $\sA$ and $\sC$, this contraction is defined by ${ \sA \!:\! \sC \!:=\! \rTr [\sA^{\rT}\sC] }$.}

In practice, ${ \phi (\bv_{i}) }$ is also given by the identity
\begin{equation}
    \phi [ \bv_i(\tau) ] = \!\int\!\! \rd \bv \, f_{\bv}^{i}(\tau) \, \phi_{\bv} .
    \label{eq:Oper_vi}
\end{equation}
With this definition, the empirical \DF\ from Eq.~\eqref{eq:def_FN} simply reads
\begin{equation}
    F_N(\bv, \tau) = m \sum_{i} f_{\bv}^{i} (\tau) .
    \label{eq:def_FN_fi}
\end{equation}

Starting from Eq.~\eqref{eq:Oper_vi},
we find that ${ \rd \phi (\bv_{i}) }$
is equivalently given by ${ \rd \phi(\bv_i) \!=\! \!\int\! \rd \bv \, \phi (\bv) \, \rd f_{\bv}^{i} (\tau) }$.
As a result, integrating Eq.~\eqref{eq:Ito_O} by parts in velocity, we obtain the \SPDE\@ satisfied by ${ \rd f_{\bv}^{i} }$. It reads
\begin{align}
    \rd f_{\bv}^{i} (\tau) = {} & - \p_{\bv} \!\cdot\! \!\big[ f_{\bv}^{i} (\tau) \, \rd \bxi_{\bv,\tau}^{i} \big] - \p_{\bv} \!\cdot\! \!\bigg[ f_{\bv}^{i} (\tau) \sum_{j \neq i} m \, \ba_{\bv\bv_j} \bigg] \rd \tau
    \nonumber
    \\
    & {} + \half \, \p_{\bv} \p_{\bv} \!:\! \bigg[ f_{\bv}^{i} (\tau) \sum_{j \neq i} m \, \sB_{\bv\bv_j} \bigg] \rd \tau .
    \label{eq:SPDE_fi}
\end{align}
Following Eq.~\eqref{eq:def_FN_fi}, we may sum Eq.~\eqref{eq:SPDE_fi} over all particles to obtain the \SPDE\@ for ${ \rd F_{N,\bv} \!:=\! \rd F_{N} (\bv) }$. To do so, we extend the sums from ${ j \!\neq\! i }$ to all $j$, using the no-self-interaction assumption ${ \sB(\bZero) \!=\! \sZero }$ and ${ \ba(\bZero) \!=\! \bZero }$.
We also replace discrete sums of the form ${ \sum_{j} m ... (\bv_j) }$ with the integral ${ \!\int\! \rd \bvp \, F_N(\bvp) ... (\bvp) }$.
Overall, we obtain
\begin{align}
    {} & \rd F_{N,\bv}(\tau) = - \p_{\bv} \!\cdot\! \bigg[ F_{N,\bv} \!\!\int\!\! \rd \bvp \, \ba_{\bv\bvp} \, F_{N,\bvp} \bigg] \rd \tau
    \nonumber
    \\
    & {} + \half \p_{\bv} \p_{\bv} \!:\! \bigg[ F_{N,\bv} \!\!\int\!\! \rd \bvp \,\sB_{\bv\bvp} F_{N,\bvp} \bigg]  \rd \tau + \rd \xi_{\bv,\tau} ,
    \label{eq:FN_evol}
\end{align}
with the noise
\begin{equation}
    \rd \xi_{\bv,\tau} := - \p_{\bv} \!\cdot\! \bigg[ \sum_{i} m \, f_{\bv}^{i} (\tau) \, \rd \bxi_{\bv,\tau}^{i}  \bigg] .
    \label{eq:def_xi_field}
\end{equation}

\subsection{Recovering the Landau operator}
\label{app:recovering_Landau}

We now show that the deterministic part of Eq.~\eqref{eq:FN_evol} corresponds to the Landau drift of \LDK\@. Using the definition from Eq.~\eqref{eq:def_drift_a}, together with ${ \p_{\bv}\!\cdot\!\sB_{\bv\bvp} \!=\! -\p_{\bvp}\!\cdot\!\sB_{\bv\bvp} }$,
we can write
\begin{equation}
    \!\int\!\! \rd \bvp \, \ba_{\bv\bvp} \, F_{N,\bvp}  = \!\int\!\! \rd \bvp \, \sB_{\bv\bvp} \!\cdot\! \p_{\bvp} F_N .
    \label{eq:identity_drift}
\end{equation}
By the very same manipulation, we obtain
\begin{equation}
    \p_{\bv} \!\cdot\! \!\int\!\! \rd \bvp \, \sB_{\bv\bvp} \, F_{N,\bvp} \!=\! \!\int\!\! \rd \bvp \, \sB_{\bv\bvp} \!\cdot\! \p_{\bvp} F_N .
    \label{eq:identity_div}
\end{equation}
Expanding the second-order term of Eq.~\eqref{eq:FN_evol} as
\begin{align}
    \half \, \p_{\bv} \p_{\bv} \!:\! \bigg[ F_{N,\bv} & \!\!\int\!\! \rd \bvp \, \sB_{\bv\bvp} \, F_{N,\bvp} \bigg] \nonumber \\
    = {} & \half \, \p_{\bv} \!\cdot\! \bigg[ \bigg( \!\int\!\! \rd \bvp \, \sB_{\bv\bvp} \, F_{N,\bvp} \bigg) \!\cdot\! \p_{\bv} F_N \nonumber \\
    & \qquad + F_{N,\bv} \, \p_{\bv} \!\cdot\!\! \int\!\! \rd \bvp \, \sB_{\bv\bvp} \, F_{N,\bvp} \bigg] ,
    \label{eq:expand_diffusion}
\end{align}
and combining it with the drift term via Eqs.~\eqref{eq:identity_drift} and~\eqref{eq:identity_div}, the deterministic part of Eq.~\eqref{eq:FN_evol} reduces to
\begin{equation}
    \mathrm{Drift} = \half \p_{\bv} \!\cdot\! \!\int\!\! \rd \bvp \sB_{\bv\bvp} \bigg[ F_{N,\bvp} \p_{\bv} F_N \!-\! F_{N,\bv} \p_{\bvp} F_N \bigg] \rd \tau .
    \label{eq:Det_Landau}
\end{equation}
This is exactly the Landau collision operator, $\mLL$, from Eq.~\eqref{eq:Landau_operator}.
As a consequence, the evolution of the empirical \DF\@ takes the form
\begin{equation}
    \rd F_N(\bv, \tau) = \mLL [F_N, F_N](\bv) \, \rd \tau + \rd \xi_{\tau} (\bv) ,
    \label{eq:LDK_unknown_noise}
\end{equation}
with ${ \rd \xi_{\tau} (\bv) }$
given by Eq.~\eqref{eq:def_xi_field}.
It now remains to show that the statistics of ${ \rd \xi_{\tau} (\bv) }$ coincides with the one of the \LDK\@ noise ${ \rd \mW_{\tau}[F_N] }$, from Eq.~\eqref{eq:LDK}.

\subsection{Statistics of the noise}
\label{app:noise_statistics}

We start from Eq.~\eqref{eq:def_xi_field}.
Since the ${ \rd \bW_{ij, \tau} }$ are forward It\^o increments, they are independent of the state of the system at time $\tau$. As a consequence, the noise is of zero mean, ${ \langle \rd \xi_{\bv,\tau} \rangle \!=\! 0 }$.

Second, exploiting the odd-parity of ${ \rd \bW_{ij,\tau} }$ and the even-parity of ${ \sB^{1/2}_{\bv_{i}\bv_{j}} }$, the double sum in Eq.~\eqref{eq:def_xi_field} can be recast as a sum over distinct pairs ${ (i \!<\! j) }$,
\begin{equation}
    \rd \xi_{\bv, \tau} = - \p_{\bv} \!\cdot\! \bigg[ \sum_{\substack{i , j \\ i < j}} m^{3/2} \sB^{1/2}_{\bv_i\bv_j} \Delta f^{ij}_{\bv} (\tau)  \rd \bW_{ij, \tau} \bigg] ,
    \label{eq:xi_i<j}
\end{equation}
where we introduced ${ \Delta f^{ij}_{\bv} (\tau) \!:=\! f_{\bv}^{i} (\tau) \!-\! f_{\bv}^{j} (\tau) }$. In that expression: (i) the divergence form ensures mass conservation; (ii) the dependence on ${ \Delta f_{\bv}^{ij} }$ ensures momentum conservation;
(iii) the conservation of the kinetic energy relies on the projector property, ${ \sB^{1/2}_{\bu} \bu \!=\! \bZero }$.

Since increments associated with distinct pairs are independent, ${ \langle \rd \bW_{ij, \tau} \!\otimes\! \rd \bW_{kl, \tau} \rangle \!=\! \delta_{ik} \delta_{jl} \, \sI \, \rd \tau }$ (for ${ i \!<\! j }$ and ${ k \!<\! l }$),
the equal-time covariance of ${ \rd \xi }$ directly follows from Eq.~\eqref{eq:xi_i<j}.
It reads
\begin{equation}
    \big\langle \rd \xi_{\bv,\tau}  \rd \xi_{\bvp,\tau} \big\rangle \!=\! \p_{\bv} \!\cdot\! \p_{\bvp} \!\cdot\! \!\bigg[ \!\sum_{\substack{i , j \\ i < j}} \! m^{3} \sB_{\bv_i\bv_j} \Delta f^{ij}_{\bv} (\tau) \Delta f^{ij}_{\bvp} (\tau) \!\bigg] \rd \tau ,
\end{equation}
where we used ${ \sB^{1/2} [\sB^{1/2}]^{\rT} \!=\! \sB }$.
We now extend the sum from ${ i \!<\! j }$ to all pairs ${ (i,j) }$.
Once again, since ${ \sB_{\bZero} \!=\! \sZero }$, the diagonal contribution ${ i \!=\! j }$ does not contribute.
One gets
\begin{equation}
    \big\langle \rd \xi_{\bv,\tau}  \rd \xi_{\bvp,\tau} \big\rangle \!=\! \half \p_{\bv} \!\cdot\! \p_{\bvp} \!\cdot\! \!\bigg[ \!\sum_{i, j} m^{3} \sB_{\bv_i\bv_j} \Delta f^{ij}_{\bv} (\tau) \Delta f^{ij}_{\bvp} (\tau) \!\bigg] \rd \tau .
    \label{eq:xi_covariance_raw}
\end{equation}

The next step of the calculation
is to expand the product
\begin{equation}
    \Delta f^{ij}_{\bv} \, \Delta f^{ij}_{\bvp} = f_{\bv}^{i} f_{\bvp}^{i} + f_{\bv}^{j} f_{\bvp}^{j}
    - f_{\bv}^{i} f_{\bvp}^{j} - f_{\bv}^{j} f_{\bvp}^{i} .
    \label{eq:expand_Delta_f}
\end{equation}
In that expression,
two types of terms emerge.
These can be evaluated using the microscopic definition ${ f_{\bv}^{i} \!:=\! \delta_{\bv_{i} \bv} }$.

\textit{Diagonal terms}. The terms ${ f_{\bv}^{i} f_{\bvp}^{i} }$ and ${ f_{\bv}^{j} f_{\bvp}^{j} }$ yield identical sums by swapping the dummy indices $i$ and $j$. This symmetry cancels the overall ${ 1/2 }$ factor in Eq.~\eqref{eq:xi_covariance_raw}. Using the identity ${ f_{\bv}^{i} f_{\bvp}^{i} \!=\! \delta_{\bv \bvp} f_{\bv}^{i} }$, the kernel ${ \sB_{\bv_i\bv_j} }$ is directly evaluated at ${ \bv_i \!=\! \bv }$. The diagonal contribution simplifies to
    \begin{align}
        \half \sum_{i, j} m^3 \sB_{\bv_i\bv_j} \big[ & f_{\bv}^{i} f_{\bvp}^{i} \!+\! f_{\bv}^{j} f_{\bvp}^{j} \big] = \sum_{i, j} m^3 \sB_{\bv_i\bv_j} f_{\bv}^{i} \delta_{\bv\bvp}
        \nonumber
        \\
        = {} & m \, \delta_{\bv\bvp} \bigg[ m \sum_{i} f_{\bv}^{i} \bigg] \bigg[ m \sum_{j} \sB_{\bv\bv_j} \bigg]
        \nonumber
        \\
        = {} & m \, \delta_{\bv\bvp} \, F_{N,\bv} \!\!\int\!\! \rd \bvpp \, \sB_{\bv\bvpp} \, F_N(\bvpp) .
        \label{eq:diagonal_contribution}
    \end{align}

\textit{Cross terms}. Similarly, the cross terms ${ f_{\bv}^{i} f_{\bvp}^{j} }$ and ${ f_{\bv}^{j} f_{\bvp}^{i} }$ in Eq.~\eqref{eq:expand_Delta_f}
contribute identically and cancel the ${ 1/2 }$ factor in Eq.~\eqref{eq:xi_covariance_raw}. The product ${ f_{\bv}^{i} f_{\bvp}^{j} \!=\! \delta_{\bv_i\bv} \delta_{\bv_j\bvp} }$ acts as a double Dirac distribution, directly forcing the evaluation of the kernel at ${ \bv_i \!=\! \bv }$ and ${ \bv_j \!=\! \bvp }$. The cross contribution becomes
\begin{align}
    - \half \sum_{i, j} m^3 \sB_{\bv_i\bv_j} \big[ & f_{\bv}^{i} f_{\bvp}^{j} \!+\! f_{\bv}^{j} f_{\bvp}^{i} \big] = - \sum_{i, j} m^3 \sB_{\bv_i\bv_j} f_{\bv}^{i} f_{\bvp}^{j} 
    \nonumber
    \\
    = {} & - m \sB_{\bv\bvp} \bigg[ m \sum_{i} f_{\bv}^{i} \bigg] \bigg[ m \sum_{j} f_{\bvp}^{j} \bigg]
    \nonumber
    \\
    = {} & - m \sB_{\bv\bvp} F_{N,\bv} F_{N,\bvp} .
    \label{eq:cross_contribution}
\end{align}

Gathering both contributions into the double divergence, we finally obtain the covariance of the noise
\begin{align}
    \big\langle \rd \xi_{\bv,\tau} \rd \xi_{\bvp,\tau} \big\rangle {} & = m \p_{\bv} \!\cdot\! \bigg\{ \p_{\bvp} \!\cdot\! \Big[ \!-\!  F_{N,\bv} F_{N,\bvp}  \sB_{\bv\bvp}
    \nonumber
    \\
 + \, {} & \delta_{\bv \bvp} F_{N,\bv} \!\!\int\!\! \rd \bvpp \, \sB_{\bv\bvpp}  F_N(\bvpp) \Big] \bigg\} \rd \tau
 \nonumber
 \\
    & {} = \tfrac{1}{N} C[F_N]_{\bv\bvp} \rd \tau ,
    \label{eq:xi_covariance}
\end{align}
with the covariance $C{ [F_{N}] }$ introduced in Eq.~\eqref{eq:covariance_kernel}.

Given the state of the system at time $\tau$,
the noise ${ \rd \xi }$ is a superposition of independent Gaussian increments.
It is therefore itself Gaussian, and its statistics is entirely characterized by its (vanishing) mean and the covariance from Eq.~\eqref{eq:xi_covariance}. Identifying ${ \rd \xi \!=\! N^{-1/2} \rd \mW_{\tau}[F_N] }$, Eq.~\eqref{eq:LDK_unknown_noise} becomes \LDK\@ (Eq.~\ref{eq:LDK}).

\section{Initial conditions}
\label{app:IC}

In this Appendix, we derive the initial conditions
for the rescaled moments, $S_{n}$ (Eq.~\ref{eq:def_Sn}).
We emphasize that at ${ \tau \!=\! 0 }$, the moments do not depend on the dynamics, so that ${ \SDK_n(0) \!=\! \SNB_n(0) }$.
Their leading-order contributions define the initial conditions of $\Smin_n$.
In this appendix, we therefore drop the superscripts and write ${ S_{n} \!=\! \Smin_{n} }$.
In the following, we first evaluate $M_n$ (Eq.~\ref{eq:def_Mn}).
Then, we keep only the leading terms in ${1/N}$,
to obtain the initial conditions of $S_n$.

\subsection{Initial conditions for \texorpdfstring{$S_2$}{S2}}
\label{app:IC_S2}

We first focus on $S_2$. From the definition of $M_2$ (Eq.~\ref{eq:def_Mn}), the initial two-point moment is given by
\begin{align}
    M_{\1\2}(0) & {} = \big\langle \delta F_{\1}(0) \, \delta F_{\2} (0) \big\rangle ,
    \label{eq:def_S2_IC}
\end{align}
where here the average ${ \langle \cdot \rangle }$ is over the initial conditions.
Since fluctuations are centered, ${ \langle \delta F \rangle \!=\! 0 }$.
Denoting ${ F_{0} (\1) \!:=\! F(\1, \tau \!=\! 0)}$
as well as ${ P_{0} (\1) \!:=\! F_{0} (\1) / \Mtot }$,
we can expand the empirical distribution as
\begin{equation}
    \delta F_\1 (0) = m \sum_{i} \deltaf^{i}_{\1} ,
\end{equation}
where we defined the initial single-particle fluctuation
\begin{equation}
    \deltaf^{i}_{\1} := \delta_{\bv_i\1} - P_0(\1) ,
    \label{eq:def_delta_i_v}
\end{equation}
recalling our notation ${ \delta_{\bv_i\1} \!:=\! \deltaD (\bv_{i} \!-\! \bv_{1}) }$.
Substituting Eq.~\eqref{eq:def_delta_i_v} into Eq.~\eqref{eq:def_S2_IC}, we have
\begin{align}
    M_{\1\2}(0) & {} =  m^2 \sum_{i,j}  \big\langle \deltaf^{i}_{\1} \deltaf^{j}_{\2} \big\rangle .
    \label{eq:calc_avg_M2}
\end{align}
Assuming that the particles are initially independent, any cross-term in Eq.~\eqref{eq:calc_avg_M2} with ${ i \!\neq\! j }$ identically vanishes since ${ \langle \deltaf^{i}_{\1} \deltaf^{j}_{\2} \rangle \!=\! \langle \deltaf^{i}_{\1} \rangle \langle \deltaf^{j}_{\2} \rangle \!=\! 0 }$. The double sum therefore reduces to its diagonal contribution reading
\begin{align}
     M_{\1\2}(0) & {} = m^2 \sum_{i} \big\langle \deltaf^{i}_{\1} \deltaf^{i}_{\2} \big\rangle .
\end{align}
Assuming that particles are identically distributed according to $F_0$, the $N$ terms in the sum are identical. Using an arbitrary particle $i$ as a reference, we obtain
\begin{align}
     M_{\1\2}(0) = m^2 N \, \big\langle \deltaf^{i}_{\1} \deltaf^{i}_{\2} \big\rangle .
     \label{eq:M2_delta_1}
\end{align}
Expanding the expectation value of the product yields
\begin{align}
    \big\langle \deltaf^{i}_{\1} \deltaf^{i}_{\2} \big\rangle = {} & \big\langle \big [\delta_{\bv_i \1} \!-\! P_0(\1)] [\delta_{\bv_i \2} \!-\! P_0(\2)] \big\rangle
    \nonumber
    \\
    = {} & \big\langle \delta_{\bv_i \1} \delta_{\bv_i \2} \big\rangle \!-\! P_0(\1) P_0(\2) .
    \nonumber
\end{align}
Using the property of the Dirac delta, ${ \delta_{\bv_i \1} \delta_{\bv_i \2} \!=\! \delta_{\1 \2} \delta_{\bv_i \1} }$, the first term simplifies to
\begin{align}
    \big\langle \delta_{\bv_i \1} \delta_{\bv_i \2} \big\rangle & {} = \delta_{\1 \2} \, \big\langle \delta_{\bv_i \1} \big\rangle
    \nonumber
    \\ 
    & {} = \delta_{\1 \2} \, P_0(\1) .
\end{align}
Recombining these terms into Eq.~\eqref{eq:M2_delta_1}, and using the rescaling relation ${ S_2 \!=\! N M_2 }$ together with ${ m^2 N^2 \!=\! \Mtot^2 }$, we finally obtain
\begin{equation}
    S_{\1\2}(0) = \Mtot \delta_{\1 \2} F_0(\1) \!-\! F_0(\1) F_0(\2) .
    \label{eq:S2_initial}
\end{equation}
The first term accounts for phase-space granularity (finite-$N$ shot noise), while the ${ -F_0(\1) F_0(\2) }$ correction arises because $S_2$ is a central moment.

\subsection{Initial conditions for \texorpdfstring{$S_3$}{S3}}
\label{app:IC_S3}

We now consider $S_3$.
At ${ \tau \!=\! 0 }$, we have
\begin{align}
    M_{\1\2\3}(0) := {} & \big\langle \delta F_{\1}(0) \delta F_{\2}(0) \delta F_{\3}(0) \big\rangle
    \nonumber
    \\
    = {} & \big\langle (F_N(\1 , 0) \!-\! F_0(\1))(F_N(\2 , 0) \!-\! F_0(\2))
    \nonumber
    \\
    \times \, & {} \,\, (F_N(\3 , 0) \!-\! F_0(\3)) \big\rangle .
\end{align}
Substituting the single-particle fluctuation ${\deltaf^{i}_{\bv}}$ defined in Eq.~\eqref{eq:def_delta_i_v} into the expression of $M_3$ yields a triple sum over the particle indices
\begin{equation}
    M_{\1\2\3}(0) = m^3 \sum_{\mathclap{i,j,k}} \big\langle \deltaf^{i}_{\1} \deltaf^{j}_{\2} \deltaf^{k}_{\3} \big\rangle .
    \label{eq:sum_ijk_S_3_init}
\end{equation}
Since ${ \langle \deltaf^{i}_{\1} \rangle \!=\! 0 }$ for all $i$, any term in this sum in which one particle index appears only once vanishes identically.
Consequently, the only non-vanishing contributions arise from the fully diagonal terms where ${ i \!=\! j \!=\! k }$. This gives
\begin{equation}
    M_{\1\2\3}(0) = m^3 \sum_{i} \big\langle \deltaf^{i}_{\1} \deltaf^{i}_{\2} \deltaf^{i}_{\3} \big\rangle .
\end{equation}
Using an arbitrary particle $i$ as a reference, we obtain
\begin{equation}
    M_{\1\2\3}(0) = \Mtot m^2 \, \big\langle \deltaf^{i}_{\1} \deltaf^{i}_{\2} \deltaf^{i}_{\3} \big\rangle .
    \label{eq:calc_M3_progress}
\end{equation}
This expression confirms that the three-point moment scales as ${ \mO(1/N^2) }$ at the initial time, as stated in Eq.~\eqref{eq:def_Sn}.

To evaluate Eq.~\eqref{eq:calc_M3_progress},
we expand using Eq.~\eqref{eq:def_delta_i_v}.
We obtain
\begin{equation}
    \big\langle \deltaf^{i}_{\1} \deltaf^{i}_{\2} \deltaf^{i}_{\3} \big\rangle \!=\! \big\langle \big[ \delta_{\bv_i \1} \!-\! P_0(\1) \big] \big[ \delta_{\bv_i \2} \!-\! P_0(\2) \big] \big[ \delta_{\bv_i \3} \!-\! P_0(\3) \big] \big\rangle .
\end{equation}
We can then write
\begin{align}
    \big\langle &\deltaf^{i}_{\1} \deltaf^{i}_{\2} \deltaf^{i}_{\3} \big\rangle = \big\langle \delta_{\bv_i \1} \delta_{\bv_i \2} \delta_{\bv_i \3} \big\rangle - \big\langle \delta_{\bv_i \1} \delta_{\bv_i \2} P_0(\3) \big\rangle
    \nonumber
    \\
    & - \big\langle \delta_{\bv_i \1} \delta_{\bv_i \3} P_0(\2) \big\rangle - \big\langle \delta_{\bv_i \2} \delta_{\bv_i \3} P_0(\1) \big\rangle
    \nonumber
    \\
    & + \big\langle \delta_{\bv_i \1} P_0(\2) P_0(\3) \big\rangle + \big\langle \delta_{\bv_i \2} P_0(\1) P_0(\3) \big\rangle
    \nonumber
    \\
    & + \big\langle \delta_{\bv_i \3} P_0(\1) P_0(\2) \big\rangle  \!-\! P_0(\1) P_0(\2) P_0(\3) .
\end{align}
Using the property ${ \langle \delta_{\bv_i \1} \rangle \!=\! P_0(\1) }$,
the triple product of Dirac distributions simplifies into
\begin{equation}
    \big\langle \delta_{\bv_i \1} \delta_{\bv_i \2} \delta_{\bv_i \3} \big\rangle = \delta_{\1 \2} \delta_{\2 \3} P_0(\3) .
\end{equation}
Similarly, the quadratic Dirac contributions reduce to
\begin{align}
    \big\langle \delta_{\bv_i \1} & \delta_{\bv_i \2} P_0(\3) \big\rangle = \delta_{\1 \2} P_0(\2) P_0(\3) .
\end{align}
Recombining all these terms and injecting the result into the relation ${ S_3 \!=\! N^2 M_3 }$, we obtain
\begin{align}
    S_{\1\2\3}(0) = & {} \Mtot^2 \delta_{\1 \2} \delta_{\2 \3} F_0(\3) \!-\! \Mtot \delta_{\1 \2} F_0(\2) F_0(\3)
    \nonumber
    \\
    - \, {} & \Mtot \delta_{\2 \3} F_0(\1) F_0(\2) \!-\! \Mtot \delta_{\1 \3} F_0(\2) F_0(\3)
    \nonumber
    \\
    + \, {} & 2 F_0(\1) F_0(\2) F_0(\3) .
    \label{eq:IC_S3}
\end{align}
The first term reflects the fully diagonal single-particle contribution associated with finite-$N$ shot noise.
The second set of terms represents the pairwise corrections induced by centering the fluctuations, and the final background term results from the recentering.

\subsection{Initial conditions for \texorpdfstring{$S_4$}{S4} and \texorpdfstring{$S_5$}{S5}}
\label{app:IC_S4_S5}

Using the same counting argument as before, the leading contributions to $M_4$ arise from the ${2 \!+\! 2}$ partitions of the particles.
These involve two independent particle labels and therefore scale as $N^{-2}$. 
With the rescaling ${ S_4 \!=\! N^2 M_4 }$, at ${ \tau \!=\! 0 }$, we obtain
\begin{align}
    S_{\1\2\3\4} \!=\! S_{\1\2} S_{\3\4} \!+\! S_{\1\3} S_{\2\4} \!+\! S_{\1\4} S_{\2\3} .
    \label{eq:S4_initial}
\end{align}

Similarly, the leading contributions to $M_5$ arise from the ${ 2 \!+\! 3 }$ partitions of the particles. With the rescaling ${ S_5 \!=\! N^3 M_5 }$, at ${ \tau \!=\! 0 }$, we obtain
\begin{align}
    S_{\1\2\3\4\5} = {} & 
    S_{\1\2} S_{\3\4\5} + 
    S_{\1\3} S_{\2\4\5} + 
    S_{\1\4} S_{\2\3\5} + 
    S_{\1\5} S_{\2\3\4}
    \nonumber
    \\
    + \, {} & S_{\2\3} S_{\1\4\5} + 
    S_{\2\4} S_{\1\3\5} +
    S_{\2\5} S_{\1\3\4} + S_{\3\4} S_{\1\2\5}
    \nonumber
    \\
    + \, {} & S_{\3\5} S_{\1\2\4} + 
    S_{\4\5} S_{\1\2\3}.
    \label{eq:S5_initial}
\end{align}

\subsection{Generalization}
\label{app:IC_Sn}

For higher-order moments, the leading initial condition can be
identified by considering the multiplicities of the particle labels
appearing in the expansion of $M_n$. 
Since each single-particle fluctuation is centered, any contribution containing a given particle label only once vanishes. Nonzero contributions are therefore associated with partitions of $n$ into blocks of size at least two.

If such a partition contains $r$ blocks, the corresponding term
involves $r$ independent sums over particle labels and scales as $N^{r-n}$.
The leading contribution is consequently obtained by maximizing the
number of blocks. For even $n$, the unique leading structure consists
exclusively of pairs,
\begin{equation}
   n=2+\cdots+2,
\end{equation}
whereas, for odd $n$, it consists of one triplet and pairs,
\begin{equation}
   n=3+2+\cdots+2.
\end{equation}
It follows that, at ${ \tau=0 }$, 
\begin{align}
    S_{\1\cdots\mathbf{2k}} & {} = \mS \big[ S_{\1\2} \!\times\! \cdots \!\times\! S_{(\mathbf{2k-1})(\mathbf{2k})} \big] ,
\label{eq:IC_S2k}
\end{align}
while for odd moments, we have
\begin{align}
    S_{\1\cdots\mathbf{2k+1}} = \mS \big[ S_{\1\2\3} \!\times\! S_{\4\5} 
    \!\times\! {} & \cdots \!\times\! S_{(\mathbf{2k})(\mathbf{2k+1})} \big] .
\label{eq:IC_S2k+1}
\end{align}
In both expressions, $\mS$ is the symmetrization operator (see Appendix~\ref{app:S6_Sn_closure} for details).
As argued in the main text,
these decompositions of the higher-order moments
remain valid throughout the long-term evolution: this is the quasi-Gaussian closure.
In Appendices~\ref{app:S4_closure} and~\ref{app:S5_closure}, we prove that the moments defined by the quasi-Gaussian closure for $S_{4}$ and $S_{5}$
coincide with the solutions $\ST_4$ and $\ST_5$ of the truncated hierarchy. 
In Appendix~\ref{app:S6_Sn_closure}, we illustrate that this proof extends similarly to moments of higher order.

\section{Derivation of the minimal hierarchy for \texorpdfstring{$F$, $S_2$ and $S_3$}{F, S2 and S3}}
\label{app:minimal_hierarchy_S2_S3}

In this Appendix, we derive the three kinetic equations needed to build the minimal hierarchy, namely the evolution equations for $F$, $\Smin_2$ and $\Smin_3$ at leading order in $1/N$ (Eqs.~\ref{eq:Landau},~\ref{eq:Lyapunov_S2} and~\ref{eq:Lyapunov_S3}). 

Let us define the exact finite-$N$ mean and the centered fluctuation as
\begin{align}
    \barF_N(\bv,\tau) & {} :=\langle F_N(\bv,\tau)\rangle, \\
    \delta F(\bv,\tau) & {} :=F_N(\bv,\tau)-\barF_N(\bv,\tau) ,
    \label{eq:def_barFN}
\end{align}
where, here, ${ \langle \cdot \rangle }$ stands for the average
over realizations of \LDK\@ (Eq.~\ref{eq:LDK}).
Unlike $F$, which solves the Landau equation, $\barF_N$ includes finite-$N$ corrections.
Throughout this appendix, $M_n$ denotes the moments of $F_N$ evolving according to \LDK\@, so that ${ \SDK_n \!=\! N^{\lceil n/2\rceil} M_n }$.

We start from \LDK\@ (Eq.~\ref{eq:LDK}).
Taking its expectation value, the exact
finite-$N$ mean $\barF_N$ satisfies
\begin{equation}
    \rd \barF_N = \mLL [\barF_N,\barF_N] \, \rd \tau + \mLL [M_2] \, \rd \tau ,
    \label{eq:Landau_1/N2}
\end{equation}
where
\begin{align}
    \mLL [M_2]_{\1} \!:=\! {} & \half \p_{\1} \!\cdot\! \!\!\int\!\! \rd \2 \, \sB_{\1 \2} \big[ \p_{\1} \!-\! \p_{\2} \big] M_{\1\2} .
\end{align}
Subtracting Eq.~\eqref{eq:Landau_1/N2} from \LDK\@, the centered fluctuation field ${ \delta F }$ satisfies
\begin{align}
    \rd \delta F_{\1} = {} & \omL_{\1} \delta F_{\1} \, \rd \tau
    + \mLL [\delta F,\delta F]_{\1} \, \rd \tau
    \nonumber
    \\
    - \, {} & \mLL [M_2]_{\1} \, \rd \tau
    + \tfrac{1}{\sqrt{N}} \, \rd \mW_{\tau, \1} ,
    \label{eq:LDK_deltaF}
\end{align}
where $\omL_{\bv}$ is the Landau operator linearized around $\barF_N$ and acting on the single variable $\bv$ in ${ \delta F }$.
It reads
\begin{align}
    \omL_{\1} \deltaF_\1 := {} & \half \p_{\1} \!\cdot\!\!\int\!\! \rd \2 \, \sB_{\1 \2} \bigg[ \p_{\1} \barF_N  \deltaF_{\2} \!-\! \p_{\2} \barF_N \deltaF_{\1} \nonumber \\
    & {} + \barF_{N,\2} \p_{\1} \delta F \!-\! \barF_{N,\1} \p_{\2} \delta F \bigg] .
    \label{eq:def_omL}
\end{align}
The nonlinear term in Eq.~\eqref{eq:LDK_deltaF} reads
\begin{align}
    \mLL [\delta F,\delta F]_\1 := {} & \half \p_{\1} \!\cdot\!\!\int\!\! \rd \2 \, \sB_{\1 \2}  \big[ \p_{\1} \!-\! \p_{\2} \big] \deltaF_\1 \deltaF_{\2} .
\end{align}
The noise increment obeys the covariance relation
\begin{equation}
    \big\langle \rd \mW_{\tau, \1}\rd \mW_{\tau, \2} \big\rangle
    = C[F_N]_{\1 \2}\,\rd \tau ,
\end{equation}
where here $\langle\cdot\rangle$ denotes the average over the Wiener increment, conditional on the \DF\@ $F_N$ at time $\tau$.
Up to this point, the evolution equations are exact.

Since ${ M_2 \!=\! \mathcal O(N^{-1}) }$, Eq.~\eqref{eq:Landau_1/N2} gives ${\barF_N \!=\! F \!+\! \mO(N^{-1})}$ at fixed time.
We distinguish the exact operator $\omL$, linearized around $\barF_N$, from the leading-order operator
\begin{equation}
    \mL_{\1}:= \omL_{\1} \big|_{\barF_N=F},
    \label{eq:def_linL_F}
\end{equation}
linearized around $F$. In deriving the minimal equations for the rescaled moments $\Smin_2$ and $\Smin_3$, replacing $\barF_N$ by $F$ in their coefficients produces only $\mathcal O(N^{-1})$ corrections.
These corrections are discarded at the order considered here.
Importantly, the centering is not changed: $\delta F$ and its moments remain defined relative to $\barF_N$.

\subsection{Landau equation}
\label{app:Landau_equation_derivation}

To recover the Landau equation (Eq.~\ref{eq:Landau}), we use the expansion ${ \barF_{N} \!=\! F \!+\! \mO (1/N) }$ in Eq.~\eqref{eq:Landau_1/N2}.
We note that the contribution involving $M_{2}$ in Eq.~\eqref{eq:Landau_1/N2} is of relative order ${ 1/N }$.
It is discarded in the minimal hierarchy.
We recover
\begin{equation}
    \rd F = \mLL [F , F] \, \rd \tau .
    \label{eq:recovering_Landau}
\end{equation}

\subsection{Two-point moment \texorpdfstring{$S_2$}{S2}}
\label{app:S2_derivation}
We now consider the dynamics of the moment
${ M_{\1\2} \!:=\! \langle \deltaF_{\1}\deltaF_{\2} \rangle }$,
where ${ \langle\cdot\rangle }$
denotes the ensemble average over both the random initial
conditions and the \LDK\@ noise up to time $\tau$.
Applying It\^o's lemma yields
\begin{equation}
    \rd(\deltaF_{\1}\deltaF_{\2})
    = \deltaF_{\2}\rd\deltaF_{\1}
    + \deltaF_{\1}\rd\deltaF_{\2}
    + \rd\deltaF_{\1}\rd\deltaF_{\2}.
\end{equation}

We first consider the term ${\deltaF_{\2}\rd\deltaF_{\1}}$.
Using Eq.~\eqref{eq:LDK_deltaF}, it reads
\begin{align}
    \deltaF_{\2}\rd\deltaF_{\1}
    = {} &
    \deltaF_{\2}\omL_{\1}\deltaF_{\1}\,\rd\tau
    +\deltaF_{\2}\mLL[\deltaF,\deltaF]_{\1}\,\rd\tau
    \nonumber\\
    - {} &
    \deltaF_{\2}\mLL[M_2]_{\1}\,\rd\tau
    +\deltaF_{\2}\tfrac{1}{\sqrt N}\rd\mW_{\tau,\1}.
    \label{eq:drift_S2}
\end{align}
We now take the expectation value of this expression.
The first term gives the linear advection
\begin{equation}
    \big\langle\deltaF_{\2}\omL_{\1}\deltaF_{\1}\big\rangle
    =\omL_{\1}M_{\1\2}.
    \label{eq:calc_term1_spde_s2}
\end{equation}
The second term in Eq.~\eqref{eq:drift_S2} gives the nonlinear
contribution, namely
\begin{align}
    \big\langle\deltaF_{\2}\mLL[\deltaF,\deltaF]_{\1}\big\rangle
    = {} & \half\p_{\1}\!\cdot\!\!\int\!\rd\bv \,
    \sB_{\1\bv}[\p_{\1}-\p_{\bv}]M_{\1\2\bv}
    \nonumber\\
    =: {} & \mN[M_3]_{\1;\2} ,
    \label{eq:def_nonl_N}
\end{align}
where the label before the semicolon identifies the active variable and emphasizes that the nonlinear operator
acts on $\1$ and ignores $\2$.

The term ${\deltaF_{\2}\mLL[M_2]_{\1}}$ in
Eq.~\eqref{eq:drift_S2} gives a null contribution. Indeed,
\begin{equation}
    \big\langle\deltaF_{\2}\mLL[M_2]_{\1}\big\rangle
    =\big\langle\deltaF_{\2}\big\rangle
     \mLL[M_2]_{\1}=0,
\end{equation}
since ${\langle\deltaF_{\2}\rangle=0}$.
The last term in Eq.~\eqref{eq:drift_S2} also vanishes because the forward It\^o increments are independent of
the system's current state at time $\tau$.
Thus, ${\langle\deltaF_{\2}\rd\mW_{\tau,\1}\rangle=0}$.
The same arguments apply to the symmetric term
${\deltaF_{\1}\rd\deltaF_{\2}}$ upon exchanging $\1$ and $\2$.

We now turn to the cross-term
${\rd\deltaF_{\1}\rd\deltaF_{\2}}$ in the It\^o product rule.
Its expectation is
\begin{equation}
    \big\langle\rd\deltaF_{\1}\rd\deltaF_{\2}\big\rangle
    =\tfrac1N\big\langle C[F_N]_{\1\2}\big\rangle\,\rd\tau.
    \label{eq:calc_crossterm_S2}
\end{equation}
Combining these results, we obtain the
exact equation within the \LDK\@ description
for $M_2$. It reads
\begin{align}
    \p_\tau M_{\1\2}
    = {} &
    [\omL_{\1}+\omL_{\2}]M_{\1\2}
    +\mN[M_3]_{\1;\2}
    \nonumber\\
    &+\mN[M_3]_{\2;\1}
    +\tfrac1N\big\langle C[F_N]_{\1\2}\big\rangle.
    \label{eq:M2_unrescaled}
\end{align}

We now retain only the leading-order terms in the equation for $M_2$.
Since ${ \barF_N \!=\! F \!+\! \mO(N^{-1}) }$, we have ${ \omL_{\1} \!=\! \mL_{\1} \!+\! \mO(N^{-1}) }$, where $\mL_{\1}$ is linearized around $F$ (Eq.~\ref{eq:def_linL_F}). 
Because ${ M_2 \!=\! \mO(N^{-1}) }$, replacing $\omL_{\1}$ by $\mL_{\1}$ changes the equation only at order $N^{-2}$.
Likewise, ${\langle C[F_N]_{\1\2}\rangle \!=\! C[F]_{\1\2} \!+\! \mO(N^{-1}) }$, so the correction to the covariance term, which carries a prefactor ${ 1/N }$, is also of order $N^{-2}$.
Finally, the nonlinear terms involving $M_3$ are of order $N^{-2}$, since ${ M_3 \!=\! \mO(N^{-2}) }$. 
All these contributions can therefore be neglected at leading order in the equation for $M_2$.

Rescaling via ${ \SDK_2 \!=\! N M_2 }$ and discarding the $\mO(N^{-1})$ corrections, the leading-order contribution $\Smin_2$ satisfies
\begin{equation}
    \p_\tau \Smin_{\1\2} = [\mL_{\1} \!+\!\mL_{\2}]\Smin_{\1\2} + C[F]_{\1\2} .
    \label{eq:S2_minimal_appendix}
\end{equation}
This is exactly Eq.~\eqref{eq:Lyapunov_S2}, the Lyapunov equation.

\subsection{Three-point moment \texorpdfstring{$S_3$}{S3}}
\label{app:S3_derivation}

For $M_{3}$, we begin with the fluctuation equation centered on
$\barF_N$ and then identify the terms that survive at leading order.
To derive the evolution of
${M_{\1\2\3}:=\langle\deltaF_{\1}\deltaF_{\2}\deltaF_{\3}\rangle}$,
we apply It\^o's lemma to the triple product. We obtain
\begin{align}
    \rd(\deltaF_{\1}\deltaF_{\2}\deltaF_{\3})
    = {} & \deltaF_{\2}\deltaF_{\3}\rd\deltaF_{\1}
    +\deltaF_{\1}\deltaF_{\3}\rd\deltaF_{\2}
    \nonumber\\
    + \, {} & \deltaF_{\1}\deltaF_{\2}\rd\deltaF_{\3}
    +\deltaF_{\3}\rd\deltaF_{\1}\rd\deltaF_{\2}
    \nonumber\\
    + \, {} & \deltaF_{\1}\rd\deltaF_{\2}\rd\deltaF_{\3}
    +\deltaF_{\2}\rd\deltaF_{\1}\rd\deltaF_{\3}
    \nonumber\\
    + \, {} & \rd\deltaF_{\1}\rd\deltaF_{\2}\rd\deltaF_{\3}.
    \label{eq:Ito_S_3}
\end{align}
The triple variation
${\rd\deltaF_{\1}\rd\deltaF_{\2}\rd\deltaF_{\3}
=\mO(\rd\tau^{3/2})}$ vanishes in the It\^o differential.
Using Eq.~\eqref{eq:LDK_deltaF}, the terms of the form
${\deltaF_{\2}\deltaF_{\3}\rd\deltaF_{\1}}$ read
\begin{align}
    \deltaF_{\2}\deltaF_{\3}\rd\deltaF_{\1}
    = {} &
    \deltaF_{\2}\deltaF_{\3}
    \omL_{\1}\deltaF_{\1}\,\rd\tau
    +\deltaF_{\2}\deltaF_{\3}
    \mLL[\deltaF,\deltaF]_{\1}\,\rd\tau
    \nonumber\\
    - \, {} &
    \deltaF_{\2}\deltaF_{\3}\mLL[M_2]_{\1}\,\rd\tau
    \nonumber \\
    + \, {} & \deltaF_{\2}\deltaF_{\3}
    \tfrac{1}{\sqrt N}\rd\mW_{\tau,\1}.
    \label{eq:23d1}
\end{align}
Taking the ensemble average, the first contribution gives
\begin{align}
    \big\langle\deltaF_{\2}\deltaF_{\3}
    \omL_{\1}\deltaF_{\1}\big\rangle
    &=
    \omL_{\1}
    \langle\deltaF_{\2}\deltaF_{\3}\deltaF_{\1}\rangle
    \nonumber\\
    &=
    \omL_{\1}M_{\1\2\3}.
    \label{eq:calc_lin_M3}
\end{align}
The second term in Eq.~\eqref{eq:23d1} gives the nonlinear
coupling ${\mN[M_4]_{\1;\2\3}}$, namely
\begin{align}
    \langle\deltaF_{\2}\deltaF_{\3}
    \mLL[\deltaF,\deltaF]_{\1}\rangle
    = {} &
    \half\p_{\1}\!\cdot\!\!\int\!\rd\bv\,
    \sB_{\bv\1}
    [\p_{\1}-\p_{\bv}]
    M_{\bv\1\2\3}
    \nonumber\\
    =: {} & \mN[M_4]_{\1;\2\3} ,
    \label{eq:def_mN_calc}
\end{align}
as already introduced in Eq.~\eqref{eq:def_nonl_N}.

The third term in Eq.~\eqref{eq:23d1} contributes via
\begin{align}
    \langle\deltaF_{\2}\deltaF_{\3}
    \mLL[M_2]_{\1}\rangle
    = {} & M_{\2\3}\mLL[M_2]_{\1}
    \label{eq:calc_M2M3_prg}\\
    = {} &
    \half M_{\2\3}\p_{\1}\!\cdot\!\!\int\!\rd\bv\,
    \sB_{\bv\1}
    [\p_{\1}-\p_{\bv}]M_{\1\bv}.
    \nonumber
\end{align}
This contribution is of the same order as $M_3$, since
${ (M_2)^2 \!=\! \mO(N^{-2}) }$, and must therefore be kept.
The stochastic term in Eq.~\eqref{eq:23d1} has zero expectation,
\begin{equation}
    \langle\deltaF_{\2}\deltaF_{\3}
    \rd\mW_{\tau,\1}\rangle=0.
\end{equation}
Indeed, $\deltaF_{\2}$ and $\deltaF_{\3}$ are evaluated
at time $\tau$, while the It\^o noise increment is of zero
mean given the state at that time.
The same calculations apply to the other two terms
containing a single ${\rd\deltaF}$ in Eq.~\eqref{eq:Ito_S_3},
upon permuting ${(\1,\2,\3)}$.

We now consider the terms containing two increments
${\rd\deltaF}$ in Eq.~\eqref{eq:Ito_S_3}. At order
${\rd\tau}$, one of them gives
\begin{equation}
    \deltaF_{\3}\rd\deltaF_{\1}\rd\deltaF_{\2}
    =\tfrac1N\deltaF_{\3}
    \rd\mW_{\tau,\1}\rd\mW_{\tau,\2}
    +o(\rd\tau).
\end{equation}
After averaging and using the conditional noise covariance,
we obtain
\begin{equation}
    \langle\deltaF_{\3}\rd\deltaF_{\1}
    \rd\deltaF_{\2}\rangle
    =\tfrac1N
    \langle\deltaF_{\3}C[F_N]_{\1\2}\rangle\,\rd\tau.
    \label{eq:S3_noise_cross}
\end{equation}
Because $C[F_N]$ depends on $F_N$, it is correlated with
$\deltaF_{\3}$. We therefore expand it around the exact
mean $\barF_N$. Since $C$ is quadratic in its argument,
this expansion is exact,
\begin{align}
    C[F_N]_{\1\2}
    = {} &
    C[\barF_N]_{\1\2}
    \nonumber\\
    + \, {} &
    \int\!\rd\bu\,
    \frac{\delta C[\barF_N]_{\1\2}}
         {\delta\barF_{N,\bu}}\,
    \deltaF_{\bu}
    \nonumber\\
    + \, {} &
    \half\!\!\int\!\rd\bv\,\rd\bu\,
    \frac{\delta^2 C[\barF_N]_{\1\2}}
         {\delta\barF_{N,\bu}\,\delta\barF_{N,\bv}}\,
    \deltaF_{\bv}\deltaF_{\bu}.
\end{align}
The functional derivatives in this expansion are evaluated
at $\barF_N$. Substituting it into
Eq.~\eqref{eq:S3_noise_cross} yields
\begin{align}
    \tfrac1N
    \langle\deltaF_{\3}C[F_N]_{\1\2}\rangle & {}
    =
    \tfrac1N\langle\deltaF_{\3}
    C[\barF_N]_{\1\2}\rangle
    \label{eq:calc_source_S3} \\
    &+\tfrac1N\!\!\int\!\!\rd\bu\,
    \frac{\delta C[\barF_N]_{\1\2}}
         {\delta\barF_{N,\bu}}\,
    M_{\bu\3}
    \nonumber\\
    & + \tfrac{1}{2N}\!\!\int\!\!\rd\bv\rd\bu
    \frac{\delta^2 C[\barF_N]_{\1\2}}
         {\delta\barF_{N,\bu}\,\delta\barF_{N,\bv}}\,
    M_{\3\bu\bv}. \nonumber
\end{align}
The first term vanishes because $C[\barF_N]$ is
deterministic and ${ \langle\deltaF_{\3}\rangle \!=\! 0 }$.
The last term is ${ \mO(N^{-3}) }$, since
${ M_3 \!=\! \mO(N^{-2}) }$.
Thus, to order $N^{-2}$, we have
\begin{align}
    \tfrac1N\langle\deltaF_{\3}C[F_N]_{\1\2}\rangle
    = {} &
    \tfrac1N\!\int\!\rd\bu\,
    \frac{\delta C[\barF_N]_{\1\2}}
         {\delta\barF_{N,\bu}}\,
    M_{\bu\3}
    \nonumber
    \\
    + \, {} & \mO(N^{-3}).
    \label{eq:exp_source_M3}
\end{align}
The other two terms containing two increments
${\rd\deltaF}$ follow by cyclic permutation of
${(\1,\2,\3)}$.

Combining Eqs.~\eqref{eq:calc_lin_M3},
\eqref{eq:def_mN_calc}, \eqref{eq:calc_M2M3_prg}
and~\eqref{eq:exp_source_M3}, we obtain, to order
$N^{-2}$,
\begin{align}
    \p_\tau M_{\1\2\3}
    = {} & \!\!\sum_{\i\in\{\1,\2,\3\}}\!\!\bigg\{
    \omL_{\i}M_{\i\j\k}
    +\mN[M_4]_{\i;\j\k}
    -M_{\j\k}\mLL[M_2]_{\i}
    \nonumber\\
    &+\tfrac1N\!\int\!\rd\bu\,
    \frac{\delta C[\barF_N]_{\j\k}}
         {\delta\barF_{N,\bu}}\,
    M_{\i\bu}
    \bigg\}
    \!+\! \mO(\tfrac{1}{N^{3}}).
    \label{eq:p_tS_3_S_4}
\end{align}
In each term of the sum, ${(\i,\j,\k)}$ runs over
the three cyclic permutations of ${(\1,\2,\3)}$.
Thus ${M_{\i\j\k}=M_{\1\2\3}}$ by symmetry.
Such a notation allows us to write compactly the sum over the three cyclic permutations, so that one has
\begin{equation}
    \sum_{\mathclap{\i\in\{\1,\2,\3\}}} A_{\i \j \k} = A_{\1 \2 \3} \!+\! A_{\2 \3 \1} \!+\! A_{\3 \1 \2} .
\end{equation}

To fully close Eq.~\eqref{eq:p_tS_3_S_4},
we use the Wick closure (Eq.~\ref{eq:closure_S4})
to rewrite $M_{4}$ as a function of $M_{2}$.
One has
\begin{equation}
    M_{\1\2\3\4}
    \!=\! M_{\1\2}M_{\3\4} \!+\! M_{\1\3}M_{\2\4} \!+\! M_{\1\4}M_{\2\3} \!+\! \mO(\tfrac{1}{N^{3}}).
\end{equation}
This closure holds at ${ \tau \!=\! 0 }$ (see Appendix~\ref{app:IC_S4_S5}) and is preserved by the truncated hierarchy, as shown in Appendix~\ref{app:S4_closure}.
Substitution into Eq.~\eqref{eq:def_mN_calc} gives
\begin{align}
    \mN&[M_4]_{\i; \j \k}= \half M_{\j \k} \p_{\i} \!\cdot\!\!\int\!\! \rd \bv \, \sB_{\i \bv} \left[ \p_{\i} \!-\! \p_{\bv} \right]
    M_{\bv \i}
    \label{eq:nonlinear_S3_full}
    \\
    & {} + \half \p_{\i}\!\cdot\!\bigg\{
    \left[\!\int\!\! \rd \bv \, \sB_{\i \bv}M_{\bv \j}\right] \p_{\i} M_{\i \k} \!-\! M_{\i \k}\!\int\!\! \rd \bv \, \sB_{\i \bv}
    \p_{\bv}M_{\bv \j} \bigg\}
    \nonumber
    \\
    & {} + \half \p_{\i} \!\cdot\!\bigg\{
    \left[\!\int\!\! \rd \bv \, \sB_{\i \bv}M_{\bv \k}\right] \p_{\i} M_{\i \j}  \!-\! M_{\i \j}\!\int\!\! \rd \bv \, \sB_{\i \bv}
    \p_{\bv}M_{\bv \k} \bigg\} .
    \nonumber
\end{align}
The contribution from the pairing $M_{\bv\i}M_{\j\k}$
cancels $M_{\j\k}\mLL[M_2]_{\i}$ exactly. The
remaining error comes from the $\mO(N^{-3})$
remainder in the Wick closure for $M_4$.

We now replace $\barF_N$ by $F$ in the coefficients of
the $M_3$ equation. Since ${ \barF_N \!-\!F \!=\! \mO(N^{-1})}$,
this introduces only $\mO(N^{-3})$ corrections to both
the linear and source terms. Thus, we may use $\mL_{\i}$
in place of $\omL_{\i}$ (as defined in Eq.~\ref{eq:def_linL_F})
and evaluate
the functional derivative of $C$ at $F$.
The leading-order equation becomes
\begin{align}
    \p_\tau M_{\1\2\3}
    = {} &
    \sum_{\i\in\{\1,\2,\3\}}
    \!\!\! \big\{
    \mL_{\i}M_{\i\j\k}
    +\mK[M_2]_{\i;\j\k}
    \big\}
    \\
    & {} +\tfrac1N
    \sum_{\i\in\{\1,\2,\3\}}\!\int\!\rd\bu\,
    \frac{\delta C[F]_{\j\k}}{\delta F_{\bu}}\,
    M_{\i\bu} +\mO(\tfrac{1}{N^{3}}),
    \nonumber
\end{align}
where
\begin{equation}
    \mK[M_2]_{\i;\j\k}
    :=\mQ_{\i;\j\k}+\mQ_{\i;\k\j},
    \label{eq:def_mN_mM}
\end{equation}
with
\begin{equation}
    \mQ_{\i;\j\k}
    :=\half\p_{\i}\!\cdot\!\!\int\!\rd\bv\,
    \sB_{\bv\i}
    [\p_{\i}-\p_{\bv}]
    M_{\bv\j}M_{\i\k}.
    \label{eq:def_mM}
\end{equation}
The order of $\j$ and $\k$ matters in
$\mQ_{\i;\j\k}$, whereas their contributions are
symmetrized in $\mK[M_2]_{\i;\j\k}$.
The cancellation above removes the disconnected
product $M_{\j\k}\mLL[M_2]_{\i}$.
This was to be expected for
the evolution of the centered third moment
since it is also the third cumulant.

Finally, rescaling via ${ \SDK_3 \!=\! N^2 M_3 }$ and ${ \SDK_2 \!=\! N M_2 }$, and discarding the resulting $\mO(N^{-1})$ corrections, the leading-order contribution $\Smin_3$ satisfies
\begin{equation}
    \p_\tau \Smin_{\1\2\3}
    =\mF_{\1\2\3}
    +\mF_{\2\3\1}
    +\mF_{\3\1\2} ,
    \label{eq:p_tS3_mF}
\end{equation}
with the final expression
\begin{widetext}
\begin{align}
    \mF_{\1\2\3}:={}
    &\half\p_{\1}\!\cdot\!\!\int\!\rd\bv\,
    \sB_{\1\bv}
    \bigg[
    \p_{\1}F_{\1}\Smin_{\bv\2\3}
    -\p_{\bv}F_{\bv}\Smin_{\1\2\3}
    +F_{\bv}\p_{\1}\Smin_{\1\2\3}
    -F_{\1}\p_{\bv}\Smin_{\bv\2\3}
    \bigg]
    \nonumber\\
    +{}&
    \half\p_{\1}\!\cdot\!\bigg\{
    \left[\!\int\!\rd\bv\,
    \sB_{\1\bv}\Smin_{\bv\2}\right]
    \p_{\1}\Smin_{\1\3}
    -\Smin_{\1\3}\!\int\!\rd\bv\,
    \sB_{\1\bv}\p_{\bv}\Smin_{\bv\2}
    \bigg\} +     \half\p_{\1}\!\cdot\!\bigg\{
    \left[\!\int\!\rd\bv\,
    \sB_{\1\bv}\Smin_{\bv\3}\right]
    \p_{\1}\Smin_{\1\2}
    -\Smin_{\1\2}\!\int\!\rd\bv\,
    \sB_{\1\bv}\p_{\bv}\Smin_{\bv\3}
    \bigg\}
    \nonumber\\
    +{}&
    \Mtot\!\int\!\rd\bv\,
    \Smin_{\bv\1}\,
    \p_{\2}\!\cdot\!\p_{\3}\!\cdot\!
    \bigg\{
    \delta_{\2\3}
    \bigg[
    F_{\2}\sB_{\2\bv}
    +\delta_{\2\bv}\!\!\int\!\rd\bvp\,
    F_{\bvp}\sB_{\2\bvp}
    \bigg]
    -
    \big[
    \delta_{\2\bv}F_{\3}
    +\delta_{\3\bv}F_{\2}
    \big]\sB_{\2\3}
    \bigg\}.
    \label{eq:def_mF_S3_full}
\end{align}
\end{widetext}
This is exactly Eq.~\eqref{eq:Lyapunov_S3},
presented in the main text.

\section{Full Landau fluctuation hierarchy}
\label{app:Sn_equation}

In this Appendix, we present the full Landau fluctuation hierarchy,
i.e.\ the exact evolution equations for all the moments $\SDK_n$.
Contrary to the truncated hierarchy, no term is discarded here:
this hierarchy is equivalent to \LDK\@ (Eq.~\ref{eq:LDK}).

In this Appendix, we will use the following notations.
For a moment of order $n$, we introduce the index set ${ A_n \!:=\! \{\1,...,\n\} }$. 
For ${ \i \!\in\! A_n }$, we denote its complementary set by ${ \widehat{\dli} \!:=\! A_n \!\setminus\! \{\i\} }$.
Similarly, for an unordered pair ${ \{\a,\b\} \!\subset\! A_n }$,
we write ${ \widehat{\a\b} \!:=\! A_n \!\setminus\! \{\a,\b\} }$.
Since all the moments are symmetric, the ordering of the indices within these complementary sets is irrelevant.
Starting from \LDK\@ and generalizing the It\^o calculus procedure as described in Appendix~\ref{app:minimal_hierarchy_S2_S3},
after lengthy manipulations,
the general evolution equation for the $n$-th moment reads
\begin{align} 
    \p_{\tau} \SDK_{\1...\n}
    = {} & \sum_{\i=\1}^{\n} \omL_{\i} \SDK_{\1... \n} + \mu_n \sum_{\i=\1}^{\n} \mN[\SDK_{n+1}]_{\i;\widehat{\dli}}
    \nonumber
    \\ 
   - {} & \mu_n \sum_{\i=\1}^{\n} \mLL[\SDK_2]_{\i} \, \SDK_{\widehat{\dli}} + \sum_{\mathclap{\{\a,\b\}\subset A_n}} D^{(n)}_{\a\b},
    \label{eq:General_Hierarchy_Sn}
\end{align}
where we introduced
\begin{equation}
\mu_n = 
\begin{cases}
\displaystyle 1 / N {} & \mathrm{if} \;\; n \;\; \mathrm{even} ,
\\
\displaystyle 1 {} & \mathrm{otherwise}.
\end{cases}
\label{eq:def_mun}
\end{equation}
The parity dependence of $\mu_n$ follows from the rescaling ${\SDK_n=N^{\lceil n/2\rceil}M_n}$. 
For even $n$, the terms involving $M_{n+1}$ and $M_2M_{n-1}$ are smaller than $M_n$ by a factor ${ 1/N }$. For odd $n$, they are of the same order as $M_n$.
We also set ${\SDK_0 \!:=\! 1}$, while ${ \SDK_1 \!=\! 0 }$ since fluctuations are centered.
In Eq.~\eqref{eq:General_Hierarchy_Sn}, the first sum contains the linearized Landau advection, $\omL_{\i}$ (Eq.~\ref{eq:def_omL}).
It is linearized around $\barF_N$ and acts exclusively on the velocity variable $\i$ in ${ \SDK_{\1... \n} }$.

The second term in Eq.~\eqref{eq:General_Hierarchy_Sn}
is the nonlinear feedback coupling to higher-order moments ($\mN[\SDK_{n+1}]_{\i;\widehat{\dli}}$). 
It couples the $n$-th order moment to the ${(n \!+\! 1)}$-th order and is defined as
\begin{align}
    \mN[\SDK_{n+1}]_{\i;\widehat{\dli}} \!:=\! {} & \half \p_{\i} \!\cdot\! \!\!\int \!\! \rd \bu \, \sB_{\i\bu} \big[ \p_{\i} \!-\! \p_{\bu} \big] \SDK_{\bu\i\widehat{\dli}} .
    \label{eq:NL_S_n}
\end{align}
This is functionally equivalent to the action of the Landau collision operator on the ${(n \!+\! 1)}$-th moment with respect to the velocity variable $\i$. 
In Eq.~\eqref{eq:General_Hierarchy_Sn}, the third sum contains ${ -\mu_n\mLL[\SDK_2]_{\i} \SDK_{\widehat{\dli}} }$.
It is a disconnected contribution arising from the centering on $\barF_N$.

Finally, the last term in Eq.~\eqref{eq:General_Hierarchy_Sn}
is the noise-induced source tensor ${ D^{(n)}_{\a \b} }$.
It gathers all finite-$N$ contributions from the covariance ${ \barC_{\a \b} \!:=\! C[\barF_N]_{\a \b} }$. It reads
\begin{align}
    D^{(n)}_{\a \b} := {} & \barC_{\a \b}\, \SDK_{\widehat{\a\b}} + \mu_n \!\!\int\!\! \rd\bu\, \frac{\delta \barC_{\a\b}}{\delta (\barF_{N,\bu})} \, \SDK_{\bu\widehat{\a\b}}
    \nonumber
    \\
   + \, {} & \half \tfrac{1}{N} \!\!\int\!\! \rd\bu\,\rd\bw\, \frac{\delta^2 \barC_{\a\b}} {\delta (\barF_{N,\bu}) \delta (\barF_{N,\bw})} \SDK_{\bu\bw\widehat{\a\b}} .
    \label{eq:def_Dab_general}
\end{align}

Equation~\eqref{eq:General_Hierarchy_Sn} is supplemented with the evolution equation for $\barF_N$ (Eq.~\ref{eq:Landau_1/N2}).
It reads 
\begin{equation}
    \p_\tau\barF_N = \mLL[\barF_N,\barF_N] + \tfrac{1}{N}\mLL[\SDK_2] .
\end{equation}
Together, these two equations form the full Landau fluctuation hierarchy associated with \LDK\@.
At this stage no term has been discarded.

Formally, solving this full hierarchy provides a complete statistical characterization of the fluctuations generated by \LDK\@. 
In principle, this hierarchy also encodes the rare fluctuations present within the Landau picture,
i.e.\ the rare fluctuations generated by the cumulative effect of weak two-body deflections. However, it does not encode the irreducible three-body effects of the Hamiltonian dynamics.

\section{Truncated Landau fluctuation hierarchy}
\label{app:truncated_hierarchy_dynamical}

Starting from the full hierarchy (Eq.~\ref{eq:General_Hierarchy_Sn}), one can derive the truncated hierarchy by discarding the terms that are subdominant in ${ 1/N }$. This has been done for $S_2$ and $S_3$ in Appendix~\ref{app:minimal_hierarchy_S2_S3}; in this Appendix, we do it for moments of any order.
To do so, we assume that the scalings ${ \SDK_n \!=\! \mO(1) }$, established for ${ \tau \!=\! 0 }$ in Appendix~\ref{app:IC}, are preserved by the dynamics.
Then we substitute $\barF_N$, $\omL_{\i}$ and $\barC$ with their respective leading-order counterparts $F$, $\mL_{\i}$ and $C$, and keep only the dominant terms.

Calling $\ST_n$ the solutions of this truncated hierarchy, we obtain different evolution equations depending on the parity of $n$, through the prefactor $\mu_n$ (Eq.~\ref{eq:def_mun}).
For even $n$, in Eq.~\eqref{eq:General_Hierarchy_Sn}, the terms involving $\mN$ and $\mLL$, as well as the terms involving ${ \delta C/\delta F }$ and ${ \delta^2 C/\delta F^2 }$, are suppressed by a factor ${ 1/N }$.
This leads to
\begin{align}
    \p_{\tau}\ST_{A_{2k}} = {} & \sum_{\mathclap{\i \in A_{2k}}} \mL_{\i} \ST_{A_{2k}} + \sum_{\mathclap{\{\a,\b\}\subset A_{2k}}} C_{\a\b}\, \ST_{\widehat{\a\b}} .
\label{eq:Minimal_Lyapunov_2k}
\end{align}
For odd moments, only the term involving ${ \delta^2 C/\delta F^2 }$ is discarded in Eq.~\eqref{eq:General_Hierarchy_Sn}.
This leads to
\begin{align}
    \p_{\tau} \ST_{A_{2k+1}} = {} & \sum_{\mathclap{\i\in A_{2k+1}}} \mL_{\i} \ST_{A_{2k+1}} + \sum_{\mathclap{\i\in A_{2k+1}}} \mN[\ST_{2k+2}]_{\i;\widehat{\dli}}
    \nonumber
    \\
    & {} - \sum_{\mathclap{\i\in A_{2k+1}}} \mLL[\ST_2]_{\i}\, \ST_{\widehat{\dli}} + \sum_{\mathclap{\{\a,\b\}\subset A_{2k+1}}} C_{\a\b}\, \ST_{\widehat{\a\b}} \nonumber \\
    & {} + \sum_{\mathclap{\{\a,\b\}\subset A_{2k+1}}} \int \!\! \rd\bu\, \frac{\delta C_{\a\b}}{\delta (F_{\bu})} \ST_{\bu\widehat{\a\b}} .
    \label{eq:Minimal_Lyapunov_2k+1}
\end{align}
Equations~\eqref{eq:Minimal_Lyapunov_2k} and~\eqref{eq:Minimal_Lyapunov_2k+1} form the truncated hierarchy. They show a simple structure.
The even moments form a closed sub-hierarchy, since $\ST_{2k}$ depends only on $\ST_{2k-2}$.
In contrast, each odd moment $\ST_{2k+1}$ is coupled to the next even moment, $\ST_{2k+2}$. By construction, $\ST_n$ is the leading-order contribution to the rescaled $n$-th order moment $\SDK_n$.
We highlight that the truncated hierarchy (Eqs.~\ref{eq:Minimal_Lyapunov_2k}
and~\ref{eq:Minimal_Lyapunov_2k+1}) is an infinite, unclosed set of equations for all the $\ST_n$.
It should a priori be distinguished from the minimal hierarchy of the main text, 
which only consists of the kinetic equations for $F$, $\Smin_2$ and $\Smin_3$, supplemented by the quasi-Gaussian closure for $\Smin_n$, with ${ n \!\geq\! 4 }$.

As expected, for ${ n \!=\! 2 }$, Eq.~\eqref{eq:Minimal_Lyapunov_2k} reduces to Eq.~\eqref{eq:Lyapunov_S2} in the main text.
For ${ n \!=\! 3 }$, Eq.~\eqref{eq:Minimal_Lyapunov_2k+1} reduces to Eq.~\eqref{eq:Lyapunov_S3} once $\ST_4$ is replaced by its Wick closure (Eq.~\ref{eq:closure_S4}, justified in Appendix~\ref{app:S4_closure}). Hence ${ \Smin_2 \!=\! \ST_2 }$ and ${ \Smin_3 \!=\! \ST_3 }$.

In Appendices~\ref{app:S4_closure} and~\ref{app:S5_closure},
we explicitly show that the proposed closure relations defining $\Smin_4$ and $\Smin_5$ satisfy the corresponding truncated hierarchy equations,
Eqs.~\eqref{eq:Minimal_Lyapunov_2k} and~\eqref{eq:Minimal_Lyapunov_2k+1}, respectively. 
In Appendix~\ref{app:S6_Sn_closure}, we then outline the recursive argument underlying the extension of this proof to an arbitrary fixed-order moment $\Smin_n$.
This will show that, for any $n$, ${ \Smin_n \!=\! \ST_n }$.

\section{Closure equation for \texorpdfstring{$\Smin_4$}{S4}}
\label{app:S4_closure}

In this Appendix, consistently with the convention introduced in the main text, we recall that $\Smin_4$ is defined for all ${ \tau \!\geq\! 0 }$ by the following closure (Eq.~\ref{eq:closure_S4}):
\begin{align}
    \Smin_{\1 \2 \3 \4} = {} & \Smin_{\1 \2} \, \Smin_{\3 \4} + \Smin_{\1 \3} \, \Smin_{\2 \4} + \Smin_{\1 \4} \, \Smin_{\2 \3}.
    \label{eq:closure4}
\end{align}
We shall show that $\Smin_4$ is the solution to the truncated hierarchy of Appendix~\ref{app:truncated_hierarchy_dynamical}, i.e.\ ${ \Smin_4 \!=\! \ST_4 }$, and hence
\begin{equation}
    \SDK_4
    =
    \Smin_4
    + \mO \big( \tfrac{1}{N} \big).
\end{equation}
By definition of $\Smin_2$, we have ${ \Smin_4 \!=\! \mO(1) }$.
To prove ${ \Smin_4 \!=\! \ST_4 }$, 
we first show that ${ \Smin_4 (\tau \!=\! 0) \!=\! \ST_4 (\tau \!=\! 0) }$, and then that $\Smin_4$ satisfies the truncated kinetic equation for $\ST_4$ derived in Appendix~\ref{app:truncated_hierarchy_dynamical}. 
From Eq.~\eqref{eq:Minimal_Lyapunov_2k}, this equation reads
\begin{align}
    \p_{\tau} \ST_{\1 \2 \3 \4} = {} & \left[ \mL_{\1} \!+\! \mL_{\2} \!+\! \mL_{\3} \!+\! \mL_{\4} \right] \ST_{\1 \2 \3 \4} \nonumber \\
    + \, {} & C_{\1\2} \, \ST_{\3 \4} + C_{\3\4} \, \ST_{\1 \2} +  C_{\1\3} \, \ST_{\2 \4}
    \nonumber
    \\
    + \, {} & C_{\2\4} \, \ST_{\1 \3} + C_{\1\4} \, \ST_{\2 \3} + C_{\2\3} \, \ST_{\1 \4},
    \label{eq:S4_Lyapunov}
\end{align}
where we used the shorthand notation ${ C_{\i\j} := C[F]_{\i\j} }$.

\subsection{Initial condition for \texorpdfstring{$\Smin_4$}{S4}}
\label{sec:G4_initial}

Appendix~\ref{app:IC_S4_S5} (Eq.~\ref{eq:S4_initial}) shows that, at leading order in ${ 1/N }$, the initial condition for $\ST_4$ is exactly
\begin{equation}
    \ST_4(\tau \!=\! 0) = \Smin_4(\tau \!=\! 0) .
\end{equation}

\subsection{\texorpdfstring{$\Smin_4$}{S4} is a solution of the truncated hierarchy}
\label{sec:G4_solution}

Now, we prove that $\Smin_4$ satisfies Eq.~\eqref{eq:S4_Lyapunov}.
The evolution equation of $\Smin_2$ is given by
\begin{equation}
    \p_{\tau} \Smin_{\i \j} =  \big[ \mL_{\i} \!+\! \mL_{\j} \big] \Smin_{\i \j} + C_{\i\j}.
\end{equation}
Let us start with the first term of $\Smin_4$.
It evolves according to
\begin{align}
    \p_{\tau} \big[ \Smin_{\1 \2}\Smin_{\3 \4} \big] {} & = \big[ \p_{\tau} \Smin_{\1 \2} \big] \Smin_{\3 \4} + \Smin_{\1 \2} \big[ \p_{\tau}  \Smin_{\3 \4} \big] \nonumber \\
    & = \big[ \big(\mL_{\1} \!+\! \mL_{\2} \big) \Smin_{\1 \2} \!+\! C_{\1\2} \big] \Smin_{\3 \4} \nonumber \\
    & \quad + \Smin_{\1 \2} \big[ \big(\mL_{\3} \!+\! \mL_{\4} \big) \Smin_{\3 \4} \!+\! C_{\3\4} \big] .
\end{align}
Because $\mL_{\1}$ and $\mL_{\2}$ only act on variables $\1$ and $\2$, we can pull out these operators. We get
\begin{align}
    \p_{\tau} \big[ \Smin_{\1 \2} \Smin_{\3 \4} \big] = {} & \big( \mL_{\1} \!+\! \mL_{\2} \!+\! \mL_{\3} \!+\! \mL_{\4} \big) \big[ \Smin_{\1 \2}\Smin_{\3 \4} \big] \nonumber \\
    + {} \, & \Smin_{\1 \2} C_{\3\4} + \Smin_{\3 \4} C_{\1\2} .
\end{align}
Now, we do the exact same computations for the other two terms in $\Smin_4$. This leads to
\begin{align}
    \p_{\tau} \Smin_{\1 \2 \3 \4} = {} & \left[ \mL_{\1} \!+\! \mL_{\2} \!+\! \mL_{\3} \!+\! \mL_{\4} \right] \Smin_{\1 \2 \3 \4} + C_{\1\2}\Smin_{\3 \4}
    \nonumber
    \\
+ \, {} & C_{\3\4}\Smin_{\1 \2}  + C_{\1\3}\Smin_{\2 \4} + C_{\2\4}\Smin_{\1 \3}  + C_{\1\4}\Smin_{\2 \3}
\nonumber
\\
+ \, {} & C_{\2\3}\Smin_{\1 \4} .
\end{align}
This is exactly Eq.~\eqref{eq:S4_Lyapunov} for $\Smin_4$. 

To summarize, we have shown that $\Smin_4$, as defined by Eq.~\eqref{eq:closure4}, satisfies the same evolution equation and initial condition as $\ST_4$. Therefore, we have ${ \Smin_{4} (\tau) \!=\! \ST_{4} (\tau) }$,
for all $\tau$.

\section{Closure equation for \texorpdfstring{$\Smin_5$}{S5}}
\label{app:S5_closure}

Following the same approach as in Appendix~\ref{app:S4_closure}, we recall that $\Smin_5$ is defined by the following closure (Eq.~\ref{eq:Closure_2k+1}):
\begin{align}
    \Smin_{\1\2\3\4\5}
    := {} &
    \Smin_{\1\2} \Smin_{\3\4\5} +
    \Smin_{\1\3} \Smin_{\2\4\5} +
    \Smin_{\1\4} \Smin_{\2\3\5} +
    \Smin_{\1\5} \Smin_{\2\3\4} \nonumber \\
    + \, {} & \Smin_{\2\3} \Smin_{\1\4\5} +
    \Smin_{\2\4} \Smin_{\1\3\5} +
    \Smin_{\2\5} \Smin_{\1\3\4} \nonumber \\
    + \, {} & \Smin_{\3\4} \Smin_{\1\2\5} +
    \Smin_{\3\5} \Smin_{\1\2\4} +
    \Smin_{\4\5} \Smin_{\1\2\3} .
    \label{eq:closure_S5}
\end{align}
We now show that ${ \Smin_5(\tau) \!=\! \ST_5(\tau) }$ for all ${\tau \!\geq\! 0}$.
The calculation relies on the closure for $\Smin_6$, presented in Appendix~\ref{app:S6_Sn_closure}.
For the rest of this Appendix, we write $S_n$ as shorthand for $\Smin_n$.

\subsection{Initial condition for \texorpdfstring{$S_5$}{S5}}
\label{sec:G5_initial}

From Appendix~\ref{app:IC_S4_S5} (Eq.~\ref{eq:S5_initial}) and the definition of $S_{5}$ in Eq.~\eqref{eq:closure_S5}, we have, at leading order ${ \mO(1) }$,
\begin{equation}
    S_{5} (\tau \!=\! 0) = \ST_{5} (\tau \!=\! 0) .
\end{equation}

\subsection{\texorpdfstring{$S_5$}{S5} is a solution of the truncated hierarchy}
\label{sec:G_5_solution}

In the following, for simplicity, we use the following notations: for an unordered pair ${ \{\a,\b\} \!\subset\! A \!:=\! \{\1,\2,\3,\4,\5\} }$ (with ${ \a \!\neq\! \b }$), we denote by $\{ \c,\d,\e \}$ its complementary set in $A$. Furthermore, the complementary set of a single element ${ \i \!\in\! A }$ is denoted by ${ \{\j,\k,\l,\m\} }$.
Armed with these notations, we can write Eq.~\eqref{eq:closure_S5} as
\begin{equation}
    S_{\1 ... \5} = \sum_{\mathclap{\{\a,\b \} \subset A}} S_{\a\b} S_{\c\d\e} .
    \label{eq:closure_S5_bis}
\end{equation}
Setting ${ n \!=\! 5 }$ in Eq.~\eqref{eq:Minimal_Lyapunov_2k+1}, we have to show that Eq.~\eqref{eq:closure_S5_bis} is a solution to the truncated equation for $\ST_5$.
Since ${ \ST_2 \!=\! \Smin_2 }$ (Appendix~\ref{app:truncated_hierarchy_dynamical}), ${ \ST_4 \!=\! \Smin_4 }$ (Appendix~\ref{app:S4_closure}) and ${ \ST_6 \!=\! \Smin_6 }$ (Appendix~\ref{app:S6_closure}), this equation can be written in terms of the $S_n$ only. It reads
\begin{align}
    \p_{\tau} S_{\1 ... \5} = {} & \sum_{\i=\1}^{\5} \mL_{\i} S_{\1 ... \5} + \sum_{\i=\1}^{\5} \mN[S_6]_{\i;\j\k\l\m} \nonumber \\
    - \, {} & \sum_{\i=\1}^{\5} \mLL [S_{2}]_{\i} \, S_{\j\k\l\m} + \sum_{\mathclap{\{\a, \b\} \subset A}} C_{\a\b} \, S_{\c\d\e} \nonumber \\
    + \, {} & \sum_{\mathclap{\{\a, \b\} \subset A}} \;\, \bigg\{ \frac{\delta C_{\a\b}}{\delta (F_{\cdot})} \, , S_{\c\d\e\cdot} \bigg\} ,
    \label{eq:S5_Lyapunov}
\end{align}
where we defined the scalar product
\begin{equation}
    \big\{ f , g \big\} := \!\!\int\!\! \rd \bv f(\bv) g(\bv) .
\end{equation}
In Eq.~\eqref{eq:S5_Lyapunov},
the ``$\cdot$'' in the subscript means that the scalar product applies to this argument, namely
\begin{equation}
    \bigg\{ \frac{\delta C_{\a\b}}{\delta (F_{\cdot})}\, , S_{\c\d\e\cdot} \bigg\} = \!\!\int\!\! \rd \bu \, \frac{\delta C_{\a\b}}{\delta (F_{\bu})} S_{\c\d\e\bu} \, .
\end{equation}

In order to check Eq.~\eqref{eq:S5_Lyapunov}, we start by computing ${ \p_{\tau} S_{5} }$ using the Leibniz rule. We get
\begin{align}
    \p_{\tau} S_{\1 ... \5} = {} & \sum_{ \{\a,\b\} \subset A} \big[ S_{\c\d\e} \p_{\tau} S_{\a\b} + S_{\a\b} \p_{\tau} S_{\c\d\e} \big] .
    \label{eq:Leibniz_G5_p_t}
\end{align}
The evolution equation of $S_{2}$ is given by (Eq.~\ref{eq:Lyapunov_S2})
\begin{equation}
    \p_{\tau} S_{\a\b} =  \left[\mL_{\a} + \mL_{\b}\right] S_{\a\b} + C_{\a\b} ,
    \label{eq:S2_ab_evolution}
\end{equation}
and for $S_{3}$ we have (Eq.~\ref{eq:Lyapunov_S3})
\begin{align}
    \p_{\tau} & {} S_{\c\d\e} = \sum_{\mathclap{\k \in \{\c,\d,\e\}}} \mL_{\k} S_{\c\d\e} \!+\!\! \sum_{\mathclap{\ip \in \{\c,\d,\e\}}} \mK[S_2]_{\ip;\jp\kp}
        \label{eq:S3_cde_evolution}
    \\
    & {} + \!\!\int\!\! \rd \bu \, \frac{\delta C_{\c\d}}{\delta (F_{\bu})} S_{\e\bu} + \!\!\int\!\! \rd \bu \, \frac{\delta C_{\c\e}}{\delta (F_{\bu})} S_{\d\bu} + \!\!\int\!\! \rd \bu \, \frac{\delta C_{\d\e}}{\delta (F_{\bu})} S_{\c\bu} , \nonumber 
\end{align}
where here ${\{ \jp, \kp \} \!=\! \{\c,\d,\e\} \!\setminus\! \{\ip\} }$.
The full expression of ${ \mK[S_2]_{\ip;\jp\kp} }$ is given in Eq.~\eqref{eq:def_mN_mM}.
By replacing Eqs.~\eqref{eq:S2_ab_evolution} and~\eqref{eq:S3_cde_evolution} into Eq.~\eqref{eq:Leibniz_G5_p_t},
we get
\begin{align}
    \p_{\tau} S_{\1 ... \5} = {} & \!\!\sum_{ \{\a,\b\} \subset A} \!\bigg[ \!\sum_{\k \in \{\c,\d,\e\}} \!\!\! \mL_{\k} S_{\c\d\e} S_{\a\b}
    \nonumber
    \\
    & {} + \sum_{\mathclap{\ip \in \{\c,\d,\e\}}} \mK[S_2]_{\ip;\jp\kp} S_{\a\b} \!+ \!\!\int\!\! \rd \bu \, \frac{\delta C_{\c\d}}{\delta (F_{\bu})} S_{\e\bu} S_{\a\b}
    \nonumber
    \\
    & {} + \!\!\int\!\! \rd \bu \, \frac{\delta C_{\c\e}}{\delta (F_{\bu})} S_{\d\bu} S_{\a\b} +\!\!\int\!\! \rd \bu \, \frac{\delta C_{\d\e}}{\delta (F_{\bu})} S_{\c\bu} S_{\a\b}
    \nonumber
    \\
    & {} + \big( \mL_{\a} \!+\! \mL_{\b} \big) S_{\a\b} S_{\c\d\e} + C_{\a\b} S_{\c\d\e} \bigg] .
    \label{eq:Leibniz_G5}
\end{align}

In order to show that $S_{5}$ satisfies the truncated hierarchy (Eq.~\ref{eq:S5_Lyapunov}), we must verify that the r.h.s.\ of Eq.~\eqref{eq:Leibniz_G5} coincides, term by term, with the r.h.s.\ of Eq.~\eqref{eq:S5_Lyapunov}, evaluated using the definitions of $S_4$ (Eq.~\ref{eq:closure_S4}) and $S_6$ (Eq.~\ref{eq:closure_S6}).
The $S_4$ closure was established in Appendix~\ref{app:S4_closure}. The $S_6$ closure can be checked independently from the even-order Eq.~\eqref{eq:Minimal_Lyapunov_2k}: 
it involves $S_4$ but not $S_5$. 
The logical sequence is therefore to establish the closures for $S_4$ and $S_6$ before $S_5$.
Indeed, proving the closure of $S_6$ does not depend on $S_5$.

Let us start with the first term in Eq.~\eqref{eq:S5_Lyapunov}: \begin{align}
    \sum_{\i=\1}^\5 & \mL_{\i} S_{\1 ... \5}
    = \sum_{\i=\1}^\5 \sum_{\{\a,\b\} \subset A} \mL_{\i}\!\left(S_{\a\b}S_{\c\d\e}\right)
        \label{eq:Linearized_S2_S3}
    \\
    = {} & \!\!\! \sum_{\{\a,\b\} \subset A} \!\!\!\! S_{\c\d\e} \big(\mL_{\a} \!+\! \mL_{\b} \big)S_{\a\b} + \!\!\!\! \sum_{\{\a,\b\} \subset A} \!\!\!\! S_{\a\b} \!\!\!\!\! \sum_{\k \in \{\c,\d,\e\}} \!\!\!\! \mL_{\k} S_{\c\d\e} , \nonumber
\end{align}
since the operator $\mL_{\i}$ distributes its action according to its index: if ${ \i \!\in\! \{\a,\b\} }$, it acts on $S_{\a\b}$ and factors out $S_{\c\d\e}$, whereas if ${ \i \!\in\! \{\c,\d,\e\} }$, it acts on $S_{\c\d\e}$ and factors out $S_{\a\b}$.
The result obtained in Eq.~\eqref{eq:Linearized_S2_S3} matches exactly the terms in the first and last lines of Eq.~\eqref{eq:Leibniz_G5}.

Let us now deal with the nonlinear term in Eq.~\eqref{eq:S5_Lyapunov}, namely ${ \sum_{\i=\1}^{\5} \mN[S_6]_{\i;\j\k\l\m} }$.
Its full expression is given by
\begin{align}
    \mN[S_6]_{\i;\j\k\l\m} &\!:=\! \half \p_{\i} \!\cdot\!\!\int\!\! \rd \bv \sB_{\bv \i} \big[ \p_{\i} \!-\! \p_{\bv} \big] S_{\bv\i\j\k\l\m} .
\end{align}
Here, we recall that the variable $\i$ is fixed in the sum $\sum_{\i=\1}^{\5}$, with its complementary set given by ${ \{\j,\k,\l,\m\} }$.
Using the definition of $S_{6}$ from Eq.~\eqref{eq:closure_S6}, this leads to
\begin{align}
    \mN[S_6]_{\i;\j\k\l\m} &= \half \p_{\i} \!\cdot\!\!\int\!\! \rd \bv \, \sB_{\bv \i}
    \big[ \p_{\i} \!-\! \p_{\bv} \big] \nonumber \\
    & {} \times \bigg\{ S_{\bv\i} \bigg[ S_{\j\k} S_{\l\m} \!+\! S_{\j\l} S_{\k\m} \!+\! S_{\k\l} S_{\j\m}  \bigg]  \nonumber \\
    & {} + \sum_{\mathclap{\{\a,\b\} \subset \{\j,\k,\l,\m\}}} S_{\a\b} \Big[ S_{\bv\c} S_{\i\d} + S_{\bv\d} S_{\i\c} \Big] \bigg\} ,
\end{align}
where we must separate the cases according to whether the dummy variable $\bv$ is paired with the main variable $\i$ or not. If they are paired, we have the term ${S_{\bv\i}}$ which multiplies all other combinations of $S_{2} S_{2}$. This leads to the first contribution, namely
\begin{equation}
    S_{\bv\i} \bigg[ S_{\j\k} S_{\l\m} \!+\! S_{\j\l} S_{\k\m} \!+\! S_{\k\l} S_{\j\m} \bigg] = S_{\bv\i} S_{\j\k\l\m} ,
    \label{eq:comput_S2_in_progress}
\end{equation}
where we recognized the Wick closure defining $S_{4}$.
If $\bv$ couples with one of the indices ${ \c \!\in\! \{\j,\k,\l,\m \} }$, then the variable $\i$ couples with one of the three remaining indices, which we denote by $\d$.
This leaves two free indices ${ \{ \a,\b \} \!=\! \{\j,\k,\l,\m\} \!\setminus\! \{\c,\d\} }$. This leads to the contribution
\begin{align}
    \sum_{\c} \sum_{\d} \sum_{\{\a,\b\}} S_{\a\b} S_{\bv\c} S_{\i\d} .
\end{align}
At fixed $\i$, this configuration yields ${ 4 \!\times\! 3 \!=\! 12 }$ terms.
Now we want to put the sum over ${ \{\a,\b\}}$ on the outside. At fixed ${ \{\a,\b\} \!\subset\! \{ \j, \k, \l, \m\} }$, its complementary set ${ \{ \c,\d\} \!=\! \{ \j,\k,\l,\m \} \!\setminus\! \{ \a,\b \} }$ is also fixed. However, the indices $\c$ and $\d$ do not play a symmetric role in ${S_{\bv\c} S_{\i\d}}$.
Therefore, summing over ${ \{\a,\b\}}$ leads to
\begin{align}
\!\!\! \sum_{\{\a,\b\} \subset \{\j,\k,\l,\m\}} \!\!\!\!\! S_{\a\b} \bigg[ S_{\bv\c} S_{\i\d} + S_{\bv\d} S_{\i\c} \bigg].
    \label{eq:S_6_v_with_c}
\end{align}
We can check that this expression also gives ${ 6 \!\times\! 2 \!=\! 12 }$ terms.
Gathering these results, the term involving $S_{\bv\i}$ gives
\begin{align}
     \half \p_{\i}\!\cdot\! \!\!\int\!\! \rd \bv \sB_{\bv \i} & \big[ \p_{\i} \!-\! \p_{\bv} \big] S_{\bv\i} S_{\j\k\l\m} \!=\! \mLL [S_{2}]_{\i} \, S_{\j\k\l\m} ,
    \label{eq:L_S2_S4}
\end{align}
while the second one in Eq.~\eqref{eq:S_6_v_with_c} gives
\begin{align}
    \half \p_{\i} & \cdot\!\!\int\!\! \rd \bv \sB_{\bv \i} \big[ \p_{\i} \!-\! \p_{\bv} \big] \! \sum_{\{\a,\b\} } \! S_{\a\b} \big[ S_{\bv\c} S_{\d\i} \!+\! S_{\bv\d} S_{\c\i} \big]
    \nonumber
    \\
    = {} & \sum_{\{\a,\b\} } \! S_{\a\b}  \half \p_{\i} \cdot\! \!\!\int\!\! \rd \bv \sB_{\bv \i} \big[ \p_{\i} \!-\! \p_{\bv} \big] \big[ S_{\bv\c} S_{\d\i} \!+\! S_{\bv\d} S_{\c\i} \big] \nonumber \\
    = {} & \sum_{\{\a,\b\} } \! S_{\a\b} \big[ \mQ_{\i;\c\d} + \mQ_{\i;\d\c} \big] \nonumber \\
    = {} & \sum_{\{\a,\b\} } \! S_{\a\b} \mK[S_2]_{\i;\c\d} ,
\end{align}
where, in the last line, we recognized the expression of ${ \mK[S_2] }$ from Eq.~\eqref{eq:def_mN_mM}.
We note that the term ${ \mLL [S_{2}]_{\i} \, S_{\j\k\l\m} }$ in Eq.~\eqref{eq:L_S2_S4} exactly cancels with the one in the second line of Eq.~\eqref{eq:S5_Lyapunov}.
It remains to check that
\begin{align}
    \sum_{\i=\1}^{\5} & \sum_{\{\a,\b\} \subset \{\j,\k,\l,\m\}} \!\!\!\!\!S_{\a\b} \, \mK[S_2]_{\i;\c\d} \nonumber \\
    & {} = \!\!\! \sum_{\{\a,\b\} \subset \{\1, ... ,\5 \}} \sum_{\ip \in \{\c,\d,\e\}} \!\!\!\!\! \mK[S_2]_{\ip;\jp\kp} S_{\a\b} ,
    \label{eq:sum_abi}
\end{align}
in order to match the nonlinear term in the second line of Eq.~\eqref{eq:Leibniz_G5}.
Here we should pay attention not to blindly switch the sum over ${ \{\a,\b\} }$ with the one over $\i$ in the l.h.s.\ term.
Indeed, the pairing ${ \{\a,\b\} }$ depends on $\i$: ${ \{\a,\b\} \!\subset\! \{ \j, \k, \l, \m\} }$. On either side, each term is uniquely labeled by a
partition of ${ \{\1, ..., \5\} }$ of type ${ (1 | 2 | 2) }$. The identification between terms here is bijective: in the l.h.s.\@, we firstly choose $\i$ and then we choose ${ \{\a,\b\} }$ such that ${ \i \!\notin\! \{ \a, \b\} }$; while in the r.h.s.\@, we firstly choose ${ \{\a,\b\} }$ and then choose $\i$ such that ${ \i \!\notin\! \{ \a, \b\} }$. In both cases, the order of summation just reflects the condition ${ \i \!\notin\! \{ \a, \b\} }$. Once ${ \a,\b,\i }$ are fixed, it forces ${ \{ \jp, \kp\} \!=\! \{\c,\d\} \!\subset\! \{\1,...,\5\} \!\setminus\! (\{\a,\b\} \cup \{\i\}) }$.
This shows the identity in Eq.~\eqref{eq:sum_abi}.

Let us now turn to the source terms in the last two lines of Eq.~\eqref{eq:S5_Lyapunov}, namely
\begin{align}
    \sum_{\{\a, \b\} \subset A} C_{\a\b} &\, S_{\c\d\e} + \sum_{\{\a, \b\} \subset A} \bigg\{ \frac{\delta C_{\a\b}}{\delta (F_{\cdot})} \, , S_{\c\d\e\cdot} \bigg\} .
\end{align}
The first one on ${ C[F] }$ matches the one in the last line of Eq.~\eqref{eq:Leibniz_G5}. For the second one, we use the definition of $S_{4}$ in Wick form. This leads to
\begin{align}
    \bigg\{ \frac{\delta C_{\a\b}}{\delta (F_{\cdot})} , S_{\c\d\e\cdot} \bigg\} \!=\! \!\!\int\!\! \rd \bu \, \frac{\delta C_{\a\b}}{\delta (F_{\bu})} \big[ S_{\bu\c} S_{\d\e} \!+\! S_{\bu\d} S_{\c\e} \!+\! S_{\bu\e} S_{\c\d} \big] .
\end{align}
After relabeling the summed pairs, this matches the $\delta C/\delta F$ terms in Eq.~\eqref{eq:Leibniz_G5}.

Collecting all these results, we conclude that $S_{5}$ satisfies exactly Eq.~\eqref{eq:S5_Lyapunov}, once the definitions for $S_{4}$ and $S_{6}$ in Wick form are taken into account. Since it has the same initial condition as $\ST_{5}$ and satisfies the same linear evolution, it follows that
\begin{equation}
    S_{5}(\tau) = \ST_{5}(\tau), \quad \forall \tau \geq 0.
\end{equation}

\section{Closure equation for \texorpdfstring{$\Smin_6$}{S6} and generalization to arbitrary \texorpdfstring{$\Smin_n$}{Sn}}
\label{app:S6_Sn_closure}

\subsection{Closure equation for \texorpdfstring{$\Smin_6$}{S6}}
\label{app:S6_closure}
Following Appendix~\ref{app:S4_closure}, we start from the definition
of $\Smin_6$:
\begin{align}
    \Smin_{\1\2\3\4\5\6} := {} & \Smin_{\1\2} \Smin_{\3\4\5\6} + \Smin_{\1\3} \Smin_{\2\4\5\6} + \Smin_{\1\4} \Smin_{\2\3\5\6}
    \nonumber
    \\
    + \, {} & \Smin_{\1\5} \Smin_{\2\3\4\6} + \Smin_{\1\6} \Smin_{\2\3\4\5} ,
    \label{eq:closure_S6}
\end{align}
where $\Smin_{4}$ is given by Eq.~\eqref{eq:closure_S4}.
Equation~\eqref{eq:closure_S6} is therefore the sum over the 15 pairings of ${ \{\1,...,\6\} }$ of products of $\Smin_2$.
This is Wick's formula.
In particular, the ${ 3 \!+\! 3 }$ partitions, which would give contributions of the form ${ \Smin_3 \Smin_3 }$, are of relative order ${ 1/N }$. They are therefore absent from the leading-order initial condition.
At ${ \tau \!=\! 0 }$, we have ${ \Smin_6 \!=\! \ST_6 }$ (see Appendix~\ref{app:IC_Sn}).

Proceeding as in Appendix~\ref{app:S4_closure}, we compute ${ \p_{\tau} \Smin_6 }$ using the Leibniz rule and the evolution equation for $\Smin_2$ (see Eq.~\ref{eq:Lyapunov_S2}).
The linear contributions recombine into ${ \sum_{\i} \mL_{\i} \Smin_6 }$, since each label appears in exactly one factor $\Smin_2$.
The source contributions, ${ C_{\a\b} }$, multiply the products of the remaining factors.
This sums to ${ \Smin_{\widehat{\a\b}} }$ through Eq.~\eqref{eq:closure_S4}.
Hence, $\Smin_6$ satisfies Eq.~\eqref{eq:Minimal_Lyapunov_2k} with the same initial condition as $\ST_6$, so that ${ \Smin_6 (\tau) \!=\! \ST_6 (\tau) }$ for all ${ \tau \!\geq\! 0 }$.
We stress that this equation only involves $\Smin_2$: contributions of the form ${ \Smin_3 \Smin_3 }$ are never generated by the truncated dynamics.

\subsection{Generalization to arbitrary order}
\label{app:Sn_closure}

In this section, we do not provide a fully rigorous proof for arbitrary $n$.
Rather, we outline the recursive argument, which generalizes the explicit calculations of Appendices~\ref{app:S4_closure}, \ref{app:S5_closure} and~\ref{app:S6_closure} for ${ n \!=\! 4, 5, 6 }$.

For an arbitrary order $n$, the initial conditions given in Eqs.~\eqref{eq:IC_S2k} and~\eqref{eq:IC_S2k+1} satisfy the quasi-Gaussian closure relations.
It remains to show that these relations are preserved by the truncated dynamics (see Eqs.~\ref{eq:Minimal_Lyapunov_2k} and~\ref{eq:Minimal_Lyapunov_2k+1}).
The proof relies on the structure of the truncated hierarchy (Appendix~\ref{app:truncated_hierarchy_dynamical}): the even moments form a closed sub-hierarchy, independent of the odd ones.
We therefore treat the even moments first, and then the odd ones.

\textit{Even moments.}
The equation for $\ST_{2k}$ (Eq.~\ref{eq:Minimal_Lyapunov_2k}) only involves $\ST_{2k-2}$.
Assuming that $\Smin_{2k-2}$, defined from Eq.~\eqref{eq:Closure_2k}, satisfies ${ \Smin_{2k-2} \!=\! \ST_{2k-2} }$, the argument of Appendix~\ref{app:S6_closure} applies unchanged, and we conclude that $\Smin_{2k}$ satisfies Eq.~\eqref{eq:Minimal_Lyapunov_2k}. Hence
${ \Smin_{2k} \!=\! \ST_{2k} }$, and, by induction on $k$, this holds at all orders, independently of the odd moments.

\textit{Odd moments.}
The equation for $\ST_{2k+1}$ (Eq.~\ref{eq:Minimal_Lyapunov_2k+1}) involves the even moments $\ST_{2k+2}$ and $\ST_{2k}$, which we already know are given by the Wick closure (Eq.~\ref{eq:Closure_2k}), and the odd moment $\ST_{2k-1}$.
We therefore proceed by induction on $k$, assuming that ${ \ST_{2k-1} \!=\! \Smin_{2k-1} }$.
Substituting the closures into Eq.~\eqref{eq:Minimal_Lyapunov_2k+1}, and comparing with the time derivative of Eq.~\eqref{eq:Closure_2k+1} computed with the Leibniz rule, the terms match exactly as in Appendix~\ref{app:S5_closure}, which considers the case ${ k \!=\! 2 }$. We conclude ${ \Smin_{2k+1} \!=\! \ST_{2k+1} }$.

\textit{Explicit form.}
We have thus shown that ${ \Smin_n \!=\! \ST_n }$ for all $n$, i.e.\ that the moments of the minimal hierarchy capture the leading order of the moments of the full hierarchy, $\SDK_n$.
Their explicit expression reads
\begin{subequations}
\begin{align}
    \Smin_{\1 ... \mathbf{2k}} {} & = \tfrac{1}{2^k k!} \sum_{\sigma \in \perm_{2k}} \prod_{j=1}^{k} \Smin_{\sigma(\mathbf{2j-1})\sigma(\mathbf{2j})} ,
    \label{eq:closure_2k_explicit} \\
    \Smin_{\1 ... \mathbf{2k+1}} {} & \!\!=\! \tfrac{1}{3! 2^{k-1} (k-1)!} \!\!\!\!\!\!\sum_{\sigma \in \perm_{2k+1} } \!\!\!\!\!\! \Smin_{\sigma(\1)\sigma(\2)\sigma(\3)} \!\prod_{j=2}^{k} \Smin_{\sigma(\mathbf{2j})\sigma(\mathbf{2j+1})} ,
    \label{eq:closure_2k+1_explicit}
\end{align}
\label{eq:closure_explicit}\end{subequations}
where $\perm_n$ is the group of permutations of $n$ elements.
The prefactors compensate for the overcounting generated by the sums over permutations. For the even case, ${ 2^k k! }$ accounts for the orderings within and among the $k$ pairs, while for the odd case, ${ 3!\, 2^{k-1} (k \!-\! 1)! }$ accounts for the orderings within the triplet and within and among the ${ k \!-\! 1 }$ pairs.

\section{Central fluctuation distribution}
\label{app:PDF}

In this Appendix, we derive the probability density from Eq.~\eqref{eq:PDF} using the minimal hierarchy.
Let ${ \Phi[F_N] }$ be a smooth macroscopic one-particle observable (Eq.~\ref{eq:def_observable}). In order to make the large-$N$ ordering explicit, we introduce its rescaled fluctuations ${\delta \tPhi := (\Phi[F_N] \!-\! \langle\Phi[F_N]\rangle)/\sqrt{\bbV[\Phi]}}$.
By construction, ${ \tbbV \!:=\! \bbV [\tPhi] \!=\! 1 }$.
We then introduce the characteristic function
\begin{equation}
    \tchi (q) := \big\langle \re^{\ri q \, \delta\tPhi} \big\rangle ,
\label{eq:def_chi}
\end{equation}
where $\langle\cdot\rangle$ denotes the expectation with respect to the probability law of $F_N$ at some given time $\tau$.
This includes both the random initial conditions and the \LDK\@ noise.
The probability density $\bbP(x)$ of the standardized centered fluctuation is obtained by the inverse Fourier transform of $\tchi$. It reads
\begin{equation}
    \bbP(x) = \!\!\int\!\! \frac{\rd q}{2 \pi} \, \re^{- \ri q x} \tchi(q) .
    \label{eq:from_chi_to_p}
\end{equation}
Expanding Eq.~\eqref{eq:def_chi} in moments gives
\begin{equation}
    \tchi (q) = \sum_{n=0}^{+ \infty} \tfrac{1}{n!} (\ri q)^n \tbbM_n ,
    \label{eq:init_exp_moments}
\end{equation}
where, to shorten the notations, we wrote ${ \tbbM_{n} \!:=\! \bbM_{n} [\tPhi] }$.
We may then substitute the closure relations from Eqs.~\eqref{eq:observable_even_moments} and~\eqref{eq:observable_odd_moments}.
At leading order,
Eq.~\eqref{eq:init_exp_moments} becomes
\begin{equation}
    \tchi (q) \!=\! \sum_{k=0}^{+ \infty} \tfrac{(-1)^k}{2^k k!} q^{2k} \!+\! \tfrac{1}{6} \tbbM_3 \sum_{k=0}^{+ \infty} \tfrac{\ri^{2k+3}}{2^k k!} q^{2k+3}
    \!+\! \mO \big( \tfrac{1}{N} \big),
    \label{eq:chi_closure}
\end{equation}
where we separated the even and odd contributions and used ${ \tbbV \!=\! 1 }$.
Both sums in Eq.~\eqref{eq:chi_closure} can be expressed in terms of the same Gaussian exponential. Indeed,
\begin{equation}
    \tchi (q) = \exp \!\big(\! -\tfrac{1}{2} q^2 \big) \big[ 1 \!-\! \tfrac{1}{6} \ri \tbbM_3 q^3 \big] + \mO \big( \tfrac{1}{N} \big) ,
    \label{eq:chi_phi}
\end{equation}
where ${\tbbM_3 \!=\! \bbM_3 / \bbV^{3/2}}$.
Taking the inverse Fourier transform of the leading terms (following Eq.~\ref{eq:from_chi_to_p}),
we find
\begin{equation}
     \bbP(x) = \tfrac{1}{\sqrt{2\pi}} \exp \!\big(\! -\tfrac{x^{2}}{2} \big) \big[ 1 \!+\! \tfrac{\tbbM_3}{6} \big( x^3 \!-\! 3x \big) \big] + \mO \big( \tfrac{1}{N} \big) .
\end{equation}
This is exactly Eq.~\eqref{eq:PDF}.
This \PDF\@ is correctly normalized to unity but may not be positive in the very far tails of the distributions, especially for small values of $N$.
This is not surprising since we built this \PDF\@ by truncating the expansion at the first non-Gaussian order, whose domain of validity is restricted to the central fluctuation regime.
We also emphasize that since ${ \bbM_3 / \bbV^{3/2} \!=\! \mO(N^{-1/2}) }$, the \PDF\@ of the standardized fluctuation ${ \delta\tPhi }$ converges to the standard Gaussian density as ${ N \!\to\! \infty }$.
This is consistent with the central limit theorem.

\section{Conservation properties}
\label{app:conservation_invariants_proof}

In this Appendix, we show that the minimal hierarchy preserves exactly the collisional invariants ${ \Phi \!\in\! \{ M, \bPtot, \Ekin \} }$ (see Section~\ref{sec:cons_law}), together with all their fluctuation moments.
Following Eq.~\eqref{eq:def_observable}, each invariant is a linear functional of the \DF\@, associated with the one-particle function ${ \phi \!\in\! \{ 1, \bv, \half |\bv|^2 \} }$.

\textit{Structure of the noise.}
The conservation laws constrain the structure of the \LDK\@ noise covariance ${ C[F] }$ (Eq.~\ref{eq:covariance_kernel}).
To make this explicit, we introduce, for an arbitrary \DF\@ $F$, the covariance operator
\begin{equation}
    \mathcal{C}_F[\phi](\bv) := \!\!\int\!\! \rd\bvp\, C[F](\bv,\bvp) \, \phi(\bvp) .
    \label{eq:def_covariance_operator}
\end{equation}
Integrating Eq.~\eqref{eq:covariance_kernel} by parts over ${ \rd \bvp }$, we obtain
\begin{align}
    \mathcal{C}_F[\phi](\bv) = & {} - \Mtot \, \p_{\bv} \!\cdot\! \bigg\{ F(\bv) \!\!\int\!\! \rd \bvp \, F(\bvp) \, \nonumber \\
    & {} \times \sB_{\bv\bvp} \big[ \p_{\bv} \phi \!-\! \p_{\bvp} \phi \big] \bigg\} .
    \label{eq:covariance_operator_explicit}
\end{align}
For the three invariants, this expression vanishes because of the identity
\begin{equation}
    \sB_{\bv\bvp} \big[ \p_{\bv} \phi \!-\! \p_{\bvp} \phi \big] = \bZero .
    \label{eq:projector_identity}
\end{equation}
Indeed, ${ \p_{\bv} \phi \!-\! \p_{\bvp} \phi }$ vanishes for $M$ and $\bPtot$, while it is equal to ${ \bv \!-\! \bvp }$ for $\Ekin$, which is annihilated by the projector property of $\sB$ (see Eq.~\ref{eq:def_sB_generic}).
Hence, for any $F$,
\begin{equation}
    \mathrm{Span} \big\{ 1, \bv, \half |\bv|^2 \big\} \subset \ker \mathcal{C}_F .
    \label{eq:Ker_C}
\end{equation}
Physically, the finite-$N$ noise can neither inject nor dissipate a conserved quantity.
In the context of \LDT\@, Eq.~\eqref{eq:Ker_C} implies that any trajectory ${ F_N(\tau) }$ violating the conservation laws has an infinite action, i.e.\ a vanishing probability~\cite{Bouchet2020}.

\textit{Full hierarchy.}
For the full hierarchy, the conservation of all the moments of $\Phi$ is immediate.
Indeed, since $\Phi$ is linear in $F_N$, there is no It\^o correction, and the \LDK\@ equation (see Eq.~\ref{eq:LDK}) gives
\begin{equation}
    \rd \Phi = \tfrac{1}{\Mtot} \!\!\int\!\! \rd \bv \, \phi \, \mLL[F_N,F_N] \, \rd \tau + \tfrac{1}{\Mtot \sqrt{N}} \!\!\int\!\! \rd \bv \, \phi \, \rd \mW_\tau[F_N] .
\end{equation}
The first term vanishes because the Landau operator conserves $\Phi$~\cite{Chavanis2012homogeneous}, and the second term has zero variance because of Eq.~\eqref{eq:Ker_C}.
As a result, $\Phi$ is conserved along each realization of \LDK\@, and so are all its moments.

\textit{Minimal hierarchy.}
This argument no longer applies to the minimal hierarchy, 
since it results from a truncation and is therefore not the exact moment hierarchy of \LDK.
Conservation must therefore be checked explicitly.
For the mean, Eq.~\eqref{eq:conservation_F} follows from the conservation properties of the Landau equation~\cite{Chavanis2012homogeneous}.

\subsection{Conservation for \texorpdfstring{$\Smin_2$}{S2}}
\label{app:conservation_S2}

We start from Eq.~\eqref{eq:Lyapunov_S2}, the evolution equation for $\Smin_2$.
Contracting it with ${ \phi_{\1}\phi_{\2} }$
and integrating over both velocity variables, we have
\begin{align}
    \frac{\rd}{\rd \tau} \!\!\int\!\! & \rd \1 \rd \2 \, \phi_{\1} \phi_{\2} \, \Smin_{\1\2} \nonumber \\
    & {} = \!\!\int\!\! \rd \1 \rd \2\, \phi_{\1} \phi_{\2} \big\{ \big[ \mL_{\1} \!+\! \mL_{\2} \big] \Smin_{\1\2} + C_{\1\2} \big\} .
\end{align}
The source term vanishes, since ${ \int \! \rd \1 \, \phi_{\1} C_{\1\2} \!=\! \mathcal{C}_F[\phi](\2) \!=\! 0 }$ (see Eq.~\ref{eq:Ker_C}).

For the linear term, we insert the definition of $\mL_{\1}$ (Eq.~\ref{eq:def_mL_1}) and integrate by parts over ${ \rd \1 }$. This gives
\begin{equation}
    \!\int\!\! \rd \1 \rd \2 \, \phi_{\1} \phi_{\2} \, \mL_{\1} \Smin_{\1\2}
    = \half \!\!\int\!\! \rd \1 \rd \2 \rd \bu \, \phi_{\2} \, \p_{\1} \phi \cdot \mA_{\bu\1\2} ,
    \label{eq:mF_p_1O_O_2}
\end{equation}
with
\begin{align}
    \mA_{\bu\1\2} := {} &  -\sB_{\1\bu} \big[ \p_{\1}F \, \Smin_{\bu\2} + F_{\bu} \p_{\1} \Smin_{\1\2} \nonumber \\
    & {} - \p_{\bu}F \, \Smin_{\1\2} - F_{\1} \p_{\bu} \Smin_{\bu\2} \big] .
\end{align}
Since $\mA$ is antisymmetric under the exchange ${ \bu \!\leftrightarrow\! \1 }$, Eq.~\eqref{eq:mF_p_1O_O_2} can be symmetrized as
\begin{equation}
    \tfrac{1}{4} \!\!\int\!\! \rd \1 \rd \2 \rd \bu \, \phi_{\2} \, \big[ \p_{\1} \phi \!-\! \p_{\bu} \phi \big] \!\cdot\! \mA_{\bu\1\2} .
    \label{eq:mF_p_1O_O_2_antisy}
\end{equation}
This term vanishes because of Eq.~\eqref{eq:projector_identity}, since $\mA_{\bu\1\2}$ is proportional to the symmetric kernel $\sB_{\1\bu}$.
The same argument applies to $\mL_{\2}$.
Overall, this establishes Eq.~\eqref{eq:conservation_S_2}.

\subsection{Conservation for \texorpdfstring{$\Smin_3$}{S3}}
\label{app:conservation_S3}

We now contract the kinetic equation for $\Smin_3$ (Eq.~\ref{eq:Lyapunov_S3}) with ${ \phi_{\1} \phi_{\2} \phi_{\3} }$. This gives
\begin{align}
    \frac{\rd}{\rd \tau} & \!\!\int\!\! \rd \1 \rd \2 \rd \3 \, \phi_{\1} \phi_{\2} \phi_{\3} \, \Smin_{\1\2\3}
    \\
    = {} &  \!\!\int\!\! \rd \1 \rd \2 \rd \3 \, \phi_{\1} \phi_{\2} \phi_{\3} \big\{ \big[ \p_{\tau} \Smin_3 \big]_{\rL} \!\!+\!\big[ \p_{\tau} \Smin_3 \big]_{\rNL} \!+\! \big[ \p_{\tau} \Smin_3 \big]_{\rS} \big\}_{\1\2\3} .
    \nonumber
\end{align}
We now treat these three contributions in turn.
In particular, since the contraction is symmetric in ${ (\1,\2,\3) }$, the terms generated by the cyclic permutations $\circlearrowright_{\1\2\3}$ all give the same contribution.
It is sufficient to consider one of them.

\textit{Linear term.}
It vanishes by the same argument as for $\Smin_2$, applied to each of the three variables in turn.

\textit{Source term.}
Using Eq.~\eqref{eq:def_p_tS_3_source}, the contraction of the source term reads
\begin{equation}
    \!\int\!\! \rd \1 \rd \bu \, \phi_{\1} \, \Smin_{\1\bu} \!\!\int\!\! \rd \3 \, \phi_{\3} \bigg[ \!\int\!\! \rd \2 \, \phi_{\2} \, \frac{\delta C_{\2\3}}{\delta F_{\bu}} \bigg] + \circlearrowright_{\1\2\3} .
    \label{eq:source_cons_S3}
\end{equation}
The bracket vanishes identically. Indeed, Eq.~\eqref{eq:Ker_C} states that for any $F$, one has
\begin{equation}
\int \! \rd\bv \, \phi(\bv) \, C[F](\bv,\bvp) = 0 .
\end{equation}
The left-hand side is therefore a functional of $F$ that
vanishes identically, so that its functional derivative with respect to
$F_{\bu}$ vanishes as well. 
Since $\phi$ does not depend on $F$, this derivative
can be taken inside the integral.
This yields
\begin{equation}
\int \! \rd\bv \, \phi(\bv) \,
\frac{\delta C[F](\bv,\bvp)}{\delta F_{\bu}} = 0 .
\end{equation}
This is precisely the term inside the bracket in Eq.~\eqref{eq:source_cons_S3}.

\textit{Nonlinear term.}
To evaluate the contraction of the nonlinear term (Eq.~\ref{eq:NL_S3}), we introduce the partially contracted moment
\begin{equation}
    \psi_{\1} := \!\!\int\!\! \rd \2 \, \phi_{\2} \, \Smin_{\1\2} .
\end{equation}
Using the definition of $\mQ_{\1;\2\3}$ (Eq.~\ref{eq:def_M123}), the contraction of ${ \Smin_{\bu\2} \Smin_{\1\3} }$ with ${ \phi_{\2} \phi_{\3} }$ is ${ \psi_{\bu} \psi_{\1} }$, so that
\begin{equation}
    \!\int\!\! \rd \2 \rd \3 \, \phi_{\2} \phi_{\3} \, \mQ_{\1;\2\3}
    = \half \p_{\1} \!\cdot\! \!\!\int\!\! \rd \bu \, \sB_{\1\bu} \big[ \psi_{\bu} \, \p_{\1} \psi_{\1} - \psi_{\1} \, \p_{\bu} \psi_{\bu} \big] .
\end{equation}
The partner term $\mQ_{\1;\3\2}$ gives exactly the same contribution.
\begin{align}
    \!\int\!\! \rd \1 \rd \2 \rd \3 \, \phi_{\1} \phi_{\2} \phi_{\3} \, \mK_{\1;\2\3} \!=\! - \!\!\int\!\! \rd \1 \rd \bu \, \p_{\1} \phi \cdot \sB_{\1\bu} \, \bJ_{\1\bu} ,
\end{align}
with ${ \bJ_{\1\bu} \!:=\! \psi_{\bu} \, \p_{\1} \psi_{\1} \!-\! \psi_{\1} \, \p_{\bu} \psi_{\bu} }$.
Since ${ \bJ_{\bu\1} \!=\! - \bJ_{\1\bu} }$ and $\sB_{\1\bu}$ is symmetric, the integrand can be symmetrized as
\begin{equation}
    - \half \!\!\int\!\! \rd \1 \rd \bu \, \big[ \p_{\1} \phi \!-\! \p_{\bu} \phi \big] \cdot \sB_{\1\bu} \, \bJ_{\1\bu} .
\end{equation}
Once again, this term vanishes because of Eq.~\eqref{eq:projector_identity}.
Once contracted, the nonlinear term takes the form of a Landau collision operator acting on ${ \psi_{\bu} \psi_{\1} }$.
Hence, it inherits its conservation properties.

Collecting the linear, source and nonlinear contributions, we have proven Eq.~\eqref{eq:conservation_S_3}.

\subsection{Conservation for \texorpdfstring{$\Smin_n$}{Sn}}
\label{app:conservation_Sn}

The conservation of the invariants extends to all higher-order moments through the quasi-Gaussian closure. Indeed, within the minimal hierarchy, for all the invariants ${ \Phi \!\in\! \{ M, \bPtot, \Ekin \} }$, the $n$-th moment is expressed in terms of ${ \bbV[\Phi] }$ and ${ \bbM_3[\Phi] }$ (Eqs.~\ref{eq:observable_even_moments} and~\ref{eq:observable_odd_moments}).
Because ${ \bbV[\Phi] }$ and ${ \bbM_3[\Phi] }$ are separately conserved (Eqs.~\ref{eq:conservation_S_2} and~\ref{eq:conservation_S_3}),
every ${ \bbM_n[\Phi] }$ is also preserved.
The minimal hierarchy therefore preserves the invariants ${ \{ M, \bPtot, \Ekin \} }$ together with all their higher-order fluctuation moments.

\section{Numerical integration of the hierarchy}
\label{app:Lyapunov_numerical_integration}

In this Appendix, we present our numerical implementation of the minimal hierarchy. 
For that, we place ourselves in ${ d \!=\! 2 }$
and discretize each velocity direction, ${ (v_{x} , v_{y}) }$, within the domain 
${ -\VMAX \!\leq\! v_{x} , v_{y} \!\leq\! \VMAX }$
in $\NV$ bins of width ${ \DV \!:=\! 2 \, \VMAX / \NV }$.
Each bin is centered around the location
\begin{equation}
v_{x}^{k} := -\VMAX + (k \!-\! \tfrac{1}{2}) \, \DV
\quad \text{with} \quad
1 \!\leq\! k \!\leq\! \NV ,
\label{eq:pos_bin_Landau}
\end{equation}
and similarly for ${ \{ v_{y}^{l} \}_{1 \leq l \leq \NV} }$.

\subsection{Invariant-preserving numerical scheme}
\label{app:EP_numerical_scheme}

The main issue with integrating the minimal hierarchy comes from the fact that the invariants can be badly preserved. 
As such, it is essential to employ a numerical scheme that preserves these invariants and their $n$-th moments up to machine precision. For this purpose, we take inspiration from~\cite{Strand1994, Castillo+1995}.

As for the mass and momentum, the kinetic energy is linear in $F$.
The exact conservation of $\Ekin$, ${ \rd \Ekin \!=\! 0 }$, reads
\begin{align}
    \rd \Ekin {} & = \bigg[ \! \int_{\Omega} \!\! \rd \bv \, \p_{\tau} F(\bv) \half v^{2} \bigg] \rd \tau
    \nonumber
    \\
    & = \bigg[ \! \int_{\Omega} \!\! \rd \bv \, \half v^{2} \p_{\bv} \!\cdot\! \bFL[F] \bigg] \rd \tau ,
    \label{eq:dE_Landau_flux}
\end{align}
where $\Omega$ is the velocity domain,
and $\bFL$ is the Landau flux (Eq.~\ref{eq:Landau}).
It is defined by
\begin{equation}
    \bFL[F](\bv) := \half \!\!\int\!\! \rd\bvp\,\sB_{\bv\bvp}
    \big[ F_{\bvp} \p_{\bv}F - F_{\bv} \p_{\bvp}F \big] .
\end{equation}
Using the Green--Ostrogradski theorem, the integral from Eq.~\eqref{eq:dE_Landau_flux} reads
\begin{equation}
    \rd \Ekin \!=\! \half \bigg[  \oint_{\p \Omega}\!\! \rd S \, \big( v^2 \bFL[F] \big) \!\cdot\! \mathbf{n} - \!\!\int_{\Omega}\!\! \rd \bv \, \bFL[F] \!\cdot\! \p_{\bv}(v^2) \bigg] \rd \tau ,
    \label{eq:Green-Ostro_dE}
\end{equation}
with ${ \p \Omega }$ the boundary of the velocity domain.
For the continuous problem on an unbounded velocity domain, the boundary contribution (first term in Eq.~\ref{eq:Green-Ostro_dE}) vanishes under suitable decay conditions on $F$ and on the Landau flux.

However, for the truncated computational domain, one cannot rely on this decay. To preserve invariants, the discrete scheme must satisfy an exact discrete integration-by-parts identity (see Eq.~\ref{eq:discrete_IBP}), as we now detail.

We start from the second term of Eq.~\eqref{eq:Green-Ostro_dE} and its discretized form. Evaluated on the grid, the 1D velocity gradient operator becomes an ${\NV \!\times\! \NV }$ matrix, ${ \sG_{1} }$, that encodes the finite difference method used to compute the gradient. In 2D, $\p_{\bv}$ becomes a $2$-tensor ${ \sG \!:=\! (\sG_x, \sG_y) }$, where the components read
\begin{equation}
    \sG_x := \sG_{1} \otimes \sI; \qquad \sG_y := \sI \otimes \sG_{1}.
    \label{eq:constraints_G}
\end{equation}
Our aim here is to construct the operator $\sG$ so as to have exact numerical conservation of the invariants.
This requires two simultaneous properties.

First, the discrete gradient must differentiate the collision invariants exactly. Phrased differently, one must have
\begin{equation}
    \sG \, \1 = \bZero; \qquad \sG \,\bv = \bI; \qquad \sG \, |\bv|^2 = 2\bv ,
    \label{eq:G_invariants}
\end{equation}
where $\1$ denotes the constant function ${\bv \mapsto 1}$.
To devise the 1D gradient matrix $\sG_{1}$, we employ a standard second-order centered finite-difference scheme for the interior grid points. 
In this case, the (infinite) matrix $\sG_{1}$ takes the form
\begin{equation}
    \sG_{1} = \tfrac{1}{2 \, \DV}
    \begin{pmatrix}
        \ddots & \vdots & \vdots & \vdots & \vdots \\
        -1 & 0 & 1 & 0 & \cdots \\
        0 & -1 & 0 & 1 & \cdots \\
        0 & 0 & -1 & 0 & \cdots \\
        \vdots & \vdots & \vdots & \vdots & \ddots
    \end{pmatrix}.
\end{equation}
For any polynomial ${ v \!\mapsto\! f(v) }$ quadratic in $v$ (Eq.~\ref{eq:G_invariants}), for all the interior points, we have
\begin{align}
    \sG_{1} f(v) & {} = \tfrac{1}{2 \, \DV} \big[f(v \!+\! \DV) \!-\! f(v \!-\! \DV) \big] \nonumber \\
    & {} = f^{\prime}(v) \!+\! \tfrac{\DV^{2}}{6} f^{\prime\prime\prime}(v) + \mO(\DV^4)
    \nonumber
    \\
    {} & = f^{\prime} (v) ,
\end{align}
since all the derivatives ${ f^{(n)} }$ with ${n \!>\! 2}$ are zero.
As a consequence, for the interior points, we have
\begin{align}
   \sG_{1} f(v) = f^{\prime}(v) .
\end{align}
We now want to define ${ \sG_{1} f(v) }$
for ${ v \!=\! \pm \VMAX }$ using the values of $f$ inside the grid,
i.e.\ using one-sided finite differences. We start with the expansion
\begin{align}
    f^{\prime}(-\VMAX) = {} & a f(-\VMAX) + b f(-\VMAX \!+\! \DV)
    \nonumber
    \\
    + \, {} & c f(-\VMAX \!+\! 2 \, \DV) ,
\end{align}
where ${a, b}$ and $c$ are (yet) undetermined coefficients.
We can find them by Taylor-expanding ${f(-\VMAX \!+\! \DV)}$ and ${f(-\VMAX \!+\! 2 \, \DV)}$ up to second order in $\DV$.
This is an exact expansion since $f$ is quadratic in $v$ (Eq.~\ref{eq:G_invariants}). 
This leads to 
\begin{align}
     f^{\prime}&(-\VMAX) {} = a f(-\VMAX)
     \\
     & {} + b \left[ f(-\VMAX) \!+\! \DV \, f^{\prime}(-\VMAX) \!+\! \tfrac{\DV^2}{2} f^{\prime\prime}(-\VMAX) \right] 
    \nonumber
     \\
     & {} + c \left[ f(-\VMAX) \!+\! 2 \, \DV f^{\prime}(-\VMAX) \!+\! \tfrac{(2\DV)^2}{2} f^{\prime\prime}(-\VMAX) \right].
     \nonumber
\end{align}
This sum should be exactly equal to ${ f^{\prime}(-\VMAX) }$ in order to eliminate the first- and second-order errors.
This leads to the linear system
\begin{equation}
\begin{cases} 
    \displaystyle a + b + c {} & = 0 ,\\ 
    \displaystyle b \, \DV + 2c \, \DV {} & = 1 , \\ 
    \displaystyle \half b \, \DV^{2} + 2c \, \DV^2 {} & = 0 ,
\end{cases}
\quad \to \quad
\begin{cases} 
    \displaystyle a {} & = - \tfrac{3}{2 \, \DV} , \\ 
    \displaystyle b {} & = \;\;\; \tfrac{2}{\DV} , \\ 
    \displaystyle c {} & = - \tfrac{1}{2 \, \DV} .
\end{cases}
\end{equation}
Overall, the resulting gradient operator takes the form
\begin{equation}
    \sG_{1} \!:=\! \tfrac{1}{2 \, \DV}
    \begin{pmatrix}
        -3 & 4 & -1 & 0 & 0 & \dots & 0 \\
        -1 & 0 & 1 & 0 & 0 & \dots & 0 \\
        0 & -1 & 0 & 1 & 0 & \dots & 0 \\
        \vdots & & \ddots & \ddots & \ddots & & \vdots \\
        0 & \cdots & 0 & -1 & 0 & 1 & 0 \\
        0 & \cdots & 0 & 0 & -1 & 0 & 1 \\
        0 & \cdots & 0 & 0 & 1 & -4 & 3
    \end{pmatrix}.
    \label{eq:G_EP}
\end{equation}
We stress that Eq.~\eqref{eq:G_EP} complies exactly
with Eq.~\eqref{eq:G_invariants}.
Naturally, the present approach is not specific to 2D and extends to arbitrary dimension of velocity.

We now turn to the second requirement stemming from Eq.~\eqref{eq:dE_Landau_flux}.
Indeed, the numerical evolution of $F$ is obtained by applying the discrete divergence $\sD$ to the Landau flux (Eq.~\ref{eq:dE_Landau_flux}).
To ensure the conservation, the discrete divergence operator $\sD$ must be constructed such that it preserves the continuous integration-by-parts identity
\begin{equation}
    \int_{\Omega}\!\!\rd \bv \, \phi_{\bv} \, \p_{\bv}\!\cdot\!\bF_{\bv} = -\!\!\int_{\Omega}\!\! \rd \bv \p_{\bv}\phi_{\bv}\!\cdot\!\bF_{\bv} ,
    \label{eq:continuous_IBP}
\end{equation}
at the grid level.
This discrete integration-by-parts identity transfers the action of the divergence on the flux $\bF$ to that of the gradient on the invariant $\phi$. 
Since $\sG$ differentiates the collision invariants exactly, the symmetries of the Landau flux and the projection property of the collision kernel $\sB$ then imply their conservation up to machine precision.
Following~\cite{Strand1994, Castillo+1995}, to reproduce Eq.~\eqref{eq:continuous_IBP} at the discrete level, we equip scalar  functions $\phi$ and $\psi$ on the 2D grid with the inner product
\begin{equation}
    \langle \phi,\psi\rangle_{\DV} := \DV^2 \, \bphi^{\rT} \bpsi ,
\end{equation}
with the vector ${ \bphi \!:=\! \{ \phi (\bv) \}_{\bv} }$ and similarly for $\bpsi$.
The discrete counterpart of Eq.~\eqref{eq:continuous_IBP} is then
\begin{equation}
    \left\langle \phi,\sD\!\cdot\!\bF \right\rangle_{\DV} = -\left\langle \sG\phi,\bF \right\rangle_{\DV} .
    \label{eq:discrete_IBP}
\end{equation}
This is required to hold for any $\phi$ and $\bF$.
Therefore, the 1D divergence operator must be chosen as the negative adjoint of the 1D gradient~\cite{Castillo+1995}, namely
\begin{equation}
    \sD_{1} := - \sG_{1}^{\rT} .
    \label{eq:D_vs_G}
\end{equation}
We emphasize that even with ${ \sD \!=\! -\sG^{\rT} }$, if $\sG$ does not differentiate quadratic polynomials exactly, as guaranteed by Eq.~\eqref{eq:G_EP}, invariants will not be conserved to machine precision.

Imposing Eq.~\eqref{eq:D_vs_G}, the conservation of the invariants holds up to round-off errors, independently of $\VMAX$.
As such, the truncation of the velocity domain only affects the accuracy of the solution, through the neglected tails of $F$ and the reduced accuracy of the discrete operators.
We checked that this effect is negligible by increasing $\VMAX$ from ${ \VMAX \!=\! 6 \, v_{0} }$ to ${ \VMAX \!=\! 7 \, v_{0} }$ at fixed $\DV$.
This produced relative changes smaller than ${ 0.2\% }$ in the observable $T_x$.

\subsection{Landau equation}
\label{app:Landau}

We now detail the numerical implementation of the Landau equation (Eq.~\ref{eq:Landau}).
The mean \DF\@ $F$, since it is a one-variable function, is represented on the grid of ${ \NV^{2} }$ elements from Eq.~\eqref{eq:pos_bin_Landau}.
To compute the rate of change ${ \p_{\tau} F }$
from Eq.~\eqref{eq:Landau},
we proceed in four successive steps.

First, we estimate the gradient ${ \p_{\bv} F }$
using the finite-difference stencil from Eq.~\eqref{eq:G_EP}.
Second, we discretize the integral over ${ \rd \bvp }$
appearing in Eq.~\eqref{eq:Landau_operator} as a Riemann sum on the grid.
Since $\sB$ only depends on ${ \bv \!-\! \bvp }$, this sum is a discrete convolution,
which we evaluate with FFTs (see below).
Third,
we compute the divergence of the resulting flux w.r.t.\ $\bv$,
using the discretized divergence operator from Eq.~\eqref{eq:D_vs_G}.
This gives us the rate of change ${ \p_{\tau} F }$ for every bin.
Finally, once all the rates of change have been estimated,
we integrate the Landau equation (Eq.~\ref{eq:Landau})
forward in time using the first-order explicit Euler method.
Given a time step $\DT$, we perform
\begin{equation}
F (\tau \!+\! \DT) = F (\tau) + \DT \, \p_{\tau} F .
\label{eq:integration_scheme_Landau}
\end{equation}
Because the Landau equation is a nonlinear diffusion equation, Eq.~\eqref{eq:integration_scheme_Landau} is subject to the standard stability constraint ${ \DT \!\lesssim\! (\DV)^2/\max(\sDL) }$, with $\sDL$ the Landau diffusion tensor~\cite{Morton+2005}.
Since the equation is integrated in the macroscopic time ${ \tau \!=\! t/N }$, this constraint is independent of $N$ at fixed $\Mtot$ and $G$.

The computation of the collision integral from Eq.~\eqref{eq:Landau} is the main bottleneck of the calculation.
For each grid point $\bv$,
it requires the computation of an integral over all other grid points $\bvp$.
These integrals are convolutions of the generic form
\begin{equation}
\!\!\int\!\! \rd \bvp \, \sB(\bv \!-\! \bvp) \, g(\bvp) ,
\label{eq:conv_generic_F}
\end{equation}
with $g$ standing either for $F$ or its gradient ${ \p_{\bv} F }$.
A naive, direct evaluation of Eq.~\eqref{eq:conv_generic_F}
for all grid points requires ${ \mO (\NV^{4}) }$ operations.
Fortunately, such an integral can be accelerated
using Fast Fourier Transforms (FFTs),
so that each convolution reduces to a simple product in Fourier space,
followed by an inverse transform. 
To avoid pollution from periodization,
we also add zero-padding.
Overall, this lowers the complexity to ${ \mO (\NV^{2} \ln \NV) }$.
In practice, we precompute once and for all
the Fourier transform of each independent component
of the kernel,
so that only the transforms of $F$ and ${ \p_{\bv} F }$
need to be recomputed at each timestep.
The discretization error of this scheme scales like ${ \mO(\NV^{-2}) }$ (Fig.~\ref{fig:cv_NV}).

\subsection{Kinetic equation for \texorpdfstring{$S_2$}{S2}}
\label{app:S2_equation}

We now turn to the integration of the kinetic equation for $S_2$
(Eq.~\ref{eq:Lyapunov_S2}).
The two-point moment ${ S_2(\bv, \bvp) }$
is represented on the two-point grid ${ (\bv, \bvp) }$,
i.e.\ on an array of size ${ \NV^{2} \!\times\! \NV^{2} }$.
Its evolution is coupled to that of the mean \DF\@ $F$, so that the integrations of $F$ and $S_{2}$
must be performed simultaneously.

The rate of change ${ \p_{\tau} S_2 }$ (Eq.~\ref{eq:Lyapunov_S2})
splits into a linear transport part,
${ (\mL_{\1} \!+\! \mL_{\2})[S_2] }$, and a source term, ${ C[F] }$.
To evaluate the linearized operator,
we note that $\mL_{\1}$ involves,
for each fixed second argument $\bvp$,
a convolution integral
of the same form as in Eq.~\eqref{eq:conv_generic_F},
where $g$ now stands for a slice ${ \bv \!\mapsto\! S_2(\bv, \bvp) }$
of the moment (or its gradient).
Since the convolution of $\sB$ with $F$ or its gradients
does not depend on $\bvp$,
we precompute these convolutions once per timestep, before looping over the slices of $S_2$.
The source term ${ C[F] }$ (Eq.~\ref{eq:covariance_kernel})
is evaluated directly on the grid.
In that definition,
the Dirac distribution is represented by
\begin{equation}
\deltaD(\bv_i\!-\!\bv_j) \longrightarrow \delta_{ij} \,/\, \DV^2,
\end{equation}
so that its discrete integral equals unity.
Once the integrand is assembled,
we take its double divergence following Eq.~\eqref{eq:D_vs_G}.
Finally,
we advance simultaneously $F$ and $S_{2}$ forward in time,
using Eq.~\eqref{eq:integration_scheme_Landau}.
As for the Landau equation, the evolution equation for $S_{2}$ is subject to a stability constraint similar to the one for $F$. 
Overall, the discretization error scales as ${ \NV^{-2} }$ (see Fig.~\ref{fig:cv_NV}).

\subsection{Kinetic equation for \texorpdfstring{$S_3$}{S3}}
\label{app:S3_equation}

The integration of the kinetic equation for $S_3$
(Eq.~\ref{eq:Lyapunov_S3})
follows the same overall strategy,
now on the three-point grid ${ (\1, \2, \3) }$,
i.e.\ on an array of size ${ \NV^{2} \!\times\! \NV^{2} \!\times\! \NV^{2} }$.
The three fields ${ (F, S_2, S_3) }$ are advanced simultaneously,
since the rate of change ${ \p_{\tau} S_3 }$
depends on the instantaneous values of both $F$ and $S_2$.
In essence, the right-hand side of Eq.~\eqref{eq:Lyapunov_S3} is similar to the one for the evolution of $S_{2}$.
As a result, we resort to the exact same integration method as for $S_{2}$: (i) convolutions over velocity space are performed with FFTs; (ii) gradients and divergences are computed with the operators from Appendix~\ref{app:EP_numerical_scheme}.
As for $S_2$, the discretization error scales as ${ \NV^{-2} }$.
Finally, we exploit the permutation symmetry of the three-point moment, ${ S_{\1\2\3} \!=\! S_{\2\1\3} \!=\! S_{\1\3\2} \!=\! \dots }$, to avoid redundant evaluations and speed up the computation of the right-hand side.
However, the full array is stored in memory, so that this symmetry does not reduce the memory footprint.

\subsection{Numerical convergence}
\label{app:Hierarchy_convergence}

The hierarchy predictions presented in Fig.~\ref{fig:money_plot} were obtained through three separate integrations. 
First, the mean quantities were computed by integrating $F$ alone using ${ \VMAX \!=\! 6 \, v_0 }$, ${ \NV \!=\! 128 }$, and ${ \DT \!=\! 5 \!\times\! 10^{-3} \Trelax}$. 
Second, the variances were evaluated via a coupled integration of ${ (F,S_2) }$ using the same domain and resolution (${ \VMAX \!=\! 6 \, v_0 }$, ${ \NV \!=\! 128 }$), but with a finer time step ${ \DT \!=\! 2.5 \times 10^{-3} \Trelax}$. Finally, the third moments were obtained from a coupled integration of the ${ (F,S_2,S_3) }$ fields using ${ \VMAX \!=\! 6 \, v_0 }$,  ${ \NV \!=\! 50 }$, and ${ \DT \!=\! 4 \times 10^{-2} \Trelax }$. 
Due to memory constraints, the integration of $S_3$ is restricted to ${ \NV \!=\! 50 }$, for which the total memory footprint reaches approximately $460$ GB.

\begin{figure}[htbp!]
\centering
\includegraphics[width=0.45\textwidth]{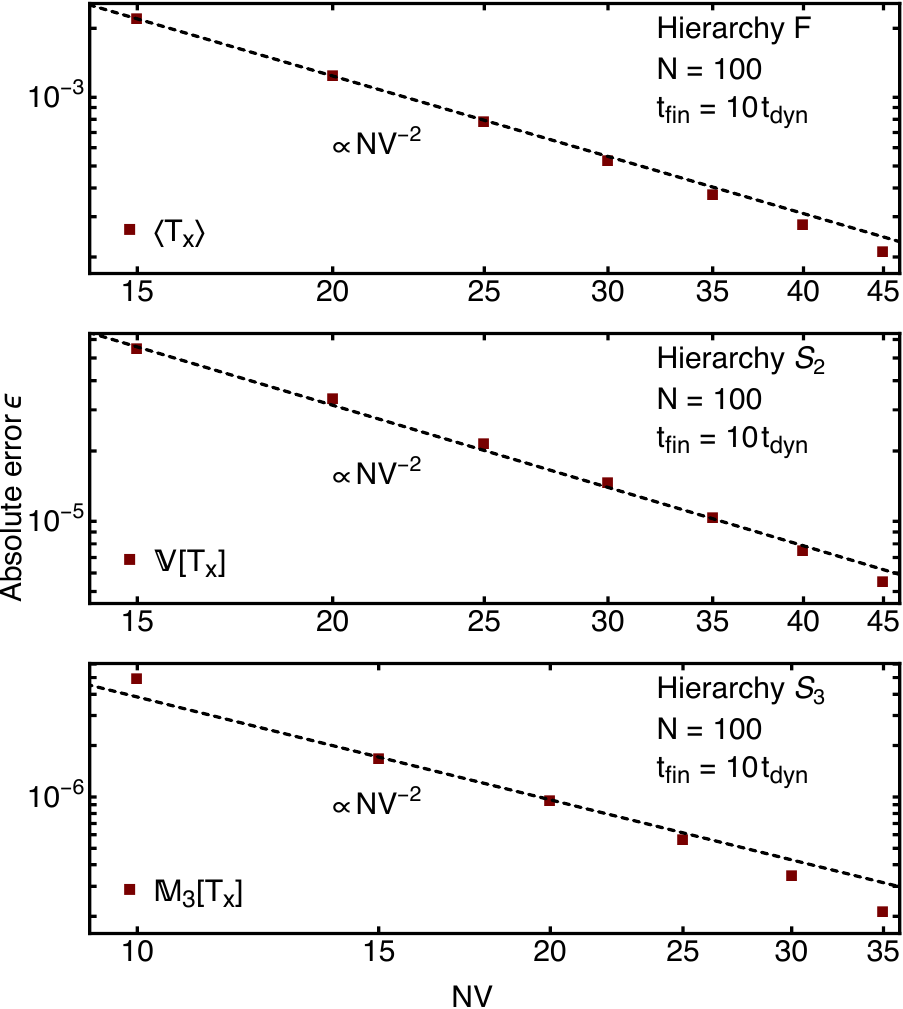}
\caption{
Absolute error in
${ \langle T_x \rangle }$ (top), ${ \bbV[T_x] }$ (middle), and ${ \bbM_{3} [T_x] }$ (bottom) evaluated at ${ \Tfin \!=\! 0.1 \, \Trelax }$. Absolute errors are computed w.r.t.\ the calculation with ${ \NVref \!=\! 128 }$ for $F$, $S_2$, and ${ \NVref \!=\! 55 }$ for $S_3$. 
For this test, all calculations use ${ N \!=\! 100 }$ and ${ \DT \!=\! 5 \times 10^{-3} \Trelax }$. The finite difference scheme from Eq.~\eqref{eq:G_EP} is second-order accurate.
}
\label{fig:cv_NV}
\end{figure}

We now assess the numerical convergence of the integration of $F$, $S_2$, and $S_3$ by monitoring ${ \langle T_x \rangle }$, ${ \bbV[T_x] }$, and ${ \bbM_3[T_x] }$ respectively. To analyze the convergence with $\NV$, each of these quantities, which we denote by $A$, is computed for several values of $\NV$ at fixed $\DT$ and $\Tfin$. We then measure the scaling in $\NV$ of the absolute error 
\begin{equation}
    \epsilon(\NV) := \big| A(\NV) \!-\! A (\NVref) \big| ,
    \label{eq:error_NV}
\end{equation}
with $\NVref$ some large reference value of $\NV$.
As shown in Fig.~\ref{fig:cv_NV}, the measured errors scale as
${ \epsilon(\NV) \!\thicksim\! \NV^{-2} }$.

\begin{figure}[htbp!]
\centering
\includegraphics[width=0.45\textwidth]{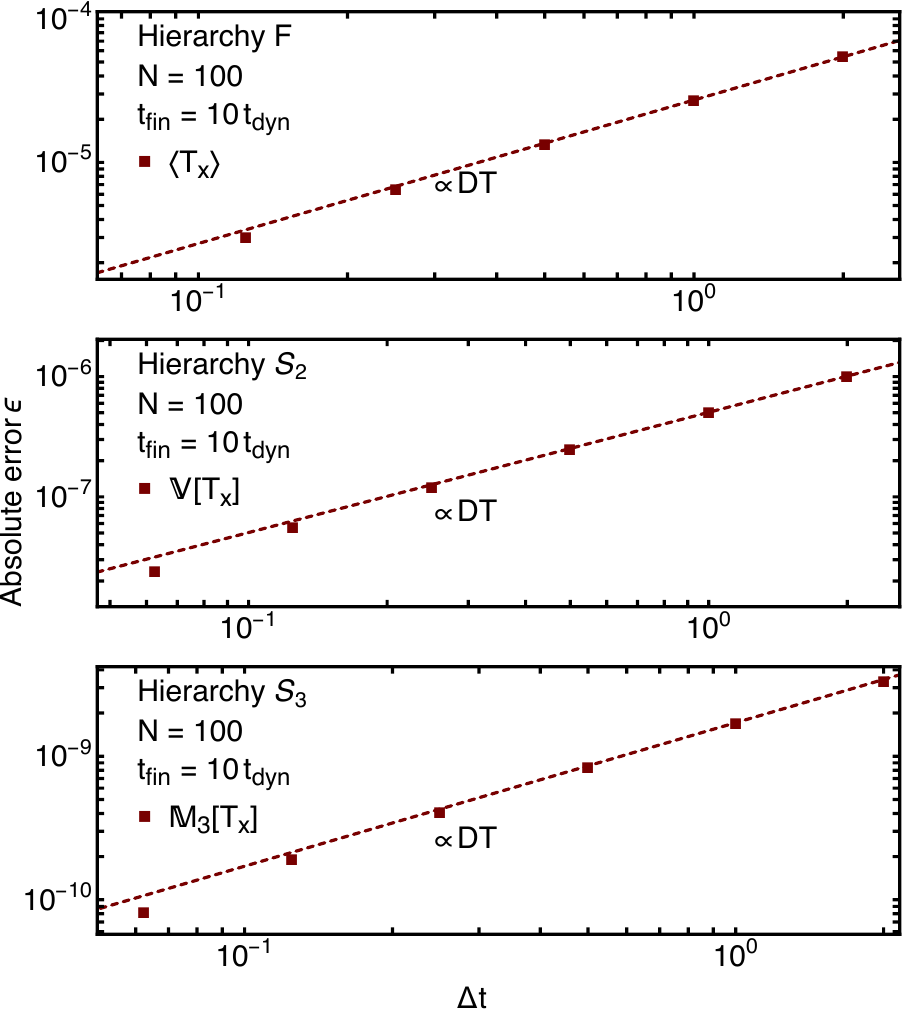}
\caption{
Same as Fig.~\ref{fig:cv_NV},
but for the integration timestep $\DT$. 
Absolute errors are computed w.r.t.\ ${ \DTref \!=\! 2^{-6} \times 10^{-2} \, \Trelax }$.
For this test, we use ${ N \!=\! 100 }$, with ${ \NV \!=\! 64, 25, 14 }$, for $F$, $S_{2}$, and $S_{3}$ respectively.
As expected, the integration scheme from Eq.~\eqref{eq:integration_scheme_Landau}
is first-order accurate in time.
}
\label{fig:cv_DT}
\end{figure}

For the time-step convergence, each of these quantities is computed for several values of $\DT$ at fixed $\NV$ and $\Tfin$. We then measure the absolute error as
\begin{equation}
    \epsilon(\DT)
    := \left| A (\DT) \!-\! A (\DTref) \right| ,
    \label{eq:error_DT}
\end{equation}
with $\DTref$ some small reference timestep.
As shown in Fig.~\ref{fig:cv_DT}, the measured errors scale linearly with $\DT$.
This is consistent with the first-order accuracy of the Euler integration scheme from Eq.~\eqref{eq:integration_scheme_Landau}.

\subsection{Conservation of the invariants}
\label{app:invariants_nbody}

\begin{figure*}[htbp!]
\includegraphics[width=0.95 \textwidth]{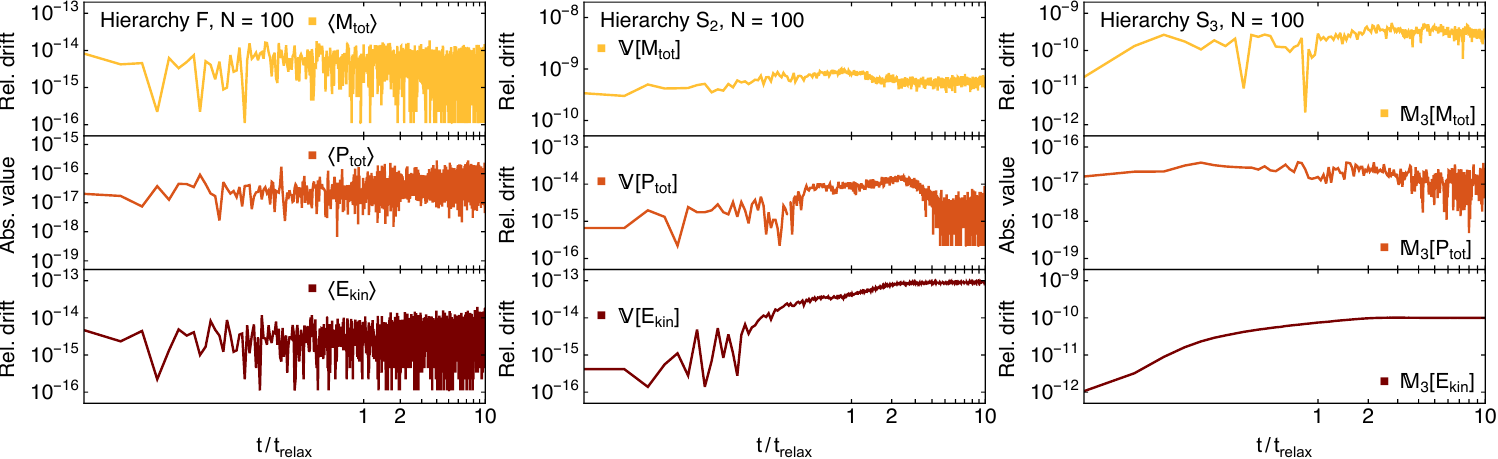}
\caption{Error on the (i) average (left panel), (ii) variance (middle panel) and (iii) third moment (right panel) of each invariant (from top to bottom, mass, total momentum and total kinetic energy). 
All simulations were performed for ${ N \!=\! 100 }$ using the same parameters as in Fig.~\ref{fig:money_plot}.
Each invariant is very well conserved by the hierarchy.
}
\label{fig:invariants}
\end{figure*}

In Fig.~\ref{fig:invariants}, we explore the conservation of the average, variance and third moment of the invariants $M$, $\bPtot$ and $\Ekin$, during the numerical integration of the minimal hierarchy.

In this figure, we note that the mean value of all the invariants is conserved at machine precision, while the variance and third moment are conserved up to $ 10^{-10}$.
We also checked that this is due to the accumulation of round-off errors when performing the numerical integrations. For example, for $S_{3}$ with ${ \NV \!=\! 50 }$, the numerical scheme computes a sum of more than ${ (\NV^2)^3 \!\thicksim\! 10^{10}}$ terms. Assuming that the round-off errors accumulate randomly, the total effective error scales as ${10^{-16} \!\times\! \sqrt{(\NV^2)^3}} \!\thicksim\! 10^{-11} $. This is roughly consistent with what is observed numerically. We checked that this error decreases when $\NV$ is reduced. To avoid this numerical pollution, one could possibly use more precise summation algorithms, such as Kahan summation, at the price of a (slight) increase in the numerical cost~\citep[see, e.g.\@, section~{VIII.5} in][]{Hairer+2006}.

In Fig.~\ref{fig:invariants}, one can also note that compared to $M$ and $\Ekin$, $\bPtot$ is conserved up to machine precision. 
A possible explanation is that, for $M$ and $\Ekin$, the triple sum involves large terms which must cancel each other, so that the accumulated round-off error is proportional to their magnitude.
Conversely, for $\bPtot$, the integrand is odd under the parity transformation ${ \bv \!\to\! -\bv }$.
This is exactly preserved by our symmetric velocity grid.
The contributions of symmetric grid points therefore cancel pairwise, and, in essence, no round-off error accumulates.

\section{\texorpdfstring{$N$}{N}-body simulations}
\label{app:Parameters_numerical_convergence}

\textit{$N$-body simulations}.
To generate the plots in Fig.~\ref{fig:money_plot}, we performed 256\,000 independent realizations of the $N$-body simulations for systems of ${N \!=\! 25, 50, 100}$, and 200 particles.
For this purpose, we use the exact same $N$-body numerical code as in~\cite{Elrhirhayi+2026}.
Each simulation was integrated up to a final time ${ \Tfin \!=\! 10 \, \Trelax }$ using a time step ${\Delta t \!=\! 0.005 \, \Tdyn}$.

\textit{Initial conditions}. 
At ${ t \!=\! 0 }$,
particles are drawn independently from one another
with a uniform angle $\bT$.
Velocities are drawn from the anisotropic Maxwell distribution
\begin{equation}
F(\bv; t \!=\!0) = \frac{\Mtot / 2 \pi}{(T_{x}^{\ri} T_{y}^{\ri})^{1/2}} \exp \bigg[- \frac{| v_{x} |^2}{2 T_{x}^{\ri}} -\frac{|v_{y} |^2}{2 T_{y}^{\ri}} \bigg] ,
\end{equation}
with the initial temperatures ${ (T^{\ri}_{x} , T^{\ri}_{y}) \!=\! (1.5 \, T_{0} , 0.5 \, T_{0}) }$.
Here, ${ T_{0} \!=\! v_0^2 }$ where $v_0$ represents the typical particle velocity at equilibrium.
Two-body relaxation drives this system towards the thermodynamic equilibrium with ${ T_x \!=\! T_y \!=\! T_{0} }$.

\textit{Bootstrap}. To estimate uncertainties in Fig.~\ref{fig:money_plot}, we use bootstraps.
More precisely, for each value of $N$, we perform the ensemble average over a sample of 256\,000 realizations,
drawn with repetitions. We perform this resampling 1\,000 times
to estimate the uncertainties on our measurements of the average, variance and higher-order moments.

\textit{Panel-wise rescaling}.
In each panel of Fig.~\ref{fig:money_plot}, all curves are jointly rescaled using a panel-specific affine transformation.
This mapping ensures that the overall minimum is mapped to 0.1 and the overall maximum to 0.9. Consequently, all displayed curves lie within the interval ${ [0.1,0.9] }$.

\section{Additional validations}
\label{app:additional_validations}

Instead of looking at the temperature $T_x$ (Fig.~\ref{fig:money_plot}), one can examine other observables.
In this Appendix, we consider the non-polynomial observable $\Phi_{\rm NP}$ (Eq.~\ref{eq:def_observable}), associated with the one-particle function
\begin{equation}
    \phi_{\rm NP} (\bv) = \tanh \big[ (v_x / v_0)^2 - (v_y / v_0)^2 \big] .
    \label{eq:NP_obs}
\end{equation}
We chose this particular observable because (i) $\phi_{\rm NP}$ is a non-polynomial function of $\bv$; (ii) unlike $T_x$ or $T_y$, it is not directly related to a collision invariant; and (iii) it probes not only the temperature anisotropy ${ T_x \!-\! T_y }$, but also higher-order anisotropies of the velocity distribution.
For these reasons, this non-polynomial observable provides a stringent test of the minimal hierarchy.

\begin{figure*}[htbp!]
\centering
\includegraphics[width=0.95 \textwidth]{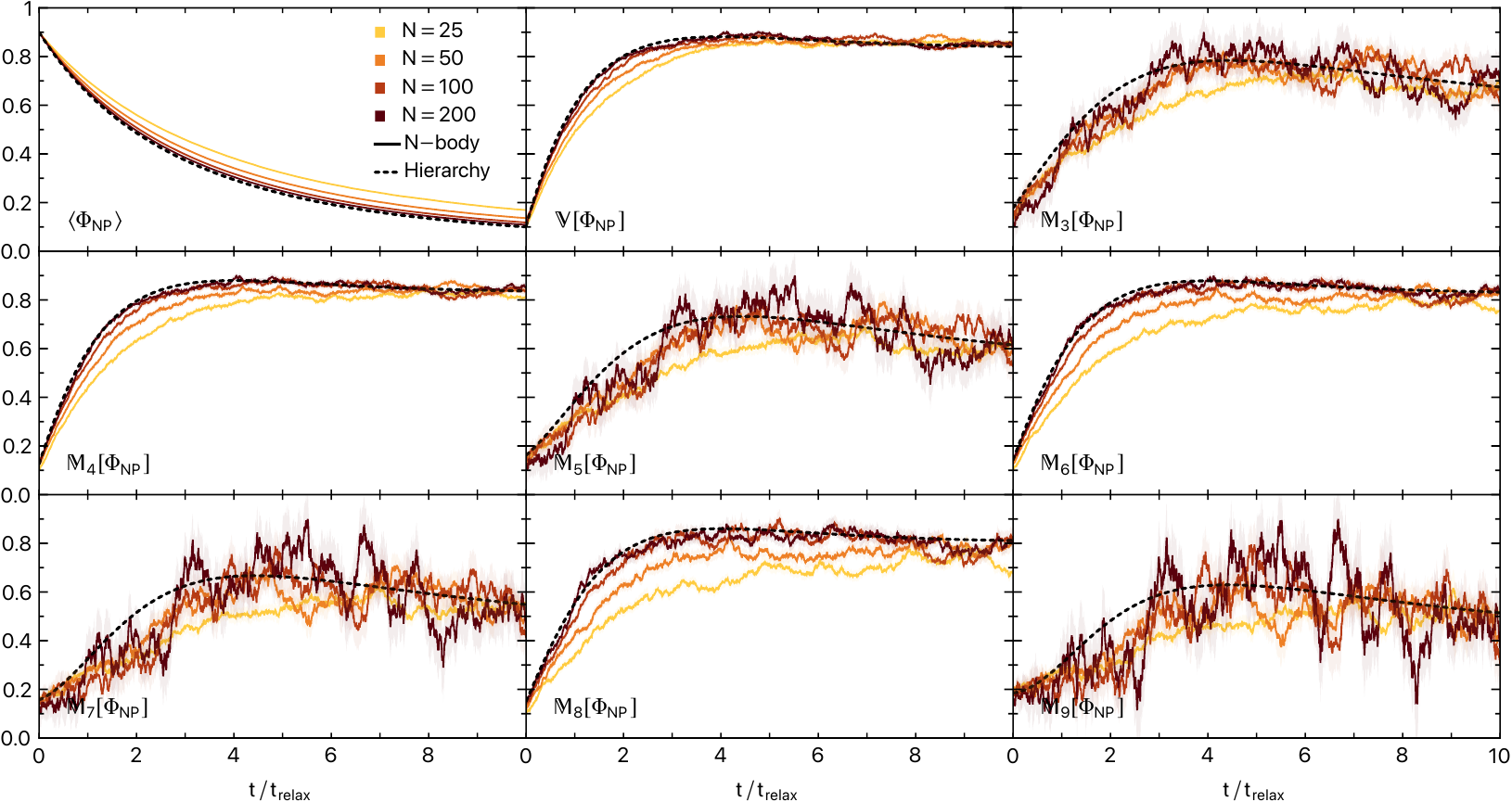}
\caption{Same as Fig.~\ref{fig:money_plot},
but for the non-polynomial observable $\Phi_{\rm NP}$ (Eq.~\ref{eq:NP_obs}).
As $N$ increases, the $N$-body measurements converge toward the minimal hierarchy solution.
}
\label{fig:NP_OBS_M_n}
\end{figure*}

In Fig.~\ref{fig:NP_OBS_M_n}, we illustrate the relaxation of the average and higher moments of $\Phi_{\rm NP}$, as predicted by the minimal hierarchy and measured in the exact same $N$-body simulations.
Compared to Fig.~\ref{fig:money_plot}, the $N$-body measurements here are noisier and their convergence toward the minimal hierarchy solution as $N$ increases is less visible. 
This is expected since $\Phi_{\rm NP}$ probes anisotropies of the velocity distribution beyond the quadratic anisotropy probed by $T_x$.
Overall, we obtain satisfactory agreement.

\end{document}